\documentclass[11pt]{article}

\usepackage{slashed,amssymb}

\newcommand{\beq}{\begin{equation}}
\newcommand{\eeq}{\end{equation}}
\newcommand{\ul}{\underline}

\newcommand{\eps}{{\epsilon}}

\newcommand{\hc}{\mathrm{h.c.}}

\newcommand{\Tr}{\mathrm{Tr }}

\newcommand{\CD}{\mathcal{D}}
\newcommand{\BCD}{\mathcal{\bar D}}

\newcommand{\CP}{{\mathcal P}}
\newcommand{\ACP}{{\bar {\mathcal P}}}

\newcommand{\bsigma}{\bar{\sigma}}
\newcommand{\dalpha}{{\dot{\alpha}}}
\newcommand{\dbeta}{{\dot{\beta}}}
\newcommand{\dgamma}{{\dot{\gamma}}}

\newcommand{\dmu}{{\dot{\mu}}}

\newcommand{\btheta}{{\bar\theta}}

\newcommand{\chE}{\mathcal E}
\newcommand{\chz}{\mathfrak z}

\newcommand{\ch}{\mathcal }

\newcommand{\W}{{\mathcal W}}

\newcommand{\bi}{{\bar i}}
\newcommand{\bj}{{\bar j}}

\newcommand{\chM}{{\mathcal M}}
\newcommand{\chN}{{\mathcal N}}

\newcommand{\eol}{\notag \\}

\newcommand{\ket}[1]{\left| #1 \right. \rangle}
\newcommand{\bra}[1]{\left. \langle #1 \right|}
\newcommand{\braket}[2]{\langle #1 | #2 \rangle}
\newcommand{\opbraket}[3]{\langle #1 |#2| #3 \rangle}

\usepackage{amsmath}
\usepackage{amsfonts}
\usepackage{wasysym}
\usepackage{graphicx}

\numberwithin{equation}{section}

\begin{document}
\thispagestyle{empty}


\hfill UCB-PTH-09/34

\hfill arXiv:0911.5426 [hep-th]

\hfill December 1, 2009

\addvspace{45pt}

\begin{center}

\Large{\textbf{One loop divergences and anomalies from chiral
superfields in supergravity}}
\\[35pt]
\large{Daniel Butter}
\\[10pt]
\textit{Department of Physics, University of California, Berkeley}
\\ \textit{and}
\\ \textit{Theoretical Physics Group, Lawrence Berkeley National Laboratory}
\\ \textit{Berkeley, CA 94720, USA}
\\[10pt] 
dbutter@berkeley.edu
\end{center}

\addvspace{35pt}

\begin{abstract}
\noindent
We apply the heat kernel method (using Avramidi's non-recursive technique)
to the study of the effective action of chiral matter in a complex representation
of an arbitrary gauge sector coupled to background $U(1)$ supergravity.
This generalizes previous methods, which restricted to 1) real
representations of the gauge sector in traditional Poincar\'e supergravity or 2) vanishing
supergravity background. In this new scheme, we identify a classical ambiguity in these
theories which mixes the supergravity $U(1)$ with the gauge $U(1)$. At the quantum
level, this ambiguity is maintained since the effective action changes only by a local
counterterm as one shifts a $U(1)$ factor between the supergravity and gauge sectors.

An immediate application of our formalism is the calculation of the one-loop
gauge, K\"ahler, and reparametrization anomalies of chiral matter coupled to
minimal supergravity from purely chiral loops. Our approach
gives an anomaly whose covariant part is both manifestly supersymmetric and
non-perturbative in the K\"ahler potential.
%
%
\end{abstract}

\setcounter{tocdepth}{2}
\newpage
\setcounter{page}{1}

\section{Introduction}
As is well known \cite{wb}, the most straightforward kinetic coupling of chiral
superfields to (old) minimal supergravity involves an exponential factor involving
the K\"ahler potential in the form
\begin{align}
S = -\frac{3}{\kappa^2} \int d^8z \, E\, e^{-\kappa^2 K/3}
\end{align}
$1/\kappa^2$ is the reduced Planck mass and in the limit of
$\kappa^2 \rightarrow 0$ and the decoupling of supergravity, the
globally supersymmetric K\"ahler term is restored with the familiar
K\"ahler invariance of
\begin{align}
K \rightarrow K + F + \bar F
\end{align}
In the locally supersymmetric case, the action is invariant under
a certain combination of K\"ahler and super-Weyl transformations under
which the determinant $E$ of the supervierbein transforms counter to the
K\"ahler potential. However, this coupling of $K$ yields a noncanonical
Einstein-Hilbert term which must be fixed either by a complicated
component-level rescaling of the various supergravity fields \cite{wb}, or via
the reformulation of the geometry of superspace to the so-called
K\"ahler superspace formulation \cite{bgg}.

In either formulation, calculating the effective action for chiral matter
coupled to supergravity in superspace itself (thus maintaining manifest
supersymmetry) is a difficult task. The K\"ahler formulation,
while being more elegant for classical calculations, makes the origin
of the supersymmetric form of the K\"ahler anomaly unclear \cite{ovrut_cardoso},
as it undoubtedly becomes intertwined with conformal transformations.
On the other hand, calculating in the original formulation
(as advocated in \cite{ovrut_cardoso}) is clearly an inelegant task.

In this paper, we advocate an alternative route. In a previous
work \cite{Butter:2009cp}, we have introduced the formulation of conformal
superspace, which encodes the superconformal algebra in the structure group
of superspace, thus realizing the tensor calculus method of
Kugo and Uehara \cite{Kugo} on an equal footing with other superspace
approaches. In this formulation, the original action would be written
\begin{align}\label{eq_action1}
S = -\frac{3}{\kappa^2} \int d^8z \, E\, \Phi_0 \bar\Phi_0 e^{-\kappa^2 K/3}
\end{align}
where $\Phi_0$ is the conformal compensator, originally introduced in
\cite{Kugo_comp} at the level of the tensor calculus.
As is well known, the original Poincar\'e formulation is found by the gauge choice $\Phi_0 = 1$
while the K\"ahler formulation is found by the choice $\Phi_0 = e^{\kappa^2 K / 6}$.
The original K\"ahler symmetry in the conformal formulation is then a classical symmetry of
the action provided we also transform
\begin{gather}
\Phi_0 \rightarrow \Phi_0 e^{\kappa^2 F / 6}
\end{gather}
We have subsequently shown \cite{paper3} how to expand generic actions
coupling supergravity, super Yang-Mills, and chiral matter to quadratic
order in quantum superfields in order to enable the calculation of one loop
effects in arbitrary locally supersymmetric models in superspace.

As a first step toward that result, in this paper we will formally construct
the one loop effective action from all chiral loops\footnote{We include the
conformal compensator but exclude any chiral fields that may (and will)
be introduced by the gauge-fixing procedure in the supergravity and super
Yang-Mills sectors.}. Our approach to the calculation is not a new one,
but constitutes a generalization and combination of 
two classic papers by McArthur \cite{McArthur} and one by
Buchbinder and Kuzenko \cite{Buchbinder:1986im}
calculating heat kernel coefficients in a Poincar\'e supergravity background\footnotemark
and another by McArthur and Osborn \cite{McArthur:1985xd} about calculating
anomalies in supersymmetric gauge theories.

\footnotetext{McArthur worked in normal coordinates, which is the
approach we will take in order to most easily apply Avramidi's non-recursive
method. Buchbinder and Kuzenko worked in a generally covariant fashion and
necessarily identified more of the interesting features of the supergeometry.
See for example their followup paper \cite{Buchbinder:1988yu} where the anomaly
term was integrated.}

This paper is divided into three sections. For the sake of
providing a self-contained description with consistent notation
throughout, we briefly review in the first section the various methods we will
use in their conventional field theory context:
the heat kernel approach, specifically a non-recursive method invented by
Avramidi \cite{avramidi}, here simplified to a normal coordinate system;
and the perturbative approach of Leutwyler \cite{Leutwyler:1985em}
for dealing with the Dirac operator in a complex representation.

In the second section, we consider the general case of chiral superfields
coupled to arbitrary background supergravity and super Yang-Mills.
The results are similar to those found in
\cite{McArthur, McArthur:1985xd},
except for the change from Poincar\'e superspace to $U(1)$
superspace, which as we have shown in \cite{Butter:2009cp},
can be understood as a gauge-fixed version of conformal superspace.

In the third section,  we apply the chiral loop calculation to
the action \eqref{eq_action1} with the addition of a superpotential term.
We find the covariant form of the reparametrization, K\"ahler, and
gauge anomalies in a form which is non-perturbative in the K\"ahler
potential, thus expanding the well-known results of \cite{ovrut_cardoso}
which restricted to a limited set of these anomalies.
The remaining non-covariant part will be dependent on the
precise choice of the definition of the effective action, and should
presumably be fixed by details of the actual UV completion of the
theory.

\section{Review: Heat kernel techniques for component fields}
\subsection{Heat kernel analysis of divergences}
The one-loop contribution to the effective action for a generic quantum
field theory usually boils down to the calculation of the regulated
quantity $\Tr\log H$ where $H$ is the second variation of the action
around the quantum fields. After an appropriate Wick rotation, $H$ usually
becomes a differential operator with a positive spectrum -- at least
perturbatively.

For example, the Euclidean effective action for a complex bosonic field $\phi$
at one-loop generically amounts to performing the path integration
\begin{align}
e^{-\Gamma_E} = \int \CD\phi \exp\left(-\int d^4x\sqrt{g}\;
	\bar\phi \left(-\Box + Q \right)\phi \right)
\end{align}
where $\Box$ is some covariant Laplacian and $Q$ is a generic matrix
which may depend on background fields.
To define the path integral requires specifying the measure. This
is usually done implicitly by specifying the meaning
of Gaussian integration. A sensible choice is
\begin{align}
\int \CD\phi \exp\left(-\int d^4x\sqrt{g}\;
	\bar\phi \phi \right) \equiv 1
\end{align}
This defines $\tilde \phi = g^{1/4} \phi$ as the path integration variable
and guarantees a manifestly diffeomorphism invariant measure.\footnote{The
measure is invariant because the ``1'' is invariant on the right side,
the integrand is invariant on the left, and so the measure should be
also.} For any internal symmetries it will often also
be manifestly invariant since $\bar \phi$ is usually in the conjugate
representation to $\phi$. For classically Weyl invariant theories
where $\phi$ has unit scaling dimension, one has $Q = -\frac{1}{6} \mathcal R + V$
where $V$ is some conformal field of dimension 2.
The Ricci scalar in $Q$ combines with $\Box$ to give
the conformally invariant Laplacian, $\Box + \mathcal R/6$.
Unfortunately, the measure is \emph{not}
conformally invariant and this leads to the familiar conformal
anomaly.\footnote{One could choose instead a different power of $g$ in
defining the measure to make it conformally invariant, but this would
trade a conformal anomaly for a diffeomorphism anomaly.}

Using the definition of Gaussian integration, the Euclidean effective
action is given by
\begin{align}
\Gamma_E = \Tr \log H = -\Gamma
\end{align}
where $H\equiv -\Box + Q$ and $\Gamma$ is the Minkowski effective action.
We would like to efficiently calculate properties of this object.
One method to calculate $\Tr\log H$ is Schwinger's
proper time technique. One makes use of the matrix equation\footnote{It is not
necessary for the function in the integral to be an exponential.
Any function $f$ with certain boundary conditions -- namely
$f(0)=1$ and $f(\infty) = 0$ sufficiently quickly -- would work. The
advantage of using the exponential is the ease of differentiating it.}
\begin{align}
\Tr \log H = -\Tr \int_0^\infty \frac{d\tau}{\tau} \exp(-\tau H).
\end{align}
which holds -- up to an infinite constant -- in the basis where $H$
is diagonal. (To prove the equality, one differentiates both sides
with respect to the eigenvalue of $H$.)

Usually $H$ is afflicted with ultraviolet divergences. Then the above definition
can be modified in several ways. One way, which is quite similar
to dimensional regularization, is to add extra powers of $\tau$ in the
definition of the trace:
\begin{align}
\left[\Tr \log H\right]_s = -\mu^{2s} \Tr \int_0^\infty \frac{d\tau}{\tau^{1-s}} \exp(-\tau H).
\end{align}
The parameter $\mu$ has dimensions of mass and is added only to make the
final result dimensionless. The integral then formally gives
\begin{align}
\left[\Tr\log H \right]_s = - \Tr \left(\frac{H}{\mu^2}\right)^{-s} \Gamma(s)
\end{align}
Since the result is proportional to $\zeta_H(s)$, the zeta-function associated
with $H$, this approach goes by the name of zeta-function regularization.
Differentiating with respect to $H$ gives
\begin{align}
\left[\Tr \frac{1}{H} \right]_s = \Tr \left\{\left(\frac{H}{\mu^2}\right)^{-s} \frac{1}{H} \Gamma(s+1)\right\}
\end{align}
with the limit agreeing as $s$ tends to zero.

Another method, which we shall adopt, is simply to introduce a small cutoff for the
parameter $\tau$:
\begin{align}
\left[\Tr \log H\right]_\eps = -\Tr \int_\eps^\infty \frac{d\tau}{\tau} \exp(-\tau H).
\end{align}
Differentiating then gives
\begin{align}
\left[\Tr \frac{1}{H}\right]_\eps = \Tr \left(e^{-\eps H} \frac{1}{H} \right)
\end{align}
The parameter $\eps$ has dimensions of length squared (or inverse energy squared).

In many problems, one can use either regulation scheme by working in
a momentum basis, performing a derivative expansion, and then doing the
resultant momentum integrals. But it is advantageous to have a formalism which
does not require doing so directly. Such an approach is the heat
kernel.\footnote{The heat kernel method has a long history, with much
of its properties worked out originally by DeWitt \cite{DeWitt:1965jb}.
A review of the heat kernel can be found in \cite{Vassilevich:2003xt}.}

The heat kernel is the formal operator $U(\tau) = \exp(-\tau H)$.
Its two point function is given by
\begin{align}
U(x,x'; \tau) = \opbraket{x}{e^{-\tau H}}{x'}.
\end{align}
and is subject to two conditions: the initial condition $U(x,x';0) = \delta(x,x')$
and the ``heat equation''
\begin{align}
\frac{dU}{d\tau} = -H U.
\end{align}
One is usually concerned with $H$'s which are perturbatively related to the
Laplacian $H_0 = - \partial^m \partial_m$ in flat space. This case
is directly solvable via Fourier transform.\footnote{This is the
only location where a momentum basis calculation is used.} The result (written
in four dimensions) is
\begin{align}
U_0(x,x';\tau) = \frac{1}{(4\pi \tau)^2} \exp\left({-|x-x'|^2/4\tau}\right)
\end{align}

This can be generalized to $H = -\partial^m \partial_m + m^2$ for constant $m^2$
in $d$ dimensions by\footnote{Zeta function regularization essentially replaces $d$
in this formula with $d-2s$, which is why it is similar to dimensional regularization.}
\begin{align}
U_0(x,x';\tau) = \frac{1}{(4\pi \tau)^{d/2}} \exp\left({-|x-x'|^2/4\tau} - \tau m^2\right)
\end{align}
but we will keep $d=4$ in all our calculations.

When the model is modified with a potential or to include a Yang-Mills gauge field,
one expects the corrections to $U$ to come in a simple perturbative way. One takes
\begin{align}
U(x,x';\tau) = \frac{1}{(4\pi \tau)^2} \exp\left({-|x-x'|^2/4\tau}\right) F(x,x'; \tau)
\end{align}
where $F(\tau)$ is assumed to be an analytic function in $\tau$ regular at $\tau=0$
and obeying $F(x,x; 0) = 1$. Applying the heat equation to this ansatz for $U$ gives
\begin{align}\label{eq_Fflat}
\frac{\partial F}{\partial \tau} +
	\frac{1}{\tau} (x^m  - x'^m) D_m F = (-D^m D_m + Q) F
\end{align}
where we have taken $H = -\Box + Q$. \footnote{If $Q$ contains a constant mass term,
one generally separates it out by positing $F$ to have an overall factor $e^{-\tau m^2}$.}
Taking $y = x-x'$, $\mathcal O = -H$,
and writing $F = \sum_{n=0}^\infty a_n \tau^n / n!$, we find a set of
recursion relations for the coefficients $a_n$
\begin{align}\label{eq_flatrecurs}
a_n + \frac{1}{n} y^m D_m a_n = \mathcal O a_{n-1}
\end{align}
for $n \geq 1$, and
\begin{align}\label{eq_flatrecurs0}
y^m D_m a_0 = 0.
\end{align}
for $n=0$. These relations can be solved as power series in $y$ for each coefficient,
using the initial condition that $[a_0]=1$, where the brakets denote taking the ``coincident
limit'' of $y=x-x' \rightarrow 0$.

The inclusion of gravity requires one to reinterpret $|x-x'|^2 = y^2$ in a coordinate-invariant
way. One makes the replacement $|x-x'|^2 / 2 \rightarrow \sigma$, where $\sigma$
is a symmetric bi-scalar function (that is, a scalar function of both $x$ and $x'$).
The heat equation becomes
\begin{align}\label{eq_Fcurved0}
-\frac{2}{\tau} F + \frac{\sigma}{2 \tau^2} F + \frac{\partial F}{\partial \tau} = 
	\frac{1}{4\tau^2} \nabla^a\sigma \nabla_a \sigma F - \frac{\Box \sigma}{2\tau} F
	- \frac{1}{\tau} \nabla^a \sigma \nabla_a F - H F
\end{align}
In order for $F$ to be analytic at $\tau=0$, the term that goes as $1/\tau^2$ must be
trivially satisfied, giving
\begin{align}\label{eq_geodet}
2 \sigma = \nabla^a \sigma \nabla_a \sigma.
\end{align}
This equation, together with $[\nabla_a \sigma]=0$ and $[\nabla_a \nabla_b \sigma] = \eta_{ab}$
uniquely determines $\sigma$ as
$\sigma = \frac{1}{2} g_{mn}(x') (x-x')^m (x-x')^n + \mathcal O((x-x'))^3$.
The remaining equation can be written in a form analogous to \eqref{eq_Fflat}
provided we rescale $F$ 
\begin{align}
F \rightarrow \Delta^{1/2} \tilde F
\end{align}
where $\Delta$ obeys
\begin{align}\label{eq_vvmdet}
\nabla^a \sigma \nabla_a \log \Delta + \Box \sigma = 4
\end{align}
with the initial condition $[\Delta]=1$. The resultant equation reads
\begin{align}\label{eq_Fcurved}
\frac{\partial \tilde F}{\partial \tau} +
	\frac{1}{\tau} \nabla^a \sigma \nabla_a \tilde F = \Delta^{-1/2} \mathcal O \Delta^{1/2} \tilde F
	\equiv \tilde {\mathcal O} \tilde F
\end{align}
where $\tilde {\mathcal O} = \Delta^{-1/2} \mathcal O \Delta^{1/2}$.

The bi-scalars $\sigma$ and $\Delta$ are well-known from the study of geodesics.
$\sigma$ is the geodetic interval -- half of the integral of $ds^2$ along the geodesic connecting
$x'$ to $x$. $\Delta$ is known as the Van Vleck-Morette determinant and represents
the Jacobian between an arbitrary coordinate system and geodesic coordinates.
The precise definitions of these objects will not concern us, since we will show
that in a suitable coordinate system both $\sigma$ and $\Delta$ take especially
simple forms.

Expanding $\tilde F$ in a power series, we find the set of recursion relations
\begin{align}\label{eq_recursn}
\tilde a_n +
	\frac{1}{n} \nabla^a \sigma \nabla_a \tilde a_n = \tilde {\mathcal O} \tilde a_{n-1}
\end{align}
for $n \geq 1$ and
\begin{align}\label{eq_recurs0}
\nabla^a \sigma \nabla_a \tilde a_0 = 0.
\end{align}
for $n=0$. These relations were first written down by DeWitt \cite{DeWitt:1965jb}
and solved recursively, using the $x \rightarrow x'$ limit of certain quantities to
derive all of them.

The importance of these coefficients lies in recalling the definition
of the regulated determinant:
\begin{align}
[\Tr \log H]_\eps &= -\Tr \int_\eps^\infty \frac{d\tau}{\tau} \exp(-\tau H)
	= -\int_\eps^\infty \frac{d\tau}{\tau} \Tr \opbraket{x}{U(\tau)}{x} \\
	&= -\int_\eps^\infty \frac{d\tau}{\tau} \frac{1}{(4\pi\tau)^2} \Tr \,\tilde F(x,x; \tau) \\
 	&= -\int_\eps^\infty \frac{d\tau}{\tau} \frac{1}{(4\pi\tau)^2}
	\sum_{n=0}^\infty \frac{\tau^n}{n!} \Tr \,\tilde [a_n]
\end{align}
where we have used $[\sigma]=0$ and $[\Delta]=1$. The total effective
action is given by the $x=x'$ limit of the coefficients $a_n$. In particular,
the divergent terms in four dimensions are
\begin{align}
[\Tr\log H]_\eps &= -\frac{1}{16\pi^2} \int d^4x \sqrt{g} \,\Tr \left(\frac{[a_0]}{2\eps^2} + \frac{[a_1]}{\eps} - \frac{[a_2]}{2} \log \eps + \textrm{finite} \right)
\end{align}
where the limit $x=x'$ has been taken.

Since the coincident limit of the heat kernel coefficients are by construction local,
the divergences in the above expression can be removed by adding local counterterms.
One can take
\begin{align}
A_{\eps}^{ct} &= +\frac{1}{16\pi^2} \int d^4x \sqrt{g} \,\Tr \left(\frac{[a_0]}{2\eps^2} + \frac{[a_1]}{\eps} - \frac{[a_2]}{2} \log \eps \right)
\end{align}
and then the regulated trace can be defined as the limit where $\eps$ tends to zero
\begin{align}
[\Tr \log H]_{reg} = \lim_{\eps \rightarrow 0} \left([\Tr\log H]_\eps + A_\eps^{ct}\right)
\end{align}
The result is explicitly $\eps$-independent and corresponds to a minimal
substraction scheme at one-loop.

This is not the only application of this method. In particular, any theory
with a potential anomaly at one-loop can be understood by the nonzero symmetry transformation
$\delta_g H$ where $g$ is an element of the potentially anomalous symmetry group.
(This can be seen to arise via the non-invariance of the path integral measure,
which was Fujikawa's perspective \cite{Fujikawa:1979ay}.) Using the proper time
regulation scheme, the transformation of the effective action is given by
\begin{align}
\delta_g [\Tr\log H]_\eps = -\Tr \int_\eps^\infty \frac{d\tau}{\tau} \delta_g \exp(-\tau H)
	= \int_\eps^\infty d\tau \Tr \left(\delta_g H \exp(-\tau H)\right)
\end{align}
where we have used cyclicity of the trace. In most cases of interest, the
anomaly has the form $\delta_g H = a \Lambda H + b H \Lambda$ for some
numerical coefficients $a$ and $b$ and some quantity $\Lambda$ which
may or may not be local. Then using cyclicity of the trace, one finds
\begin{align}
\delta_g [\Tr\log H]_\eps
	&= (a+b) \int_\eps^\infty d\tau \Tr \left(\Lambda H \exp(-\tau H)\right)
	= (a+b) \Tr\left(\Lambda e^{-\eps H} \right) \eol
	&= \frac{(a+b)}{16\pi^2} \int d^4x \sqrt g \, \Tr\left(
		\frac{[\Lambda a_0]}{\eps^2}+ \frac{[\Lambda a_1]}{\eps}
		+ \frac{[\Lambda a_2]}{2} + \mathcal O(\eps)
	\right)
\end{align}
In the case of the conformal anomaly for a conformally invariant action,
$a=3$ and $b=-1$ (that is, $H' = e^{3\Lambda} H e^{-\Lambda}$) and $\Lambda$
is a local function, one finds
\begin{align}
\delta_c [\Tr\log H]_\eps
	&= \frac{1}{16\pi^2} \int d^4x \sqrt g \, \Tr\left(
		\frac{2\Lambda [a_0]}{\eps^2}+ \frac{2 \Lambda [a_1]}{\eps}
		+ \Lambda [a_2] + \mathcal O(\eps)
	\right)
\end{align}
Usually (and we will demonstrate this) the coefficients $[a_n]$ are such
that the conformal transformation of the counter terms cancels the effect
of the two leading divergences. Then we may take $\eps \rightarrow 0$ for the
finite regulated action and find the finite conformal anomaly depends only
on $[a_2]$.

\subsection{Heat kernel analysis in normal coordinates}
DeWitt's original analysis of the heat kernel coefficients was performed
using the recursion relations and the differential equations for $\sigma$ and $\Delta$.
This approach works reasonably well for the first few coefficients
but quickly becomes unwieldy. A much more efficient method was
developed by Avramidi \cite{avramidi}, who was also the first to evaluate
the coefficient $[a_4]$ in curved space. We will review 
how his approach works here using normal coordinates. A short summary of how
to efficiently work out normal coordinate expressions for various quantities
is given in Appendix \ref{app_nc}.

In normal coordinates, one would expect the geodetic interval to
take the simple form
\begin{align}
\sigma = \frac{y}{2}
\end{align}
where $y$ is the normal coordinate for $x$ centered at $x'$. In order
for this choice to obey the required equation \eqref{eq_geodet}, one must have
\begin{align}
\nabla_a \sigma = {e_a}^m y_m = {\delta_a}^m y_m = y_a
\end{align}
Normal coordinates, as defined in Appendix \ref{app_nc},
possess the property that $y^m {e_m}^a = y^a$ as well as $y^a {e_a}^m = y^m$,
but the condition we require is slightly different. It can be shown that if the stucture
group is Riemannian plus some internal degrees of freedom, normal
coordinates possess also this additional quality.\footnote{For Einstein-Cartan
geometry with torsion, one can define normal coordinates using a Riemannian
connection and then relate the results with Riemannian curvatures and derivatives
to the torsioned quantities.}

The Van Vleck-Morette determinant is also quite simple in this coordinate system:
\begin{align}
\Delta = \det({e_a}^m) = \det({e_m}^a)^{-1}
\end{align}
which is essentially the Jacobian between $x$ and the normal coordinates $y$.
It is straightforward to show this obeys \eqref{eq_vvmdet}.

The recursion relation for the coefficients now reads
\begin{align}
\left(1 + \frac{D}{n} \right) \tilde a_n = \tilde {\mathcal O} \tilde a_{n-1}
\end{align}
where $D \equiv \nabla^a \sigma \nabla_a = y^m \partial_m$
with the special case $D \tilde a_0 = 0$.
These can be formally solved by taking $\tilde a_0 = 1$ and
\begin{align}
\tilde a_n = \left(1 + \frac{D}{n} \right)^{-1} \tilde {\mathcal O} \left(1+\frac{D}{n-1}\right)^{-1}
	\tilde {\mathcal O} \cdots \left(1 + D \right)^{-1} \tilde {\mathcal O}.
\end{align}
The operator $D \equiv y^m \partial_m$ can be thought of as the derivative along
the Riemannian geodesic. It is formally a one-dimensional derivative and possesses
eigenvalues $\ket{n}$, which are the totally symmetric $n$-tensors
\begin{align}
\ket{n} = \ket{b_1, \ldots, b_n} \equiv \frac{1}{n!} y^{b_1} \cdots y^{b_n}
\end{align}
where $D \ket{n} = n \ket{n}$.
Provided we are concerned only with quantities which are analytic in $y$ (i.e.
only those quantities which admit an analytic normal coordinate expansion) this
set of eigenvalues forms a basis. Associated with these tensors are the dual
tensors
\begin{align}
\bra{m} = \bra{a_1, \ldots, a_m} = \partial_{a_1} \cdots \partial_{a_n}.
\end{align}
The inner product $\braket{m}{n}$ is defined in the obvious way with
$y=0$ taken at the end:
\begin{align}
\braket{m}{n} = \delta_{mn} \delta_{a_1 \ldots a_n}^{b_1 \ldots b_n}.
\end{align}
We can therefore solve for $\tilde a_n$ as a power series in $y$. In the
language of the bras and kets,
\begin{align}
\tilde a_n = \sum_{k=0}^\infty \ket{k} \braket{k}{\tilde a_n}
\end{align}
where
\begin{align}\label{eq_recursA}
\braket{k}{\tilde a_n} = \sum_{j_1,\ldots,j_{k-1} \geq 0} &\left(1 + \frac{k}{n}\right)^{-1}
	 \left(1 + \frac{j_{n-1}}{n-1}\right)^{-1} \cdots \left(1 + j_1 \right)^{-1}
	\times \eol & \opbraket{k}{\tilde {\mathcal O}}{j_{n-1}} \opbraket{j_{n-1}}{\tilde {\mathcal O}}{j_{n-2}}  \cdots \opbraket{j_1}{\tilde {\mathcal O}}{0}
\end{align}
The $y=0$ limit of $\tilde a_n$ is given by $\braket{0}{\tilde a_n} = [\tilde a_n]$
and its $k$th order derivative given by the $k$-tensor $\braket{k}{\tilde a_n}$.

The essence of \eqref{eq_recursA} is that the heat kernel coefficients are
given by matrix elements of the operator $\tilde {\mathcal O}$. To evaluate such elements,
we first write $\tilde {\mathcal O}$ in terms of normal coordinates as
\begin{align}
\tilde {\mathcal O} = X^{mn} \partial_m \partial_n + Y^m \partial_m + Z
\end{align}
For the case $\tilde {\mathcal O} = \Delta^{-1/2} (\nabla^a \nabla_a - Q) \Delta^{1/2}$,
we find
\begin{align}
X^{mn} &= g^{mn} \eol
Y^m &=  - 2 g^{mn} h_n  + \partial_n g^{nm} \eol
Z &= g^{mn} h_m h_n - \partial_n g^{nm} h_m - g^{mn} \partial_m h_n - Q
	+ \Delta^{-1/2} \nabla^a \nabla_a \Delta^{1/2}
\end{align}
where $h_m$ is the connection found in $\nabla_m \equiv \partial_m - h_m$.
$Z$ can be rewritten as
\begin{align}
Z &= g^{mn} h_m h_n - \partial_n g^{nm} h_m - g^{mn} \partial_m h_n - Q\eol
	& \;\;\;\;\;
	- \frac{1}{2} \partial_n g^{nm} \partial_m \log e
	- \frac{1}{2} g^{nm} \partial_m \partial_n \log e
	- \frac{1}{4} g^{nm} \partial_m \log e \partial_n \log e
\end{align}
which shows that the original operator $\tilde {\mathcal O}$ could have
been written
\begin{align}
\tilde {\mathcal O} = g^{-1/4} \nabla_m g^{mn} \sqrt{g} \nabla_n g^{-1/4}
\end{align}
This is an indicator that we have essentially used the scalar
density $g^{1/4} \phi$ as the path integral variable.
Moreover this operator is manifestly symmetric. We will
encounter a similar structure when we deal with chiral superfields.

The divergences and anomalies are related to the $y=0$ limits of the first three
heat kernel coefficients. The zeroth coefficient is the simplest,
$\braket{0}{\tilde a_0} = 1$, and gives the quartic divergence.

The quadratic divergence is given by the first coefficient
\[
\braket{0}{\tilde a_1} = \opbraket{0}{\tilde {\mathcal O} }{0} = [Z]
\]
To evaluate $[Z]$, first note that
\[
[Z] = -Q -\frac{1}{2} [\partial^m \partial_m \log e]
\]
in normal coordinates as $y\rightarrow 0$. (Clearly $[\partial_m \log e]$
vanishes since there are no covariant vectors of the right dimension to
correspond to it.) We need the expansion of $\log e$ to $y^2$.
The vierbein in normal coordinates is given by
\[
{e_m}^a = {\delta_m}^a + \frac{1}{6} {R_{y m y}}^a + \mathcal O(y^3)
\]
where we have used the notation that a $y$ in an index slot means a $y$ is contracted
with that index. Thus $\log e$ is given by
\[
\log e = \frac{1}{6} \mathcal R_{yy} + \mathcal O(y^3)
\]
where $\mathcal R_{ab}$ is the Ricci tensor. One easily finds
\begin{align}
\braket{0}{\tilde a_1} = -Q - \frac{1}{6} \mathcal R.
\end{align}

The logarithmic divergences are given by the $y=0$ limit of $\tilde a_2$:
\[
\braket{0}{\tilde a_2} = \sum_{j_1=0} \left(1 + j_1\right)^{-1}\, 
	\opbraket{0}{\tilde {\mathcal O} }{j_1} \times \opbraket{j_1}{\tilde {\mathcal O} }{0}
\]
Although the sum is over all values of $j_1$, the first matrix element
vanishes for $j_1 \geq 3$. We easily find
\[
\braket{0}{\tilde a_2} = [Z]^2 + \frac{1}{2} [Y^m] [\partial_m Z] + \frac{1}{3} [X^{mn}] [\partial_m \partial_n Z]
\]
Using $[X^{mn}] = \eta^{mn}$ and $[Y^m] = 0$,
\[
\braket{0}{\tilde a_2} = \left(Q + \frac{1}{6} \mathcal R \right)^2 + \frac{1}{3} [\partial^m \partial_m Z]
\]
The remaining term is a little complicated to evaluate. Begin by expanding
it out, using $[h_m] = 0$ and $[\partial_p g_{mn}] = 0$:
\begin{align*}
\frac{1}{3} [\partial^m \partial_m Z]
	&= \frac{2}{3} [g^{pq} \partial^m h_p \partial_m h_q]
	- \frac{2}{3} [\partial^m \partial_p g^{pq} \partial^m h_q]
	- \frac{1}{3} [g^{pq} \partial^m \partial_m \partial_p h_q] \\
	&\;\;\;\;
	- \frac{1}{3} [\partial^m \partial_m \Delta^{-1/2} \nabla^a \nabla_a \Delta^{1/2}]
	- \frac{1}{3} [\partial^m \partial_m Q]
\end{align*}
where we have used that $[h_m]=0$ and $[\partial_p g_{mn}] = 0$.
Most of the terms can be evaluated by noting
\begin{gather*}
g^{mn} = \eta^{mn} -\frac{1}{3} R_y{}^m{}_y{}^n + \mathcal O(y^3), \;\;\;
h_m = \frac{1}{2} \mathcal F_{ym} + \mathcal O(y^2)
\end{gather*}
These give
\begin{align*}
\frac{1}{3} [\partial^m \partial_m Z]
	&= \frac{1}{6} \mathcal F^2 - \frac{1}{3} \Box Q
		- \frac{1}{3} [\partial^2 \partial^m h_m] + \frac{1}{3} [\partial^2 \Delta^{-1/2} \Box \Delta^{1/2}]
\end{align*}
The gauge field $h$ is given to cubic order by
\begin{align*}
h_n &= \frac{1}{2} \mathcal F_{yn} + \frac{1}{3} \nabla_y \mathcal F_{yn}
	+ \frac{1}{8} \nabla_y^2 \mathcal F_{yn} - \frac{1}{4!} {R_{yny}}^b \mathcal F_{by}
	+ \mathcal O(y^4)
\end{align*}
and one easily finds $[\partial^m \partial_m \partial^n h_n] = 0$.

The remaining term is significantly more messy. After some work, we find
\begin{align*}
[\partial^m \partial_m \Delta^{-1/2} \nabla^a \nabla_a \Delta^{1/2}]
	&= -\frac{1}{5} \nabla^2 \mathcal R - \frac{1}{30} \mathcal R^{ab} \mathcal R_{ab}
		+ \frac{1}{45} R^{abcd} \left(R_{abcd} + R_{adcb}\right) \\
	&= -\frac{1}{5} \nabla^2 \mathcal R - \frac{1}{30} \mathcal R^{ab} \mathcal R_{ab}
		+ \frac{1}{30} R^{abcd} R_{abcd}
\end{align*}
using the symmetry properties of the Riemann tensor. The second heat kernel
coefficient (and the logarithmic divergences) is then given by
\begin{align}
\braket{0}{\tilde a_2} = \left(Q + \frac{1}{6} \mathcal R \right)^2
	+ \frac{1}{6} \mathcal F^2
	-\frac{1}{15} \Box \mathcal R - \frac{1}{90} \mathcal R^{ab} \mathcal R_{ab}
		+ \frac{1}{90} R^{abcd} R_{abcd} - \frac{1}{3} \Box Q
\end{align}

It is useful to rewrite some of the quantities appearing here.
The square of the conformal Weyl tensor can be written
\begin{align}
C^{abcd} C_{abcd} = R^{abcd} R_{abcd} - 2 \mathcal R^{ab} \mathcal R_{ab}
	+ \frac{1}{3} \mathcal R^2
\end{align}
This quantity (and $C_{abcd}$ itself) transforms covariantly.
The four dimensional Gauss-Bonnet term 
\begin{align}
L_\chi &= R^{abcd} R_{abcd} - 4 \mathcal R^{ab} \mathcal R_{ab} + \mathcal R^2 \eol
	&= C^{abcd} C_{abcd} - 2 \mathcal R^{ab} \mathcal R_{ab} + \frac{2}{3} \mathcal R^2
\end{align}
is topological, its integral being invariant under arbitrary local
(including conformal) deformations of the metric.

We can thereby rewrite $[a_2]$ as
\begin{align}
\braket{0}{\tilde a_2} = \left(Q + \frac{1}{6} \mathcal R \right)^2
	+ \frac{1}{6} \mathcal F^2
	- \frac{1}{3} \Box \left(Q + \frac{1}{6} \mathcal R \right)
	+ \frac{1}{90} \left(
	\frac{3}{2} C^{abcd} C_{abcd} - \frac{1}{2} L_\chi - \Box \mathcal R
	\right)
\end{align}

It is worth noting that if we wanted $H$ to transform covariantly under
conformal transformations, we would choose $Q = -\frac{1}{6} \mathcal R + V$
where $V$ transforms conformally. Then $[a_1]$ and $[a_2]$ would be
\begin{align}
[a_1] &= -V \\
[a_2] &= V^2
	+ \frac{1}{6} \mathcal F^2
	- \frac{1}{3} \Box V
	+ \frac{1}{90} \left(
	\frac{3}{2} C^{abcd} C_{abcd} - \frac{1}{2} L_\chi - \Box \mathcal R
	\right)
\end{align}
and $[a_1]$ would be conformal (with dimension 2) and $[a_2]$ would be
conformal (with dimension 4) up to total derivatives.

Thus if we calculate the conformal transformation of the counter-terms, we find
\begin{align}
\delta_c A_{\eps}^{ct} &= +\frac{1}{16\pi^2} \int d^4x \sqrt{g} \,\Tr \left(4\Lambda \frac{[a_0]}{\eps^2} + 2\Lambda \frac{[a_1]}{\eps}\right)
\end{align}
and the regulated trace anomaly is finite and given by
\begin{align}
\delta_c [\Tr\log H]_{reg} &= \frac{1}{16\pi^2} \int d^4x \sqrt g \, \Tr\left(
	\Lambda [a_2]\right)
\end{align}

\subsection{Heat kernel for Dirac operators}
A common Dirac fermion model is
\begin{align}
S = \int d^4x \sqrt{g} \left(\bar\Psi i \slashed{\nabla} \Psi + \bar\Psi \mu \Psi\right)
\end{align}
where $\mu$ is a generic mass term and $\slashed{\nabla} = \gamma^a \nabla_a$ is
a covariant derivative. Written in two-component notation, the Lagrangian is
\begin{align}
\left(
\begin{array}{cc}
\chi^\alpha & \bar\psi_\dalpha
\end{array}
\right)
\left(
\begin{array}{cc}
\mu \delta_{\alpha}{}^\beta & i \sigma^a_{\alpha \dbeta} \nabla_a \\
i \bsigma_a^{\dalpha \beta} \nabla^a & \mu \delta^{\dalpha}{}_\dbeta 
\end{array}
\right)
\left(
\begin{array}{c}
\psi_\beta \\
\bar\chi^\dbeta
\end{array}
\right)
\end{align}
We assume $\Psi$ and $\bar\Psi$ to transform in conjugate representations.
This means that the Weyl fermion $\psi$ is gauge conjugate not only to $\bar\psi$
but also to $\chi$.

One can define the path integral of a Gaussian in the obvious way:
\begin{align}
\int \CD\Psi \exp\left(-\int d^4x\sqrt{g}\,
	\bar\Psi \Psi \right) \equiv 1
\end{align}
This definition is clearly diffeomorphism, Lorentz, and gauge invariant and so we
expect these symmetries to be non-anomalous. The (Euclidean) effective action is
\begin{align}
\Gamma_E = -\Tr\log D
\end{align}
where
\begin{align}
D = i \slashed{\nabla} + \mu
\end{align}
One normally proceeds using the standard fermion doubling trick, arguing
that $\Gamma_E$ cannot depend on the sign of $\mu$. Equivalently, one could argue
that $\Gamma_E$ cannot depend on the convention for the gamma matrices.
Either way, one can introduce a new operator with a relative sign flip between the
kinetic and mass terms
\begin{align}
\tilde D = -i \slashed{\nabla} + \mu
\end{align}
which should yield the same determinant as $D$. Then one may define
\begin{align}
\Gamma_E = -\frac{1}{2} \Tr\log D - \frac{1}{2} \Tr \log \tilde D
	= -\frac{1}{2} \Tr\log (\tilde D D)
\end{align}
where\footnotemark
\begin{align}
\tilde D D = \mu^2 - i [\slashed \nabla, \mu] - \mathcal F_{ab} S^{ab} - \Box 
\end{align}

\footnotetext{The same basic approach holds if we replace $\mu \rightarrow \mu + i \nu \gamma_5$.
The only major modification is that one of the terms generated is linear
in a derivative, $\nu \gamma_5 \gamma^a \nabla_a$, which must be treated as
a matrix connection. One absorbs it into a new definition of the
derivative $\nabla'$ and again proceeds as before.}

A greater level of sophistication is required when the model
of interest is chiral. Taking the above model with $\chi=\bar\chi=0$
we find in two-component notation
\begin{align}
S = \int d^4x \sqrt{g} \left(i \bar\psi_\dalpha \bsigma_b^{\dalpha \alpha} \nabla^b \psi_\alpha\right)
\end{align}
A Majorana mass term may be included:
\begin{align}
S = \int d^4x \sqrt{g} \left(i \bar\psi_\dalpha \bsigma_b^{\dalpha \alpha} \nabla^b \psi_\alpha + \frac{1}{2} \psi^\alpha \mu \psi_\alpha + \frac{1}{2} \bar\psi_\dalpha \bar \mu \bar\psi^\dalpha \right)
\end{align}
The difficulty with this model arises because the simplest
Lorentz invariant definition for the Gaussian path integration is
\begin{align}
\int \CD\psi \exp\left(-\frac{1}{2}\int d^4x\sqrt{g}\left(
	\psi^2 + \bar\psi^2 \right) \right)\equiv 1
\end{align}
For the massless case, the classical action is gauge invariant but
the measure is not.\footnote{The measure used here has the structure of a
Majorana mass term, which in four dimensions joins objects of the same
chirality. In $d = 2 + 4n$ dimensions, both the Majorana mass term and
the Dirac mass term join objects of \emph{opposite} chirality and so
there is no Lorentz invariant way to define Gaussian integration. This is
one way of explaining the celebrated gravitational (or Lorentz)
anomaly found by Alvarez-Gaum\'e and Witten.}

Explicit two-component notation can be avoided by combining $\psi$ and
$\bar\psi$ into a Majorana fermion $\Psi_M$ where
\begin{gather*}
\bar\Psi_M = \left(
\begin{array}{cc}
\psi^\alpha & \bar\psi_\dalpha
\end{array}
\right), \;\;\;
\Psi_M = \left(
\begin{array}{c}
\psi_\beta \\
\bar\psi^\dbeta
\end{array}
\right)
\end{gather*}
Then the action reads
\begin{align}
S = \frac{1}{2} \int d^4x \sqrt{g} \bar\Psi_M \left(i \hat{\slashed{\nabla}} + \hat \mu \right)\Psi_M
\end{align}
with measure
\begin{align}
\int \CD\psi \exp\left(-\frac{1}{2}\int d^4x\sqrt{g} \bar\Psi_M \Psi_M \right)\equiv 1
\end{align}
where $\hat \mu = \textrm{Re}\,\mu + i \gamma_5 \, \textrm{Im}\,\mu$
is the Majorana mass and the Majorana derivative is
\begin{align}
\hat{\slashed{\nabla}} = \left(
\begin{array}{cc}
0 & \sigma^a_{\alpha \dbeta} \tilde \nabla_a \\
\bsigma_a^{\dalpha \beta} \nabla^a & 0
\end{array}
\right)
\end{align}
where $\nabla_a$ is the derivative in the representation of $\psi$ and
$\tilde\nabla_a$ is the derivative in the conjugate representation of $\bar\psi$.
This is problematic even in the massless case since the
square of this object involves operators like $\tilde\nabla_a \nabla_b$
which do not transform covariantly and therefore make calculation
especially difficult.

We restrict ourselves now to the case of vanishing Majorana mass.
Defining $D \equiv i \hat{\slashed{\nabla}}$, path integration
yields a Pfaffian, which can be interpreted as the square root
of a determinant:
\begin{align}
\Gamma_E = -\log \textrm{Pf} \,D = -\frac{1}{2} \Tr\log D
\end{align}
The properties of the effective action are then related to the properties
of the determinant of the operator $D$. This operator can be thought of
as a mapping
\begin{align}
D : C_+(\mathbf{r}) \oplus C_-(\mathbf{\bar r}) \rightarrow
	C_+(\mathbf{\bar r}) \oplus C_-(\mathbf{r})
\end{align}
where $\mathbf{r}$ is the representation of $\psi$, $\mathbf{\bar r}$
is that of $\bar\psi$, and $+$ and $-$ denote the positive and
negative chirality sectors. As a formal operator, its determinant
is ill-defined since the domain and range are different spaces;
this is just another way of saying that its determinant does not transform in a
gauge-invariant manner. One way of making sense of this object is to note that when the
gauge coupling vanishes, $D$ ceases to a problematic operator since there is no longer
a distinction between a representation and its conjugate.
Varying the trace with respect to the coupling, we find
\begin{align}\label{eq_dd}
\delta \Tr\log D = \Tr\left(D^{-1} \delta D \right)
\end{align}
If this expression can be suitably regulated and then integrated, we
are left with a reasonable definition of the effective action. This
approach was pioneered by Leutwyler \cite{Leutwyler:1985em} in the case of fermions
and by McArthur and Osborn for the case of chiral superfields in
background Yang-Mills \cite{McArthur:1985xd}.

Following Leutwyler, we regulate \eqref{eq_dd} by introducing the
dual operator
\begin{align}
\tilde D = \left(
\begin{array}{cc}
0 & -i \sigma^a_{\alpha \dbeta} \nabla_a \\
-i \bsigma_a^{\dalpha \beta} \tilde \nabla^a & 0
\end{array}
\right)
\end{align}
so that
\begin{align}
H = \tilde D D = \left(
\begin{array}{cc}
-\Box - \mathcal F_{ab} \sigma^{ab} & 0 \\
0 & -\tilde \Box - \tilde {\mathcal F}_{ab} \bsigma^{ab}
\end{array}\right)
\end{align}
$\mathcal F_{ab} = -[\nabla_a, \nabla_b]$ is the field strength associated with
the covariant derivative and $\sigma^{ab} = \frac{1}{4} (\sigma^a \bsigma^b - \sigma^b \bsigma^a)$
in the conventions of \cite{bgg}.

We define
\begin{align}
L_\eps = \Tr \left(e^{-\eps H} D^{-1} \delta D \right)
	= \Tr\int_\eps^\infty d\tau \left(e^{-\tau H} \tilde D \delta D\right)
\end{align}
This operator can be separated into parts which are even and odd
under parity: $L_\eps = L_\eps^+ + L_\eps^-$ where
\begin{align}
L_\eps^+ &= \frac{1}{2} \Tr\int_\eps^\infty d\tau \left(e^{-\tau H} \tilde D \delta D + e^{-\tau \tilde H} D \delta \tilde D\right) \\
L_\eps^- &= \frac{1}{2} \Tr\int_\eps^\infty d\tau \left(e^{-\tau H} \tilde D \delta D - e^{-\tau \tilde H} D \delta \tilde D\right)
\end{align}
The operator $\tilde H = D \tilde D$ is the conjugate of $H$.
Using cyclicity of the trace, one can immediately deduce that
\begin{align}
L_\eps^+ &= \frac{1}{2} \Tr\int_\eps^\infty d\tau \left(e^{-\tau H} \delta H\right)
	= \delta\left(-\frac{1}{2} \Tr\int_\eps^\infty \frac{d\tau}{\tau} e^{-\tau H}\right) = \frac{1}{2} \delta [\Tr\log H]_\eps
\end{align}
which is trivially integrable. In retrospect, the even part is
certainly integrable since it corresponds to introducing a
Weyl spinor $\bar\chi$ transforming as $\psi$; then one can
simply combine $\psi$ and $\bar\chi$ into a Dirac fermion.
A straightforward calculation shows that
\begin{align}
\frac{1}{2} [\Tr\log H]_\eps = 
	-\frac{1}{32\pi^2} \left(
	\frac{\Tr [a_0^D]}{2\eps^2} + \frac{\Tr [a_1^D]}{\eps}
	- \frac{1}{2} \log \eps \, \Tr [a_2^D] + \textrm{finite}
	\right)
\end{align}
where
\begin{align}
\Tr [a_0^D] &= 4 \\
\Tr [a_1^D] &= \frac{1}{3} \mathcal R \\
\Tr [a_2^D]
	&= -\frac{4}{3} \Tr(F^{ab} F_{ab})
	- \frac{1}{10} C^{abcd} C_{abcd}
	+ \frac{11}{180} L_\chi
	+ \frac{1}{15} \Box \mathcal R
\end{align}

The odd part is not generally integrable. If it were, then $L_\eps^-$
would be the variation of the odd part of the effective action. Interpreting
the $\delta$ in $L_\eps^-$ as a differential operator, $L_\eps^-$ would be an
exact form and would obey $\delta L_\eps^- = 0$. However, one can show that
\begin{align}
C_\eps \equiv \delta L_\eps^- = \eps \int_0^1 d\lambda \, \Tr\left(
	\delta D e^{-\eps \lambda H} \delta \tilde D e^{-\eps \tilde\lambda \tilde H}
	\right)
\end{align}
(where $\tilde \lambda = 1-\lambda$) does not vanish in the limit of
vanishing $\eps$ due to singularities in the small $\eps$ limit
of the heat kernel operators appearing in the expression. Since
$\delta D = -\omega$ and $\delta \tilde D = -\tilde\omega$ are
local operators, we can perform the trace with a single insertion
of a complete set of states, giving
\begin{align}
C_\eps = \eps \int d^4x \,d^4x' \sqrt{g} \sqrt{g'}
	\int_0^1 d\lambda \,\Tr\left(
	\omega(x) U(x,x'; \eps \lambda)
	\tilde \omega(x') \tilde U(x',x; \eps \tilde \lambda)
	\right)
\end{align}
Since $\sigma(x,x') = \sigma(x',x)$ and $\Delta(x,x') = \Delta(x',x)$,
the above can be written as
\begin{align}
C_\eps = \frac{1}{(16 \pi^2)^2 \eps^3} \int_0^1 d\lambda \frac{1}{(\lambda \tilde\lambda)^2}
	\int d^4x \,d^4x' \sqrt{g} \sqrt{g'} e^{-\sigma / 2 \eps \lambda \tilde \lambda}
	\Delta(x,x') \eol
	\;\;\;\;\;\; \Tr\left(
	\omega(x) F(x,x'; \eps \lambda)
	\tilde \omega(x') \tilde F(x',x; \eps \tilde \lambda)
	\right)
\end{align}
One chooses $x'$ to be expanded in a normal coordinate system $y'$ about $x$.
Then rescaling $y' = y \times 2\sqrt{\eps \lambda \tilde \lambda}$
\begin{align}
C_\eps = \frac{1}{16 \pi^4 \eps} \int d^4x \sqrt{g}
	\int_0^1 d\lambda 
	\int d^4y \, e^{-y^2} \,
	\Tr\left(
	\omega(x) F(x,y'; \eps \lambda)
	\tilde \omega(y') \tilde F(y',x; \eps \tilde \lambda)
	\right)
\end{align}
One generally finds that $\Tr (\omega \tilde \omega)$ vanishes
(it certainly does in this case) and the triviality of $[a_0]$
guarantees that the only contribution comes from the two
$a_1$ coefficients:
\begin{align}
C = \lim_{\eps \rightarrow 0} C_\eps = \frac{1}{32 \pi^2} \int d^4x \sqrt{g}\,
	\Tr\left(
	\omega [a_1] \tilde \omega
	+ \omega \tilde \omega [\tilde a_1]
	\right)
	= \frac{i}{8 \pi^2} \int d^4x \sqrt{g}\,
	\Tr\left(
	\omega_a \omega_b \mathcal F_{cd}
	\right) \eps^{abcd}
\end{align}
where $\delta \mathcal A_b = \omega_b$. This vanishes precisely when the
symmetrized trace of three generators vanishes. This is the standard
anomaly cancellation condition and implies that the odd part of the
effective action can indeed be defined.

Since $C$ is by construction an exact local term, it can generally be
represented as the variation of a local finite counterterm $-\ell$ (defined up
to a closed form). Then one may add this counterterm to the $L_\eps^-$
and define (schematically)
\begin{align}
\delta[\Tr\log D]_\eps \equiv \frac{1}{2} \delta [\Tr\log H]_\eps + \left(L_\eps^- + \ell\right)
\end{align}
$\Tr\log H$ is generally free of gauge (but not conformal) anomalies,
and so the gauge anomaly is found in the two terms $L_\eps^-$
and $\ell$ by considering $\delta D$ to have the form of a gauge
transformation. Then $L_\eps^-$ gives the covariant gauge anomaly
and $\ell$ a finite piece which ensures that the sum has the form of a
consistent gauge anomaly. Since $\ell$ is defined only up to a closed
form, the consistent gauge anomaly is defined only up to the gauge variation
of some local term. The definition of $[\Tr\log D]_\eps$ so arrived at is not likely to
coincide with what we would have found by naively squaring
the operator, since the regulation method we have used
here damps out the high energy spectrum of the gauge invariant operator
$H$, whereas damping the high energy spectrum of $D^2$ does not have a
gauge invariant meaning. The method used here is to be preferred
since $C$ is generally free of divergences and therefore the divergent part
of $[\Tr\log D]_\eps$ is straightforwardly integrable. This procedure
is quite analogous to the normal perturbative calculation, where one
finds that the triangle diagram is not itself divergent but when
regulated produces an ambiguity in the effective action which requires
a prescription (which can be interpreted as the addition of a finite local
counterterm) in order to be defined.

\section{The case of chiral superfields}
We turn now to our actual interest: path integrals involving
chiral superfields in gravitational and gauge backgrounds.

The standard textbook coupling of supergravity to chiral matter
can be described by the conformal action\footnotemark
\begin{align}
S &= -3 \int d^4 \theta E \bar\Phi_0 \Phi_0 e^{-K/3}
	+ \left(\int d^2\theta \mathcal E \Phi_0^3 W + \hc\right) \eol
	&= -3 \left[\bar\Phi_0 \Phi_0 e^{-K/3}\right]_D
	+ \biggl(\left[\Phi_0^3 W\right]_F + \hc\biggr) \label{eq_chsugra}
\end{align}
In this expression, $K$ is the K\"ahler potential, a Hermitian function
of the chiral superfields $\Phi^i$ and their antichiral conjugates $\bar\Phi^\bi$;
$W$ is the superpotential, a chiral function of only $\Phi^i$; and
$\Phi_0$ is the conformal compensator, the only chiral superfield with
non-vanishing conformal and $U(1)_R$ weights, which are 1 and 2/3, respectively.
We denote the conformal and $U(1)_R$ weights of superfields by the ordered
pair $(\Delta, w)$, so $\Phi_0$ has weight $(1,2/3)$ and $\bar\Phi_0$ has
weight $(1,-2/3)$.
\footnotetext{For simplicity, we have neglected to include the possibility of a nontrivial holomorphic gauge
coupling for the Yang-Mills sector.} The action is invariant to
redefinitions of $\Phi_0 \rightarrow \Phi_0 e^{F/3}$ provided
$K$ and $W$ transform as $K \rightarrow K + F + \bar F$
and $W \rightarrow e^{-F} W$. When $\Phi_0$ is absorbed into the
frame of superspace, its reparametrization becomes the super Weyl
symmetry of Howe and Tucker \cite{howetucker} and the combined transformation is
the K\"ahler transformation.

Because the conformal requirements of the action are satisfied by $\Phi_0$,
$K$ and $W$ are allowed to be arbitrary. To retrieve the original
minimal supergravity formulation, one fixes the conformal gauge
by taking $\Phi_0=1$. The formulation of Cremmer et al \cite{Cremmer:1978hn},
found by taking $\Phi_0 = W^{-1/3}$, is strictly valid only when $W$ nowhere
vanishes. The formulation of Binetruy, Girardi and Grimm \cite{bgg} corresponds to
$\Phi_0 = e^{K/6}$. Yet in each of these formulations, the \emph{quanta}
of $\Phi_0$ remain in the Poincar\'e supergravity sector. Therefore,
we will avoid explicitly fixing the gauge of $\Phi_0$ until after path integrals
are taken.

This is not the only way to define a supergravity theory in superspace.
Another possibility is to allow the fields $\Phi^i$ to have non-vanishing
conformal dimension. One is immediately led to the more general form
\begin{align}
S &= \left[Z\right]_D
	+ \left[P\right]_F + \left[\bar P\right]_{\bar F}
\end{align}
where $Z$ is a weight $(2,0)$ function of chiral superfields $\Phi^I$
and their conjugates, and $P$ is a weight $(3, 2)$ purely chiral function.
In the gauge where $Z=-3$, the Einstein-Hilbert term has the standard
normalization. This more arbitrary choice is classically equivalent
to the previous one by choosing to single
out a particular chiral superfield of weight $(1,2/3)$ and rescaling all of the
other fields by it, turning them into projective variables. The K\"ahler
symmetry is then a redefinition of the projective coordinates \cite{Butter:2009cp}.

One may also choose to allow more general superfields than chiral ones.
A linear superfield of weight $(2,0)$ allows one to formulate new minimal
supergravity, where the matter couplings can be described by
\begin{align}
S \ni \left[L K\right]_D
\end{align}
Here $K$ is a Hermitian function of chiral superfields $\Phi^i$ of vanishing
weight. This theory is classically dual to \eqref{eq_chsugra} in the absence of a
superpotential, which cannot be posed because $\Phi^i$ have
vanishing $U(1)_R$ weight and so there is no way to formulate a function
of them with the necessary dimension. Allowing non-vanishing dimension for the
chiral superfields leads immediately to the more general form
\begin{align}
S &= \left[\mathcal Z\right]_D + \left[P\right]_F + \left[\bar P\right]_{\bar F}
\end{align}
where $\mathcal Z$ is weight $(2,0)$ and $P$ is $(3,2)$. One can suppose
$\mathcal Z$ to be linear in $L$, as $\mathcal Z = L K$, but there is no reason
(beyond simplicity) to impose this constraint. (In fact, one may even
introduce several linear superfields.)

These different conformal theories, even when classically dual, are not
necessarily quantum mechanically equivalent. The major stumbling block is to
formulate the Gaussian path integration for a quantum chiral superfield
$\eta$ of conformal dimension $\Delta$. Only for $\Delta=3/2$ (and therefore
$U(1)_R$ weight $w=1$) is the chiral Gaussian
\begin{align}
\int \CD\eta \CD\bar\eta \exp\left(-\int d^2\theta\, \mathcal E\, \eta^T \eta + \hc \right) \equiv 1
\end{align}
conformal and $U(1)_R$ invariant. These last invariances are necessary
for the chiral action to be supersymmetric. It is further evident that
this definition of the measure is only gauge invariant if $\eta$ is in
a real representation of the gauge group.

For more general $\eta$, it is possible to construct a gauge invariant
measure through the introduction of a field $M$\footnotemark
\begin{align}
\int \CD\eta \CD\bar\eta \exp\left(-\int d^2\theta\, \mathcal E\, \eta^T M \eta + \hc \right) \equiv 1
\end{align}
$M$ here is assumed to have the appropriate transformation properties
to render the measure gauge invariant. If an appropriate $M$ is naturally
furnished by the theory (as a function, perhaps, of the background
fields) then it may be used, but more often no such object exists. Inserting
a spurion field by hand does render a gauge invariant path integral, but
this does not eliminate the anomaly. Instead of having an effective action
which changes under a gauge transformation, one has an effective action which
changes if a different $M$ is chosen. These are, of course, the same
thing. \footnotetext{That the
measure integral has the same structure as a mass term is not coincidental; one way to
regulate the effective action we will discuss involves using this
measure field $M$ in a way analogous to a Pauli-Villars field.}

For the original supergravity and chiral matter model \eqref{eq_chsugra},
the conformal and $U(1)_R$ symmetries are effectively removed from the
theory through the use of $\Phi_0$ as a compensator field. All of the
other fields $\Phi^i$ and their quanta $\eta^i$ are chosen to have
vanishing conformal and $U(1)_R$ weights, and $\Phi_0^3$ is placed in
all chiral superspace integrations. In this way, the chiral measure
essentially becomes $\chE \Phi_0^3$. These theories amount then to
the choice $M = \Phi_0^3$. Any fields in complex representations
of gauge groups must have their path integration defined using some
other method, usually a perturbative method such as in \cite{McArthur:1985xd}.

This effectively converts the conformal theory with background $\Phi_0$ into
a Poincar\'e theory. The independent conformal and $U(1)_R$ symmetries of the
original theory survive as K\"ahler transformations of the Poincar\'e
theory. We note that if $\Phi_0$ is used in this way, the choice
$\Phi_0=1$ seems the simplest and most reasonable Gaussian path integration
for the Poincar\'e theory, but the choice for the overall factor of the measure
should presumably be equivalent to the choice of how precisely to regulate the
theory.

We will be concerned with calculating anomalies and divergences involving chiral loops.
Using the background field formalism, we split all chiral fields into a background piece
$\Phi^i$ and a quantum variation $\eta^i$,
\begin{align}\label{eq_bqsplit}
\Phi^i \rightarrow \Phi^i + \eta^i
\end{align}
All of the above theories we have mentioned have a common structure
for the part of the action quadratic in the quantum chiral superfield
$\eta^i$:
\begin{align}
S^{(2)} = \left[\bar\eta^\bi Z_{\bi j}\eta^j \right]_D + \frac{1}{2} \left(\left[\eta^i \mu_{ij} \eta^j\right]_F + \hc \right)
\end{align}
Any D-terms of the form $\eta^i Z_{ij} \eta^j$ have been chirally projected
and absorbed into $\mu_{ij}$. In performing the splitting \eqref{eq_bqsplit},
we have broken any manifest reparametrization invariance. In many classical
theories, chiral superfields parametrize a K\"ahler manifold with the
reparametrization symmetry
\begin{align}
\Phi^i \rightarrow \Lambda^i(\Phi)
\end{align}
This symmetry is manifested on the $\eta$ as
\begin{align}
\eta'^i = \frac{\partial \Lambda^i}{\partial \Phi^j} \eta^j + \mathcal O(\eta^2)
	= \Lambda^i{}_j \eta^j + \mathcal O(\eta^2)
\end{align}
In order to consistently truncate the expansion at the first term,
one would need to introduce a chiral connection for the coordinates
$\Phi$ \cite{Honerkamp:1971sh}.
Unfortunately, there is no natural object in the theory to play this role,
(the K\"ahler affine connection being non-chiral). However, provided
we work on shell, this will not be an
issue.\footnote{Alternatively, one could choose to introduce a chiral metric by hand
(which would presumably correspond to a ``chiral measure metric'' $M_{ij}$).
But this only cloaks the anomaly in a different form.}

These concerns are not major ones at the moment. As far as we are concerned,
the index $i$ can be interpreted as a gauge index; hence we
regard $S^{(2)}$ as simply
\begin{align}
S^{(2)} = \left[\bar\eta Z \eta \right]_D + \frac{1}{2} \left(\left[\eta^T \mu \eta\right]_F + \hc \right)
\end{align}
Writing this in Majorana form,
\begin{align}
S^{(2)} = \frac{1}{2} \left(
\begin{array}{cc}
\int \chE \eta^T & \int \bar\chE \bar \eta
\end{array}
\right)
\left(
\begin{array}{cc}
\mu & \CP Z^T \\
\ACP Z & \bar \mu
\end{array}
\right)
\left(
\begin{array}{c}
\eta \\
\bar \eta^T
\end{array}
\right)
\end{align}
The ``column vector'' on the right is an element of
$C_+(\mathbf r) \oplus C_-(\mathbf{\bar r})$, where
$C_+$ and $C_-$ denote respectively the spaces of chiral
and antichiral superfields and $\mathbf{r}$ and $\mathbf{\bar r}$ denote the
representations. The matrix in the center can be thought of as an operator
mapping $C_+(\mathbf{r}) \oplus C_-(\mathbf{\bar r})$ to the dual space
$C_+(\mathbf{s}) \oplus C_-(\mathbf{\bar s})$.
$\mathbf{r}$ and $\mathbf{s}$ are ``dual'' in the following way:
their index structures are conjugate in the normal Yang-Mills
sense, but their conformal and $U(1)_R$ charges are dual in the sense that
they add to $3$ and $2$, respectively.

We can introduce some suitable measure by requiring that the path integral of
\begin{align}
S_M = \frac{1}{2} \left(
\begin{array}{cc}
\int \chE \eta^T & \int \bar\chE \bar \eta
\end{array}
\right)
\left(
\begin{array}{cc}
M & 0\\
0 & \bar M
\end{array}
\right)
\left(
\begin{array}{c}
\eta \\
\bar \eta^T
\end{array}
\right)
\end{align}
be unity. Then path integration of the action $S^{(2)}$ involves calculating the
formal determinant of the operator
\begin{align}
\left(
\begin{array}{cc}
M & 0\\
0 & \bar M
\end{array}
\right)^{-1}
\left(
\begin{array}{cc}
\mu & \CP Z^T \\
\ACP Z & \bar \mu
\end{array}
\right)
= \left(
\begin{array}{cc}
M^{-1} \mu & M^{-1} \CP Z^T \\
\bar M^{-1} \ACP Z & \bar M^{-1} \bar \mu
\end{array}
\right)
\end{align}
on the space $C_+(\mathbf{r}) \oplus C_-(\mathbf{\bar r})$.
This is an endomorphism by construction (i.e. its domain and range are
the same space), so its determinant is at least formally sensible.
Equivalently, one could also calculate
\begin{align}
\left(
\begin{array}{cc}
\mu & \CP Z^T \\
\ACP Z & \bar \mu
\end{array}
\right)
\left(
\begin{array}{cc}
M & 0\\
0 & \bar M
\end{array}
\right)^{-1}
= \left(
\begin{array}{cc}
\mu M^{-1} & \CP Z^T \bar M^{-1} \\
\ACP Z M^{-1} & \bar \mu \bar M^{-1} 
\end{array}
\right)
\end{align}
on the space $C_+(\mathbf{s}) \oplus C_-(\mathbf{\bar s})$.

The above structure can be clarified by the example of a chiral superfield
in a background Yang-Mills field. We transform from the
space of covariantly chiral superfields $\Phi$ (which obey $\nabla^\dalpha \Phi=0$)
to the space of conventionally chiral superfields $\phi$ (which obey $D^\dalpha \phi=0$).
The transformation to the conventionally chiral notation involves the introduction
of the gauge prepotential $V$ and the action reads
\begin{align}
S = \left[\bar\eta e^V \eta \right]_D + \frac{1}{2} \left(\left[\eta^T \mu \eta\right]_F + \hc \right)
\end{align}
where $\mu$ is some chiral Majorana mass term.
The path integral measure can be defined by requiring the Gaussian integration of
\begin{align}
S_M = \frac{1}{2} \left[\eta^T \eta\right]_F + \hc
\end{align}
to yield unity. This amounts to choosing the spurionic measure field $M$ to be unity
in this particular gauge. The operator corresponding to $S^{(2)}$ is
\begin{align}
\left(
\begin{array}{cc}
\mu & -\frac{1}{4} \bar D^2 e^{V^T} \\
-\frac{1}{4} D^2 e^V & \bar \mu
\end{array}
\right)
\end{align}
and maps the space $C_+ \oplus C_-$ to itself. By ``degauging'' the theory, we
can define an operator whose determinant is at least sensible, however
it it not particularly calculable. Its square yields operators like
$\bar D^2 e^{V^T} D^2 e^V$ which are difficult to deal with unless in a real
representation, and there is no clear reason that the action should be invariant
under gauge transformations.\footnote{Even if a series of $\eta$ are chosen to have
vanishing anomaly coefficients, the determinant defined above will \emph{still}
give an anomalous effective action. In this case, though, the anomaly will be
cohomologically trivial: it can be removed by the addition of a local counterterm
to the effective action.}

In classical supergravity with a conformal compensator, the above action we considered
would instead have the form
\begin{align}
S^{(2)} = \left[\bar\Phi_0 \Phi_0 e^{-K/3} \, \bar\eta e^V \eta \right]_D + \frac{1}{2} \biggl(\left[\Phi_0^3 \eta^T \mu \eta\right]_F + \hc \biggr)
\end{align}
with the measure
\begin{align}
S_M = \frac{1}{2} \biggl(\left[\Phi_0^3 \eta^T \eta\right]_F + \hc \biggr).
\end{align}
This yields the operator
\begin{align}\label{eq_op1}
\left(
\begin{array}{cc}
\mu & -\frac{1}{4} \Phi_0^{-3} \bar \nabla^2 \bar\Phi_0 \Phi_0 e^{-K/3} e^{V^T} \\
-\frac{1}{4} \bar \Phi_0^{-3} \nabla^2 \bar\Phi_0 \Phi_0 e^{-K/3} e^V & \bar \mu
\end{array}
\right)
\end{align}
where $\nabla$ is the conformally covariant derivative \cite{Butter:2009cp}.
Note this approach involves degauging the Yang-Mills structure but leaving the
chiral superfields covariant with respect to the superconformal group.
Thus the operator acts on the space $C_+(0) \oplus C_-(0)$ where $0$ denotes
the conformal weight of $\eta$. Different choices for the conformal gauge of $\Phi_0$
give superficially different forms of the off-diagonal terms, but they are all conformally
equivalent.

Another approach is to absorb a factor of $\Phi_0^{3/2}$ into
$\eta$, or equivalently, split the measure factor onto both
sides of the operator. This gives
\begin{align}
\left(
\begin{array}{cc}
\mu & -\frac{1}{4} \bar \nabla^2 (\bar\Phi_0 \Phi_0)^{-2} e^{-K/3} e^{V^T} \\
-\frac{1}{4} \nabla^2 (\bar\Phi_0 \Phi_0)^{-2} e^{-K/3} e^V & \bar \mu
\end{array}
\right)
\end{align}
which acts on $C_+(3/2) \oplus C_-(3/2)$, but has the same
determinant as \eqref{eq_op1}. We will use this approach in
what follows.

The structure of these operators is quite generic in conformal
theories (or Poincar\'e theories with conformal compensators).
One generally finds
\begin{align}\label{eq_genop}
\left(
\begin{array}{cc}
\mu & \CP e^{V^T} X^{-1/2} \\
\ACP e^V X^{-1/2} & \bar \mu
\end{array}
\right)
\end{align}
acting on the space $C_+(3/2) \oplus C_-(3/2)$. 
The projectors $\CP = -\frac{1}{4} \bar\nabla^2$ and $\ACP = \frac{1}{4} \nabla^2$
are conformally covariant, $X$ is Hermitian function of conformal dimension two,
and $V$ is some generalized internal symmetry matrix.
We will henceforth interpret $V$ as a background gauge prepotential.

There is a classical invariance where a factor in $e^V / X^{1/2}$ may be considered
either as a contribution to the $U(1)$ part of $V$ or as a contribution to $X$.
We will refer to this as the ``$U(1)$ ambiguity.'' This classical symmetry
is broken by our definition of the effective action, which treats $e^V$ and $X$ in
an asymmetric way, and naturally an anomaly is introduced. It turns out that
this anomaly term is cohomologically trivial -- it is the variation of a local
counterterm -- and so the anomaly isn't truly physical.

In the operator \eqref{eq_genop}, the dimension two object $X$ could be
eliminated by fixing the conformal gauge so that $X$ is constant. There is
an equivalent way of proceeding which does \emph{not} explicitly fix the
conformal symmetry. We may introduce conformally compensated derivatives $\CD$ along with
superfields $R$, $G_c$ and $X_\alpha$ defined in terms of $X$ so that $X$ becomes covariantly
constant \cite{paper3} and the derivatives become those of Poincar\'e $U(1)$ supergravity.
Then $\CP = -\frac{1}{4} (\BCD^2 - 8 R)$ and $\ACP = -\frac{1}{4} (\CD^2 - 8 \bar R)$,
where we use the supergravity conventions of \cite{bgg}. This gives a structure that is
formally identical to gauging $X$ to be a constant, but because the conformal symmetry has
only been hidden as opposed to fixed, it is a bit more aesthetically appeasing.
Note that in this approach the $U(1)_R$ structure remains.\footnote{It
is possible to remove even the $U(1)_R$ symmetry by introducing another compensator $Y$
with weight $(0,1)$. The combination of $X$ and $Y$ can then be combined into a complex
compensator $\Psi$ of weight $(1, w)$ for arbitrary nonzero $w$. When $w=2/3$,
$\Psi$ may be further restricted to be chiral, and the original Poincar\'e
supergravity of \cite{wb} is recovered.}

The similarity of the structure of \eqref{eq_genop} to the Dirac operator is
compelling. We may define $D$ as this operator in the massless limit
\begin{align}\label{eq_Dop}
D \equiv \left(
\begin{array}{cc}
0 & \CP e^{V^T} X^{-1/2} \\
\ACP e^V X^{-1/2} & 0
\end{array}
\right)
\end{align}
and define its conjugate operator
\begin{align}
\tilde D = \left(
\begin{array}{cc}
0 & -\CP e^{-V} X^{-1/2} \\
-\ACP e^{-V^T} X^{-1/2} & 0
\end{array}
\right)
\end{align}
In choosing $\tilde D$ to enable a Leutywler-like quantization,
we have explicitly broken the classical $U(1)$ ambiguity since
$e^{-V} / X^{1/2}$ is not invariant under the same exchange of
$U(1)$ factors as its conjugate.

The Hermitian operator $H$ is
\begin{align}
H = \tilde D D = 
\left(
\begin{array}{cc}
-\CP e^{-V} X^{-1/2} \ACP e^V X^{-1/2} & 0\\
0 & -\ACP e^{-V^T} X^{-1/2} \CP e^{V^T} X^{-1/2}
\end{array}
\right)
\end{align}
Note that since $\tilde D$ is conjugate to $D$, the operators appearing
in $H$ are actually gauge covariant. We may absorb the various factors
of $e^V$ into gauge covariant derivatives (as well as commuting various
factors of $X$ past the derivatives) to yield
\begin{align}\label{eq_Hop}
H = X^{-1}
\left(
\begin{array}{cc}
-\frac{1}{16} (\BCD^2 - 8 R) (\CD^2 - 8\bar R) & 0\\
0 & -\frac{1}{16} (\CD^2 - 8\bar R)(\BCD^2 - 8 R) 
\end{array}
\right)
\end{align}
where we should properly interpret the space this acts on as
$C_+(1, \mathbf{r}) \oplus C_-(-1, \mathbf{s})$, the $1$ and
$-1$ denoting just the $U(1)_R$ charges now, since the
conformal structure has been hidden. (Before the conformal
and $U(1)_R$ charges were related so we needed only specify the
former.) Note that $X$ appears only as an overall factor,
compensating the conformal scale of the rest of the operator.
In actual calculations, $X$ can be presumed to be unity during
calculations and then restored in the final results using
dimensional analysis.

As we found in the case of the Dirac operator, the heat kernel
expansion of this operator encodes a great deal of information,
so we turn next to a derivation of that. Operators such as that
above have been considered many times in the literature
before \cite{McArthur, Buchbinder:1998qv},
but usually in the limit where the supergravity $U(1)_R$ was absent.
This corresponds to the case where $X$ is simply the product
of a chiral and an antichiral superfield (i.e. $X = \Phi_0 \bar\Phi_0$).
As the $U(1)_R$ is quite necessary for our purposes, we will
rederive similar results as those done before, but in the
case where $X$ is arbitrary and so the supergravity
$U(1)_R$ field strength $X_\alpha$ does not necessarily vanish.
Our results will therefore differ slightly from the literature
by terms involving $X_\alpha$.

\subsection{Heat kernel for a generic chiral superfield}
In deriving the heat kernel for a generic chiral superfield, we follow
closely the setup of Buchbinder and Kuzenko from their classic paper
\cite{Buchbinder:1988yu} as summarized in their textbook \cite{Buchbinder:1998qv}.
We refer the interested reader to their treatment of the subject.
The major difference here is that we work in $U(1)$ supergravity and
utilize normal coordinates in superspace in order to more easily apply
Avramidi's non-recursive technique.

The first step in deriving anomalies and divergences of \eqref{eq_Dop}
is to analyze the heat kernel structure of \eqref{eq_Hop}. Recall that the heat kernel
for a generic chiral superfield is the gauge and $U(1)_R$ covariant operator
$e^{\tau \mathcal O_+}$ where
\begin{align}
\mathcal O_+ \equiv \frac{1}{16} (\BCD^2 - 8 R) (\CD^2 - 8 \bar R)
\end{align}
acts on a chiral superfield of unit $U(1)_R$ weight.
This generalizes the global supersymmetric $\frac{1}{16} \bar D^2 D^2$.
Since the operator $\mathcal O_+$ acts only on chiral superfields, we may
expand it out as
\begin{align}
\mathcal O_+ \phi =&
	\Box \phi + W^\alpha \CD_\alpha \phi + \frac{1}{2} (\CD^\alpha W_\alpha) \phi
	- i G^{\dalpha \alpha} \CD_{\alpha \dalpha} \phi \eol
	& + \frac{1}{2} \CD^\alpha R \CD_\alpha \phi + \frac{1}{2} R \CD^2 \phi
	- \frac{1}{2} \BCD^2 \bar R \phi + 4 R\bar R \phi \eol
	& + \frac{1}{2} (1-w) X^\alpha \CD_\alpha \phi - \frac{1}{4} w (\CD^\alpha X_\alpha) \phi
\end{align}
where $\phi$ is assumed to be a chiral field of $U(1)_R$ weight $w$.
Our concern will be the case $w=1$, but we quote the general formula for reference.
With the exception of the two terms involving $W_\alpha$, which is specific to
the gauge group of $\phi$, all of the other terms in this expression are generic
supergravity terms.

One begins with the chiral heat kernel for the free theory
\begin{align}
U_0(\chz,\chz';\tau) = \frac{1}{(4\pi \tau)^2} \exp\left({-|y-y'|^2/4\tau}\right) (\theta-\theta')^2
\end{align}
in chiral coordinates $\chz = (y,\theta)$, where $\BCD^\dalpha = \partial^\dalpha$.
The additional factor of $(\theta-\theta')^2$ is to reproduce the chiral delta function:
$U_0(\chz,\chz'; 0) = \delta^4(y-y') \delta^2(\theta-\theta') = \delta^4(y-y') (\theta-\theta')^2$.
We generalize this to 
\begin{align}
U(\tau) = \frac{1}{(4\pi \tau)^2} \exp\left({-\Sigma/2\tau}\right) F
\end{align}
where $U(\chz, \chz'; \tau)$ (and $F$) is formally a bi-tensor chiral field of $U(1)_R$ weight 1 at \emph{both}
of its spacetime points. That is, for operators acting on $\chz$, $U$ is $U(1)_R$ weight 1.
However, under a \emph{global} $U(1)_R$ phase transformation, $U$ transforms with a total weight
of 2, just as $U_0$ does. The chiral bi-scalar $\Sigma$ has no chiral weight.

We demand $U(\tau)$ obey the heat equation
\begin{align}
\frac{\partial U}{\partial \tau} = \mathcal O_+ U
\end{align}
where $\mathcal O_+ = \frac{1}{16} (\BCD^2 - 8 R)(\CD^2 - 8 \bar R)$.

Before proceeding further, it is helpful to work out various operators we will encounter. 
The first is $\Box_+$, which is the chiral generalization of the d'Alembertian:
\begin{align}
\Box_+ \phi \equiv& \,\frac{1}{16} (\BCD^2 - 8 R) \CD^2 \phi \eol
	=& \,\Box \phi + W^\alpha \CD_\alpha \phi + \frac{1}{2} (\CD^\alpha W_\alpha) \phi
	- i G^{\dalpha \alpha} \CD_{\alpha \dalpha} \phi \eol
	& + \frac{1}{2} \CD^\alpha R \CD_\alpha \phi + \frac{1}{2} R \CD^2 \phi
	+ \frac{1}{2} (1-w) X^\alpha \CD_\alpha \phi - \frac{w}{4} (\CD^\alpha X_\alpha) \phi
\end{align}
This is related to $\mathcal O_+$ by
\begin{align}
\mathcal O_+ = \Box_+ - \frac{1}{2} (\BCD^2 - 8 R)\bar R
\end{align}
Note that $\Box_+$ vanishes on a covariantly constant $\phi$, while $\mathcal O_+$
includes an extra supergravity ``mass'' term.

Also of use will be the chiral generalization of $\CD^a \Sigma \CD_a \phi$, which
following Buchbinder and Kuzenko, we denote $\Sigma \ast \phi$:
\begin{align}
\Sigma \ast \phi \equiv& \,\frac{1}{16} (\BCD^2 - 8 R) \left(\CD^\alpha \Sigma \CD_\alpha \phi\right) \eol
	=& \,\CD^a \Sigma \CD_a \phi + \frac{R}{2} \CD^\alpha \Sigma \CD_\alpha \phi
	- \frac{1}{4} w \CD^\alpha \Sigma X_\alpha \phi + \frac{1}{2} \CD^\alpha \Sigma W_\alpha \phi
\end{align}

In terms of these new operations, the chiral heat equation takes the form
\begin{align}\label{eq_recursCHF}
-\frac{2}{\tau} F + \frac{\Sigma}{2\tau^2} F + \frac{\partial F}{\partial \tau} =
	\mathcal O_+ F - \frac{1}{2\tau} \Box_+ \Sigma \, F + \frac{1}{4\tau^2} \left(\Sigma \ast \Sigma\right)\, F
	- \frac{1}{\tau} \Sigma \ast F
\end{align}
which should be compared to the corresponding bosonic equation \eqref{eq_Fcurved0}.
As before, we demand the $1/\tau^2$ term yield an identity
\begin{align}
2 \Sigma = \Sigma \ast \Sigma
\end{align}
This equation is consistent with the chirality requirement of $\Sigma$.
The remaining term for $F$ can be simplified if we rescale $F$ by
$F  = \Delta^{1/2} \tilde F$ where $\Delta$ is some chiral determinant.
The result is
\begin{align*}
-\frac{2}{\tau} \tilde F + \frac{\partial \tilde F}{\partial \tau} =
	\tilde {\mathcal O}_+ \tilde F
	- \frac{1}{2\tau} \Box_+ \Sigma \, \tilde F
	- \frac{1}{\tau} \Sigma \ast \tilde F
	- \frac{1}{2\tau} \left(\Sigma \ast \log \Delta\right) \tilde F
\end{align*}
where $\tilde {\mathcal O}_+ = \Delta^{-1/2} \mathcal O_+ \Delta^{1/2}$.
We require $\Delta$ to obey the chiral equation
\begin{align}
4 = \Box_+\Sigma +  \Sigma \ast \log \Delta.
\end{align}
Provided there is no barrier to finding a chiral $\Sigma$ and $\Delta$
which obey these properties, we find the simple chiral equation
\begin{align}
\frac{\partial \tilde F}{\partial \tau}
	+ \frac{1}{\tau} D \tilde F = \tilde {\mathcal O}_+ \tilde F
\end{align}
where we have introduced the chiral operator $D \tilde F \equiv \Sigma \ast \tilde F$
to mimic the final form of the bosonic expression \eqref{eq_Fcurved}. Given the similarity between the
above formulae and the bosonic formulae, we expect their solution to take roughly the
same form. Aside from some complications and some simplifications, this will be the
case.

Note that we have not yet specified the chiral weight of $\Delta$ and $\tilde F$.
In the non-supersymmetric case, $\Delta$ was given in normal coordinates by $e^{-1}$;
we expect the chiral $\Delta$ to be given in normal coordinates by $\chE^{-1}$.
Thus we shall take $\Delta$ to have chiral weight $2$ on its $\chz$ coordinate
and $-2$ on its $\chz'$ coordinate, and so $\tilde F$ has vanishing
chiral weight on $\chz$ but weight 2 on $\chz'$.

\subsection{Chiral normal coordinates}
Before proceeding to a comprehensive analysis of the chiral heat kernel,
we need to construct a useful set of normal coordinates as in the non-supersymmetric case.
Here the procedure is a little more sophisticated, since we have coordinates associated
with $P$, $Q$, and $\bar Q$ and so several ways one might define a normal coordinate
system.

Recall that normal gauge in bosonic cooridnates was defined by requiring that the Taylor
expansion $\phi(y) = e^{y\partial} \phi$ match the covariant Taylor expansion
$\phi(y) = e^{y\cdot P} \phi$ where $P$ was the formal parallel transport operator 
(i.e. the covariant derivative). In superspace, there are three distinct coordinates
$(x,\theta,\btheta)$ and -- even in flat superspace -- several different ways of
constructing a normal coordinate system. Within global supersymmetry, Hermitian (or vector)
superspace is defined by
\begin{align}
\Psi(x,\theta,\btheta) = \exp(x P + \theta Q + \btheta \bar Q) \Psi
\end{align}
whereas chiral superspace is defined by
\begin{align}
\Psi(y,\theta,\btheta) = \exp(y P + \theta Q) \exp(\btheta \bar Q) \Psi.
\end{align}
where $\Psi$ is an arbitrary superfield.
The advantage of chiral superspace is that the chirality condition reduces to
independence of the coordinate $\btheta$ (since formally $\bar Q$ annihilates any chiral
superfield). Thus $D^\dalpha = \partial^\dalpha$ and the antichiral vierbein $E^{\dmu A}$
and its inverse $E^{\dalpha M}$ are especially simple.

We require a chiral set of normal coordinates so we shall follow suit in placing
$\exp(\btheta \bar Q)$ to the far right. However, there are several ways in which one
might define the remainder. The simple Lorentz invariant options are
\[
\exp(y P + \eta Q), \;\;\;
\exp(y P) \exp(\eta Q), \;\;\;
\textrm{or}\,\,\,
\exp(\eta Q) \exp(y P)
\]
where we introduce $\eta$ to denote the normal coordinate difference between
$\theta$ and $\theta'$. Within global supersymmetry, these are equivalent since $[Q,P]$ vanishes,
but not so in curved superspace. The first is the most symmetric and yields
a normal mode expansion in $y$ and $\eta$ completely analogous to the bosonic
case. The second is the one most useful when the spinor connections need to be
simplified. In fact, in converting an $F$-term integral to a component $x$-space
integral, one works in a coordinate system that amounts to having extracted
$\exp(\eta Q)$ to the far right. That this is suitable for components is
clear by noting that the expansion of $\phi(y,\eta)$ then looks like
\[
\CD_y \cdots \CD_y \CD_\eta \cdots \CD_\eta \phi.
\]
which is how one would naturally order these derivatives when projecting to
lowest components.

However, both of these latter two coordinate systems turn out to lack the
properties we will need. It turns out that the best system for our purposes
is the third. We define therefore
\begin{align}
G \equiv \exp(\eta Q) \exp(y P) \exp(\bar\eta \bar Q)
\end{align}
The connections are then found by first differentiating $G$,
\begin{align}
G^{-1} \partial_M G = \tilde E_M{}^A P_A + \tilde H_M{}^{\ul b} X_{\ul b}
\end{align}
and then operating with $G$ on the result:\footnote{One can simplify the last
step by reinterpreting the tilded connections as
having an extra implicit $y$ dependence in all the covariant terms, replacing
each with their covariant Taylor expansion.}
\[
E_M{}^A \equiv G \tilde E_M{}^A, \;\;\;
H_M{}^{\ul b} \equiv G \tilde H_M{}^{\ul b}.
\]
Here $P_A$ represents the formal translation operator (which is
represented on fields by the covariant derivative) and the set of
$X_{\ul b}$ consists of Lorentz, $U(1)_R$, and Yang-Mills generators.
$H_M{}^{\ul b}$ are the connections corresponding to the $X_{\ul b}$.

One immediately finds for $M=\dmu$ the connections take the rather simple form
\begin{gather}
E^{\dmu A} = \delta^{\dmu A} \left(1 - \bar\eta^2 R\right), \;\;\;
\omega^{\dmu}(M) = \frac{1}{2} \bar\eta_\dalpha R^{\dalpha \dmu}(M), \;\;\;
A^{\dmu} = 0 ,\;\;\;
{\mathcal A}^{\dmu} = 0
\end{gather}
Here we use an italicized $\mathcal A$ for the Yang-Mills connection to
distinguish it from the supergravity $U(1)_R$ connection $A$.
The inverse vierbein is easily found and allows us to write the connections
with a Lorentz form index
\begin{gather}
E^{\dalpha M} = \delta^{\dalpha M} \left(1 + \bar\eta^2 R\right), \;\;\;
\omega^{\dbeta}(M) = \frac{1}{2} \bar\eta_\dalpha R^{\dalpha \dbeta}(M), \;\;\;
A^{\dalpha} = 0 ,\;\;\;
{\mathcal A}^{\dalpha} = 0
\end{gather}
from which it is straightforward to show that when acting on an arbitrary superfield $\Psi$
without any dotted spinor indices,
\begin{align}
(\BCD^2 - 8 R) \Psi = \partial_\dmu\partial^\dmu (1 + 2 \bar\eta^2 R) \Psi
\end{align}
and so the result is explicitly independent of $\bar \eta$ and therefore chiral.

For $M=m$, the connections are given by
\begin{align}
\tilde W_m = \exp(-\bar\eta \bar Q) e^{-y P} \partial_m e^{y P} \exp(\bar\eta \bar Q)
\end{align}
Defining
\begin{align}
\hat W_m = e^{-y P} \partial_m e^{y P}
\end{align}
we then have
\begin{align}
\tilde W_m &= \exp(-\bar\eta \bar Q) \hat W_m{}^A \exp(\bar\eta \bar Q) \times
	\exp(-\bar\eta \bar Q) X_A \exp(\bar\eta \bar Q) \eol
	&= \exp(-\bar\eta \bar Q) \hat W_m{}^B \exp(\bar\eta \bar Q) \times X(\bar\eta)_B{}^A X_A
\end{align}
The final result is
\begin{align}
W_m{}^A &= \left(e^{\eta Q} e^{y P} \hat W_m{}^A\right) \times G X(\bar\eta)_B{}^A
\end{align}
Note that $X(0)_B{}^A = \delta_B{}^A$.

For $M=\mu$, the connections are given by
\begin{align}
\tilde W_\mu = \exp(-\bar\eta \bar Q) e^{-y P} e^{-\eta Q} \partial_\mu e^{\eta Q} e^{y P} \exp(\bar\eta \bar Q)
\end{align}
We first define
\begin{align}
\hat W_\mu = e^{-\eta Q} \partial_\mu e^{\eta Q}
\end{align}
which is rather simple. One finds
\begin{gather}
\hat E_{\mu}{}^A = \delta_\mu{}^A \left(1 - \eta^2 \bar R\right), \;\;\;
\hat \omega_{\mu}(M) = \frac{1}{2} \eta^\alpha R_{\alpha \mu}(M), \;\;\;
\hat A_{\mu} = 0 ,\;\;\;
{\hat {\mathcal A}}_{\mu} = 0
\end{gather}
Defining $G_y = \exp(y P)$ and $G_{\bar\eta}=\exp(\bar\eta\bar Q)$, we then have
\begin{align}
\tilde W_\mu &= G_{\bar\eta}^{-1} G_y^{-1} \hat W_\mu{}^A  G_y G_{\bar \eta} \times
	 G_{\bar\eta}^{-1} G_y^{-1} X_A G_y G_{\bar \eta} \eol
	&= G_{\bar\eta}^{-1} G_y^{-1} \hat W_\mu{}^B  G_y G_{\bar \eta} \times X(y, \bar\eta)_B{}^A X_A
\end{align}
which gives
\begin{align}
W_\mu{}^A &= \left(G_\eta \hat W_\mu{}^B \right) \times G X(y, \bar\eta)_B{}^A
\end{align}

We are most interested in the case where $\bar\eta=0$, since our heat kernel has
$\bar\theta'$ equal to $\bar\theta$. Following the non-supersymmetric case, we would like
to define $\Sigma = y^2/2$. For this to work requires $E_a{}^m y_m = y_a$ as well as
$E_\alpha{}^m y_m = 0$ -- both of which we take when $\bar\eta$ vanishes but for
arbitrary $y$ and $\eta$. Note that if we define $Y^M = (y^m, 0, 0)$, then
the above conditions -- along with $E^{\dalpha m} = 0$ which always holds in
chiral coordinates -- lead to
\[
E_A{}^M Y_M = Y_A \; \Longleftrightarrow \,Y_M = E_M{}^A Y_A
\]
so we require $E_m{}^a y_a = y_m$ and $E_\mu{}^a y_a = 0$. The first is easy
to see. It follows from $\hat E_m{}^a y_a = y_m$, which is true just as in
the non-supersymmetric case. Any term generated in $\hat E_m{}^a$ past the leading term
arose from commuting a $P$ with a $P$ or with an $M$. (No $P$ can be generated
by commuting a $P$ with a $Q$ or $\bar Q$.) Thus all the terms with a free
index $a$ will be of the form $T_{c y}{}^a$ or $R_{DC y}{}^a$. The latter
vanishes by antisymmetry of the final two indices and the former vanishes since
in the space we have, the bosonic torsion $T_{cba}$ is totally antisymmetric.
(It is proportional to $G^d \eps_{dcba}$.)

The condition for $E_\mu{}^a y_a = 0$ follows for essentially the same reason.
One notes that since the only nonzero hatted connections are
$\hat E_\mu{}^\alpha$ and $\hat \omega_\mu(M)$, we need only show that
$X_\alpha{}^a y_a = 0$ and $X_{(M)}{}^a y_a = 0$. The Lorentz term vanishes since
conjugating $M_{cd}$ by $e^{-y P}$ only gives a $P$ from terms that look like
$[M, y P]$ or $[P, y P]$ -- these both vanish as in the non-supersymmetric case.
The $Q_\alpha$ term vanishes since the only way to generate a $P$ from
commuting several $y P$'s with the initial $Q_\alpha$ is to first generate
an $M$, then commute $[M, y P]$. (This is because $[Q,P]$ by itself does not
generate a $P$.)

Thus we are free to define $\Sigma = y^2/2$. This then obeys
\begin{align}
2 \Sigma = \Sigma \ast \Sigma = \CD^a \Sigma \CD_a \Sigma + 0
	= y^a y_a
\end{align}
trivially. Note this result is consistently chiral.

Next we turn to our definition of $\Delta$. We define $\Delta = \det (E_{\ch A}{}^{\ch M}) = \chE^{-1}$
where we understand the indices $\ch A$ and $\ch M$ in $\chE$ to be only over $(a,\alpha)$
and $(m,\mu)$. We require
\begin{align}
4 = \Box_+\Sigma +  \Sigma \ast \log \Delta.
\end{align}
which amounts to
\begin{align*}
4 = \Box \Sigma - i G^{\dalpha \alpha} \CD_{\alpha \dalpha} \Sigma
	+ \frac{1}{2} \CD^\alpha R \CD_\alpha \Sigma + \frac{1}{2} R \CD^2 \Sigma
	+ \CD^a \Sigma \CD_a \log \Delta + \frac{R}{2} \CD^\alpha \Sigma \CD_\alpha \log\Delta
\end{align*}
Proceeding in a way analogous to the non-supersymmetric case, we consider taking a derivative of $\log \Delta$:
\begin{align*}
\CD_{\mathcal A} \log \Delta &= \CD_{\mathcal A} E_{\mathcal B}{}^{\mathcal M} E_{\mathcal M}{}^{\mathcal B}
	= E_{\mathcal M}{}^{\mathcal B} \CD_{\mathcal B} E_{\mathcal A}{}^{\mathcal M}
		- T_{\mathcal A \mathcal B}{}^{\mathcal M} E_{\mathcal M}{}^{\mathcal B}
\end{align*}
Here we are using an implicit grading for the indices \cite{Butter:2009cp}.
Since $E^{\dmu \mathcal B}$ vanishes, the last term becomes a trace of the torsion tensor
in the chiral space. The remaining terms become
\begin{align*}
\CD_{\mathcal A} \log \Delta
	&= \CD_{\mathcal M} E_{\mathcal A}{}^{\mathcal M}
	- E_{\mathcal M \dbeta} \CD^{\dbeta} E_{\mathcal A}{}^{\mathcal M}
		- T_{\mathcal A \mathcal B}{}^{\mathcal B} \\
	&= \CD_{\mathcal M} E_{\mathcal A}{}^{\mathcal M}
	+ E_{\mathcal M \dbeta} T^\dbeta{}_{\mathcal A}{}^{\mathcal M} - T_{\mathcal A \mathcal B}{}^{\mathcal B} \\
	&= \CD_{\mathcal M} E_{\mathcal A}{}^{\mathcal M}
	+ \left(T_\dbeta{}_{\mathcal A}{}^\dbeta - T_\dbeta{}_{\mathcal A}{}^D E_{D \dmu} E^\dmu{}^\dbeta \right)
		- T_{\mathcal A \mathcal B}{}^{\mathcal B} \\
	&= \CD_{\mathcal M} E_{a}{}^{\mathcal M} - T_{\mathcal A \mathcal B}{}^{\mathcal B}
		+T_{\mathcal A}{}_{\dbeta}{}^{\mathcal D} E_{\mathcal D}{}^{\dbeta}
\end{align*}
This gives (using $T_{a}{}^\beta{}_\beta = 2i G_a$)
\begin{align*}
4 = \CD_{\mathcal M}
	\left(\CD^a \Sigma E_a{}^{\mathcal M} + \frac{R}{2} \CD^\alpha \Sigma E_\alpha{}^{\mathcal M} \right)
\end{align*}
Since the result in the parentheses is invariant under all symmetry operations, we
can replace the overall $\CD_{\mathcal M}$ by $\partial_{\mathcal M}$. Since the
derivative involves only $y$ and $\eta$ derivatives, we can cleanly set
$\bar\eta=0$ within the parentheses, which leave behind a single factor of $y^m$ within,
giving the result.

For the calculation of the chiral heat kernel, we will need the vierbein to
second order in the coordinates $y$ and $\eta$. Omitting the details, the
result is
\begin{align}
E_m{}^a &= \delta_m{}^a + \frac{1}{2} T_{ym}{}^a + \frac{1}{3} \CD_y T_{ym}{}^a
	+ \frac{1}{2} \CD_\eta T_{ym}{}^a - \frac{1}{6} T_{ym}{}^b T_{b y}{}^a
	+ \frac{1}{6} R_{ymy}{}^a \eol
E_m{}^\alpha &= \frac{1}{2} T_{ym}{}^\alpha + \frac{1}{3} \CD_y T_{ym}{}^\alpha
	+ \frac{1}{2} \CD_\eta T_{ym}{}^\alpha - \frac{1}{6} T_{ym}{}^B T_{B y}{}^\alpha \eol
E_m{}_\dalpha &= \frac{1}{2} T_{ym}{}_\dalpha + \frac{1}{3} \CD_y T_{ym}{}_\dalpha
	+ \frac{1}{2} \CD_\eta T_{ym}{}_\dalpha - \frac{1}{6} T_{ym}{}^B T_{B y}{}_\dalpha \eol
E_\mu{}^\alpha &= \delta_\mu{}^\alpha + T_{y \mu}{}^\alpha + \frac{1}{2} \CD_y T_{y \mu}{}^\alpha
	- \frac{1}{2} T_{y \mu}{}^\beta T_{\beta y}{}^\alpha
	- \frac{1}{2} T_{y \mu}{}_\dbeta T^\dbeta{}_y{}^\alpha
	+ \CD_\eta T_{y \mu}{}^\alpha - \eta^2 \bar R \,\delta_\mu{}^\alpha \eol
E_\mu{}^a &= \frac{1}{2} R_{y \mu y}{}^a + \frac{1}{2} R_{\eta\mu y}{}^a \eol
E_{\mu \dalpha} &= T_{y \mu \dalpha} + \frac{1}{2} \CD_y T_{y \mu \dalpha}
	- \frac{1}{2} T_{y \mu}{}^\beta T_{\beta y \dalpha}
	- \frac{1}{2} T_{y \mu\dbeta} T^\dbeta{}_{y \dalpha}
\end{align}
We will need the following inverses to second order:
\begin{align}
E_a{}^m &= \delta_a{}^m - \frac{1}{2} T_{ya}{}^m - \frac{1}{3} \CD_y T_{ya}{}^m
	- \frac{1}{2} \CD_\eta T_{ya}{}^m - \frac{1}{12} T_{ya}{}^b T_{b y}{}^m
	- \frac{1}{6} R_{yay}{}^m \eol
E_a{}^\mu &= -\frac{1}{2} T_{ya}{}^\mu - \frac{1}{3} \CD_y T_{ya}{}^\mu
	- \frac{1}{2} \CD_\eta T_{ya}{}^\mu - \frac{1}{12} T_{ya}{}^b T_{b y}{}^\mu
	- \frac{1}{3} T_{ya}{}^\beta T_{\beta y}{}^\mu + \frac{1}{6} T_{y a \dbeta} T^{\dbeta}{}_y{}^\mu \eol
E_\alpha{}^\mu &= \delta_\alpha{}^\mu - T_{y \alpha}{}^\mu - \frac{1}{2} \CD_y T_{y \alpha}{}^\mu
	- \frac{1}{2} T_{y \alpha}{}^\beta T_{\beta y}{}^\mu
	+ \frac{1}{2} T_{y \alpha}{}_\dbeta T^\dbeta{}_y{}^\mu
	- \CD_\eta T_{y \alpha}{}^\mu + \eta^2 \bar R \,\delta_\alpha{}^\mu \eol
E_\alpha{}^m &= -\frac{1}{2} R_{y \alpha y}{}^m - \frac{1}{2} R_{\eta\alpha y}{}^m
\end{align}
One specific combination which we will use a great deal is
\begin{align}
X^\mu{}_\mu =& E^{a \mu} E_{a \mu} - \frac{1}{2} R E^{\alpha \mu} E_{\alpha \mu} \eol
	=& \frac{1}{4} T_y{}^{a \mu} T_{y a \mu} - R + R T_{y \alpha}{}^\alpha
	+ \frac{1}{2} \CD_y T_{y\alpha}{}^\alpha + R \CD_\eta T_{y \alpha}{}^\alpha
	- 2 \eta^2 R \bar R \eol
	& \;\;\;\;\;\; - \frac{R}{2} T_{y \alpha \dbeta} T^\dbeta{}_y{}^\alpha
	- \frac{R}{2} T_y{}^{\alpha \mu} T_{y \alpha \mu}
	+ \frac{R}{2} T_y{}^{\alpha \beta} T_{y \beta \alpha}
\end{align}
The explicit $R$ terms in the above are to be understood as $R(y,\eta)$
where
\begin{align}
R(y,\eta) = R + \CD_y R + \CD_\eta R + \frac{1}{2} \CD_y \CD_y R
	+ \frac{1}{2} \CD_\eta \CD_\eta R + \CD_\eta \CD_y R + \ldots
\end{align}

\subsection{Chiral heat kernel analysis}
The remaining differential equation for our heat kernel reads
\begin{align}
\frac{\partial \tilde F}{\partial\tau} + \frac{D \tilde F}{\tau}= 
	 \tilde {\mathcal O_+} \tilde F
\end{align}
for
\begin{align}
D \equiv \CD^a \Sigma \CD_a + \frac{R}{2} \CD^\alpha \Sigma \CD_\alpha
	+ \frac{1}{2} \CD^\alpha \Sigma W_\alpha
\end{align}
(Recall that $\tilde F$ has $U(1)_R$ weight 0 on its $\chz$ coordinate,
where $D$ acts.) In normal coordinates at $\bar\eta=0$, the above simplifies
drastically. We end up with
\begin{align}
D = y^a \CD_a = y^m \partial_m
\end{align}

We assume $F$ can be expanded as a power series in $\tau$
with $\tilde F = \sum_{n=0} A_n \tau^n / n!$, which gives recursion
relations which we can solve just as before. (We neglect placing
tildes on the coefficients $A$ for notational simplicity.) We fix
\begin{align}
A_0 = \eta^2
\end{align}
to obey both the differential equation and the necessary $\tau=0$ boundary
condition. The rest of the coefficients follow via the formal solution
of Avramidi \cite{avramidi}
\begin{align}
A_n = \left(1 + \frac{D}{n} \right)^{-1} \tilde {\mathcal O}_+ \left(1+\frac{D}{n-1}\right)^{-1}
	\tilde {\mathcal O}_+ \cdots \left(1 + D \right)^{-1} \tilde {\mathcal O}_+ \eta^2
\end{align}
As before we seek analytic power series solutions, except now the power series are
in $\eta$ as well as $y$, giving a generic ket $\ket{n,\nu}$.
Since the $\eta$ series terminates for $\nu\geq 3$, we have the the generic kets
\begin{gather}
\ket{n,0} = \ket{n}, \;\;\;
\ket{n,1} = \ket{n} \times \eta^{\beta_1}, \;\;\;
\ket{n,2} = \ket{n} \times \eta^2
\end{gather}
where $\ket n$ is as defined in the non-supersymmetric case.
We define the corresponding bras by
\begin{gather}
\bra{n,0} = \bra{n}, \;\;\;
\bra{n,1} = \bra{n} \times \partial_{\alpha_1}, \;\;\;
\bra{n,2} = -\frac{1}{4} \,\bra{n} \times \partial^{\alpha}\partial_\alpha
\end{gather}
It then follows easily as in the non-supersymmetric case
\begin{align}\label{eq_recursASF}
\braket{k,\kappa}{A_n} = \sum_{j_1,\ldots,j_{k-1} \geq 0} \;
	& \sum_{2 \geq \gamma_1,\ldots,\gamma_{k-1} \geq 0}\;
	\left(1 + \frac{k}{n}\right)^{-1}
	 \left(1 + \frac{j_{n-1}}{n-1}\right)^{-1} \cdots \left(1 + j_1 \right)^{-1} 
	\times \eol
	& \opbraket{k,\kappa}{\tilde {\mathcal O}_+}{j_{n-1},\gamma_{n-1}}
	\opbraket{j_{n-1},\gamma_{n-1}}{\tilde {\mathcal O}_+}{j_{n-2},\gamma_{n-2}}
	\cdots \opbraket{j_1,\gamma_1}{\tilde {\mathcal O}_+}{0,2}
\end{align}

We turn now to the structure of $\tilde{\mathcal O}_+$. One finds after a great
deal of work
\begin{align*}
\tilde{\mathcal O}_+ \tilde F =&\,
	\CD_{\mathcal M} \left(E^{a \mathcal M} \CD_a \tilde F + \frac{1}{2} R E^{\alpha \mathcal M} \CD_\alpha \tilde F\right)
	+ W^\alpha \CD_\alpha \tilde F + \frac{1}{2} (\CD^\alpha W_\alpha) \tilde F
	+ \frac{1}{2} W^\alpha (\CD_{\mathcal M} E_\alpha{}^{\mathcal M}) \tilde F \\
	& + \left(\Delta^{-1/2} \mathcal O_+ \Delta^{1/2}\right)\tilde F
\end{align*}
This operator can be rewritten in the manifestly symmetric form
\begin{align}
\tilde{\mathcal O}_+ =& \,
	\CD_{\ch M} X^{\ch M \ch N} \CD_{\ch N}
	+ \frac{1}{2} W^\alpha \chE_\alpha{}^{\ch M} \CD_{\ch M} + \frac{1}{2} \CD_{\ch M} \chE^{\alpha \ch M} W_\alpha
	+ \left(\Delta^{-1/2} \mathcal O_+ \Delta^{1/2}\right)
\end{align}
We have used $\mathcal E_{\ch A}{}^{\ch M}$ in place of $E_{\ch A}{}^{\ch M}$ since all
$\bar\eta$ derivatives have been removed and so we may take $\bar\eta$ to vanish without
incident. The above form is particularly striking since the operator is clearly
self-adjoint up to a change in the representation of the gauge field strength:
\begin{align}
\tilde{\mathcal O}_+^T(W_\alpha) = \tilde{\mathcal O}_+(-W_\alpha^T)
\end{align}
This is sensible since $\mathcal O_+$ appears naturally
acting between a chiral superfield $\Phi_1$ and its conjugate $\Phi_2$,
\begin{align}\label{eq_opherm}
\int \chE\, \Phi_2^T \mathcal O_+ \Phi_1 =
	\int \chE\, (\mathcal O_+ \Phi_2)^T \Phi_1 = 
	\int \chE\,  \Phi_1^T \mathcal O_+ \Phi_2
\end{align}
which is a gauge invariant expression only if $\Phi_2$ is in the representation
conjugate to $\Phi_1$.

We have introduced the ``chiral metric''
\begin{align}
X^{\ch M \ch N} = \chE^{a \ch M} \chE_a{}^{\ch N} + \frac{1}{2} R \chE^{\alpha \ch M} \chE_{\alpha}{}^{\ch N}
\end{align}
where $\ch M$ and $\ch N$ are only the chiral spinor and bosonic index.
In all these formulae an implicit grading has been used.

In general $\tilde{\mathcal O}_+$ has the form
\begin{align}
\tilde{\mathcal O_+} = 
	X^{\chM \chN} \partial_\chN \partial_\chM + Y^\chM \partial_\chM + Z
\end{align}
We have
\begin{align}
Y^\chM =& -2 X^{\chM\chN} H_\chN + \partial_\chN X^{\chN\chM} + W^\alpha E_\alpha{}^\chM \eol
Z =& -X^{\chM\chN} \partial_\chN H_\chM + X^{\chM\chN} H_\chN H_\chM - (\partial_\chM X^{\chM\chN}) H_\chN \eol
	&\, + \frac{1}{2} \CD^\alpha W_\alpha + \frac{1}{2} (\partial_\chM \chE^{\alpha \chM}) W_\alpha - W^\alpha E_\alpha{}^\chM H_\chM + \Delta^{-1/2} \mathcal O_+ \Delta^{1/2} 
\end{align}
Aside from the terms involving $W^\alpha$, the above form is strikingly similar to the
non-supersymmetric case, with $X^{\mathcal M \mathcal N}$ replacing $g^{mn}$.
The connection $H$ is really just the Yang-Mills connection $\mathcal A$; the
heat kernel function $\tilde F$ has only a Yang-Mills structure since all its
$U(1)_R$ weight is on the $\chz'$ coordinate, not the $\chz$ coordinate. If we
were to generalize our approach to include chiral superfields with Lorentz indices,
the Lorentz connection would appear here as well.

Before proceeding further, we should note the projections to $y=0$ and $\eta=0$ of the
terms given above:
\begin{gather*}
[X^{mn}] = \eta^{mn}, \;\;\; [X^{m \nu}] = 0, \;\;\; [X^{\mu \nu}] = \frac{1}{2} R \eps^{\mu \nu} \\
[Y^m] = 0, \;\;\; [Y^\mu] = \frac{1}{2} \CD^\mu R + W^\mu \\
[Z] = \frac{1}{2} \CD^\alpha W_\alpha + [\Delta^{-1/2} \mathcal O_+ \Delta^{1/2}]
\end{gather*}

The quartic divergence is proportional to $\braket{0,0}{A_0}$ which vanishes 
as required by supersymmetry.

The quadratic divergence is proportional to
\begin{align}
\braket{0,0}{A_1} = \opbraket{0,0}{\tilde {\mathcal O}_+}{0,2} = 2 [X^{\mu}{}_\mu] = - 2 R
\end{align}
This is an F-term, so the corresponding D-term would simply be 1.
In a sense, the quadratic divergence in superspace is most like the quartic divergence in
normal space.

The logarithmic divergence is given by
\[
\braket{0,0}{A_2} = \sum_{j_1, \gamma_1} \left(1 + j_1 \right)^{-1}
	\opbraket{0,0}{\tilde {\mathcal O}_+}{j_1, \gamma_1}
	\opbraket{j_1,\gamma_1}{\tilde {\mathcal O}_+}{0,2}
\]
The first matrix element vanishes trivially unless $\gamma_1 + j_1 \leq 2$.
Those satisfying this requirement are
\begin{align*}
\opbraket{0,0}{\tilde {\mathcal O}_+}{0,0} &= [Z] \\
\opbraket{0,0}{\tilde {\mathcal O}_+}{0,1} &= [Y^{\beta_1}] = W^{\beta_1} + \frac{1}{2} \CD^{\beta_1} R \\
\opbraket{0,0}{\tilde {\mathcal O}_+}{1,0} &= [Y^{b_1}] = 0 \\
\opbraket{0,0}{\tilde {\mathcal O}_+}{0,2} &= [2 X^{\alpha}{}_\alpha] = -2 R \\
\opbraket{0,0}{\tilde {\mathcal O}_+}{1,1} &= [X^{b_1 \beta_1}] = 0 \\
\opbraket{0,0}{\tilde {\mathcal O}_+}{2,0} &= [X^{b_1 b_2}] = \eta^{b_1 b_2}
\end{align*}
We require the product of these with $\opbraket{j_1,\gamma_1}{\tilde {\mathcal O}_+}{0,2}$
for $(j_1, \gamma_1) = \{(0,0), (0,1), (0,2), (2,0)\}$. The first case we've already
found. The second is
\begin{align*}
\opbraket{0,1}{\tilde {\mathcal O}_+}{0,2} &= [2 \partial_{\alpha_1} X^\phi{}_\phi] + 2 [Y_{\alpha_1}]
\end{align*}
It is straightforward to show $[\partial_{\alpha_1} X^\phi{}_\phi] = -\CD_\alpha R$, giving
\begin{align*}
\opbraket{0,1}{\tilde {\mathcal O}_+}{0,2} &= -\CD_{\alpha_1} R + 2 W_{\alpha_1}
\end{align*}
The third term is
\begin{align*}
\opbraket{0,2}{\tilde {\mathcal O}_+}{0,2} &= [-\frac{1}{2} \partial^\alpha \partial_\alpha X^\beta{}_\beta]
	- [\partial^\alpha Y_\alpha] + [Z] \\
	&= -\CD^\alpha W_\alpha + [\partial_\nu \partial_m X^{m \nu}] + [Z]
\end{align*}
but a straightforward calculation shows the middle term vanishes, leaving
\[
\opbraket{0,2}{\tilde {\mathcal O}_+}{0,2} = -\CD^\alpha W_\alpha + [Z]
\]

The fourth and final term is
\begin{align*}
\opbraket{2,0}{\tilde {\mathcal O}_+}{0,2} &= +2 \partial_{a_1} \partial_{a_2} X^{\beta}{}_\beta
\end{align*}
For simplicity, we note that only the contracted part of this is necessary, so we focus on
\begin{align*}
\partial^b \partial_b X^{\alpha}{}_\alpha =&
	\frac{1}{2} T^{c b \alpha} T_{cb \alpha} - \Box R + 2 \CD^b R T_{b \alpha}{}^\alpha
	+ R \CD^b T_{b\alpha}{}^\alpha \\
	& \;\;\;\;\;\; - R T_{b \alpha \dbeta} T^{\dbeta b \alpha}
	- R T^{b \alpha \mu} T_{b \alpha \mu}
	+ R T^{b \alpha \beta} T_{b \beta \alpha} \eol
	=& W^{\gamma \beta \alpha} W_{\gamma \beta \alpha} + \frac{1}{4} \CD^\alpha R \CD_\alpha R
	+ \frac{1}{2} X^\alpha \CD_\alpha R - \frac{1}{12} X^\alpha X_\alpha
	+ \frac{1}{2} \CD_\dalpha G^b \CD^\dalpha G_b \eol
	& - \Box R - 4i \CD^b R G_b
	-2i R \CD_b G^b
	+ 8 R^2 \bar R
	+ 4 R G^2
\end{align*}
Several terms can be collected into manifestly chiral terms, using
\begin{align*}
\Box_+ R =& \Box R + 2i  G^b \CD_b R
	+ \frac{1}{2} \CD^\alpha R \CD_\alpha R + \frac{1}{2} R \CD^2 R
	- \frac{1}{2} X^\alpha \CD_\alpha R - \frac{1}{2} (\CD^\alpha X_\alpha) R
\end{align*}
as well as
\begin{align*}
\frac{1}{4} (\BCD^2 - 8 R) G^2
	&= \frac{1}{2} \CD_\dalpha G_b \CD^\dalpha G^b -2i G^b \CD_b R + 4 G^2 R
\end{align*}
to give
\begin{align*}
\partial^b \partial_b X^{\alpha}{}_\alpha
	=& W^{\gamma \beta \alpha} W_{\gamma \beta \alpha} + \frac{1}{4} (\BCD^2 - 8 R) G^2 - \Box_+ R
	- \frac{1}{12} X^\alpha X_\alpha \\
	& + \frac{3}{4} \CD^\alpha R \CD_\alpha R + 8 R^2 \bar R + \frac{1}{2} R \BCD^2 \bar R
	- \frac{1}{2} R \CD^\alpha X_\alpha
\end{align*}

Putting all of this together gives
\begin{align*}
\opbraket{0,0}{\tilde {\mathcal O}_+}{0,0} \opbraket{0,0}{\tilde {\mathcal O}_+}{0,2} &= -2 R [Z] \\
\opbraket{0,0}{\tilde {\mathcal O}_+}{0,1} \opbraket{0,1}{\tilde {\mathcal O}_+}{0,2} &= 
	2 W^\alpha W_\alpha - \frac{1}{2} \CD^\alpha R \CD_\alpha R \\
\opbraket{0,0}{\tilde {\mathcal O}_+}{0,2} \opbraket{0,2}{\tilde {\mathcal O}_+}{0,2} &=
	-2 R [Z] + 2 R \CD^\alpha W_\alpha \\
\frac{1}{3} \opbraket{0,0}{\tilde {\mathcal O}_+}{2,0} \opbraket{2,0}{\tilde {\mathcal O}_+}{0,2} &=
	\frac{2}{3} \partial^a \partial_a X^{\beta}{}_\beta \\
	&= \frac{2}{3} W^{\gamma \beta \alpha} W_{\gamma \beta \alpha} + \frac{1}{6} (\BCD^2 - 8 R) G^2
	- \frac{2}{3} \Box_+ R
	- \frac{1}{18} X^\alpha X_\alpha \\
	& \;\;\;\;\; + \frac{1}{2} \CD^\alpha R \CD_\alpha R + \frac{16}{3} R^2 \bar R + \frac{1}{3} R \BCD^2 \bar R
	- \frac{1}{3} R \CD^\alpha X_\alpha
\end{align*}
the sum of which is
\begin{align*}
[A_2] =& 2 W^\alpha W_\alpha
	+ \frac{2}{3} W^{\gamma \beta \alpha} W_{\gamma \beta \alpha} + \frac{1}{6} (\BCD^2 - 8 R) G^2
	- \frac{2}{3} \Box_+ R - \frac{1}{18} X^\alpha X_\alpha \\
	& -4 R [Z] + 2 R \CD^\alpha W_\alpha + \frac{16}{3} R^2 \bar R + \frac{1}{3} R \BCD^2 \bar R
	- \frac{1}{3} R \CD^\alpha X_\alpha
\end{align*}
We must still evaluate $[Z]$. Begin by noting
\[
[Z] = \frac{1}{2} \CD^\alpha W_\alpha + [\Delta^{-1/2} \mathcal O_+ \Delta^{1/2}]
	= \frac{1}{2} \CD^\alpha W_\alpha - \frac{1}{2} (\BCD^2 - 8 R) \bar R + [\Delta^{-1/2} \mathcal \Box_+ \Delta^{1/2}]
\]
Evaluating the term involving $\Box_+$ is a somewhat laborious task. The most straightforward
way of doing it is to expand out all the terms so that they involve $\log \Delta$ and then
to work out the expansion of $\log \Delta$ to the necessary order. The expansion of
$\log \Delta$ to second order is
\[
\log \Delta = -2i G_y + \frac{1}{24} T_y{}^{ba} T_{yba} - \frac{1}{6} R_{ymy}{}^m
	- i \CD_y G_y + \frac{1}{2} T_{\alpha y \dbeta} T^\dbeta{}_y{}^\alpha - 2 \eta^2 \bar R
\]
Note that there are no terms linear in $\eta$. The $\Box_+$ term yields
\begin{align*}
[\Delta^{-1/2} \mathcal \Box_+ \Delta^{1/2}] =&
	- \frac{1}{4} \CD^\alpha X_\alpha + \frac{1}{4} R \partial^\alpha \partial_\alpha \log \Delta
	+ i G^b \partial_b \log \Delta + \frac{1}{2} \partial^b \partial_b \log \Delta
	+ \frac{1}{4} \partial^b \log \Delta \partial_b \log\Delta \\
	=& - \frac{1}{4} \CD^\alpha X_\alpha + 2 R \bar R
	+ \frac{1}{24} T^{cba} T_{cba} - \frac{1}{6} R_{ab}{}^{ab}
	- i \CD_b G^b  + \frac{1}{2} T_{\alpha c \dbeta} T^{\dbeta c \alpha} + G^2
\end{align*}
Using
\begin{gather*}
R_{ab}{}^{ab} = -\CD^\beta X_\beta - \frac{3}{2} (\CD^2 R + \BCD^2 \bar R) + 48 R \bar R \\
T^{cba} T_{cba} = -24 G^2, \;\;\;
T_{\alpha c \dbeta} T^{\dbeta c \alpha} = 8 R \bar R
\end{gather*}
we find that
\begin{align*}
[Z] = \frac{1}{2} \CD^\alpha W_\alpha - \frac{1}{12} \CD^\alpha X_\alpha + 2 R\bar R
\end{align*}
which gives the net result of
\begin{align}
[A_2] &= 2 W^\alpha W_\alpha
	+ \frac{2}{3} W^{\gamma \beta \alpha} W_{\gamma \beta \alpha} + \frac{1}{6} (\BCD^2 - 8 R) G^2
	- \frac{2}{3} \mathcal O_+ R - \frac{1}{18} X^\alpha X_\alpha \eol
	&= 2 W^\alpha W_\alpha
	+ \frac{2}{3} W^{\gamma \beta \alpha} W_{\gamma \beta \alpha}
	- \frac{1}{18} X^\alpha X_\alpha
	-\frac{1}{4} (\BCD^2 - 8 R) \left(-\frac{2}{3} G^2 + \frac{1}{6}(\CD^2 - 8 \bar R) R \right)
\end{align}

The divergences associated with the heat kernel of this operator are
\begin{align}
[\Tr \log H]_\eps &= -\frac{1}{16\pi^2} \int \ch E \,\Tr \left(\frac{[A_0]}{2 \eps^2} + \frac{[A_1]}{\eps} - \frac{[A_2]}{2} \log \eps + \textrm{finite} \right) + \hc
\end{align}
which we may write as
\begin{align}
[\Tr \log H]_\eps = &+\frac{1}{16\pi^2 \eps} \int E
	+\frac{\log \eps}{16\pi^2}
		\int \chE \left(W^\alpha W_\alpha + \frac{1}{3} W^{\gamma\beta\alpha} W_{\gamma\beta\alpha}
			- \frac{1}{36} X^\alpha X_\alpha \right) \eol
	& +\frac{\log \eps}{16\pi^2} \int E \left(-\frac{1}{3} G^2 -\frac{2}{3} \bar R R\right)
	+ \hc + \textrm{finite}
\end{align}
where we have dropped a total derivative.
This result for $U(1)$ supergravity agrees with the traditional calculation
(up to factors of two in the definition of the supergravity superfields)
in Poincar\'e supergravity when $X^\alpha$ vanishes \cite{McArthur}.

In the non-supersymmetric calculation (provided only a classically conformal
action was used) there was a striking feature where the logarithmic divergent
term consisted solely of conformal or topological terms. Since we could have
written our result here in terms of the heat kernel of a conformally
coupled bosonic scalar and fermionic superpartner, it should have the
same property.

Consider a small shift in the choice of compensator $X$
of the form $\delta X = X \delta U$ where $\delta U$ is a dimension zero superfield.
First note that $\delta E$ and $\delta \ch E$ both vanish if $X$ is changed a small amount.
This is because the choice of $X$ while redefining $E_a{}^M$ does so only by
shifting the spinor derivative part of the bosonic derivative. That is,
$\delta E = - E \delta E_a{}^M E_M{}^a = -E \delta E_a{}^a$ vanishes.
Similarly $\delta \chE$ vanishes.

It is straightforward to work out that
\begin{align}
\delta X_\alpha = \frac{3}{8} \bar\nabla^2 \nabla_\alpha \delta U
 = \frac{3}{8} (\BCD^2 - 8 R) \CD_\alpha \delta U
\end{align}
We similarly may calculate
\begin{align}
\delta R = -\frac{1}{8} \bar\nabla^2 \delta U - \frac{1}{4X} \nabla_\dalpha X \nabla^\dalpha \delta U
	= -\frac{1}{8} \BCD^2 \delta U
\end{align}
and
\begin{align}
\delta G_{\alpha \dalpha} = -\frac{1}{4} [\CD_\alpha, \CD_\dalpha] \delta U
\end{align}
It is straightforward to check that the specific combination
\begin{align}
\left[G^2 + 2 R \bar R\right]_D + \frac{1}{6}\left[X^\alpha X_\alpha \right]_F
\end{align}
is invariant to \emph{any} deformation of the compensator. It corresponds at the component
level to the expression
\[
-\frac{1}{6} F^{ab} F_{ab} - \frac{1}{8} \mathcal R^{ab} \mathcal R_{ab} + \frac{1}{24} \mathcal R^2
	+ \textrm{fermions}
\]
where $F_{ab}$ is the field strength of the $U(1)_R$. Noting that
\begin{align}
\left[W^{\gamma \beta \alpha} W_{\gamma \beta \alpha}\right]_F = \frac{1}{6} F^{ab} F_{ab}
	+ \frac{1}{16} C^{abcd} C_{abcd} + \textrm{fermions}
\end{align}
we find
\begin{align}
\left[G^2 + 2 R \bar R\right]_D + \frac{1}{6}\left[X^\alpha X_\alpha \right]_F
	+ \left[W^{\gamma \beta \alpha} W_{\gamma \beta \alpha}\right]_F =
	\frac{1}{16} L_\chi + \textrm{fermions}
\end{align}
up to total derivatives, where $L_\chi$ is the topological Gauss-Bonnet term.
Since $W^{\alpha \beta \gamma}$ is $X$-independent automatically,
this combination must be independent under deformations of both the compensator $X$
\emph{and} the conformal supergravity structure. Showing this directly at the superspace
level is straightforward, but requires solving the constraint structure of
supergravity. This can be done using the formulae given in \cite{paper3},
which we leave as an exercise to the interested reader.

This superfield topological combination will appear several times, so it is useful
to introduce a label for the superfield expression. We choose to define
the Hermitian combination
\begin{align}
S_{\chi} \equiv &
	\left[G^2 + \CP \bar R + \ACP R - 2 R \bar R \right]_D
	+ \frac{1}{12}\left[X^\alpha X_\alpha \right]_F
	+ \frac{1}{12}\left[\bar X_\dalpha \bar X^\dalpha \right]_{\bar F} \eol
	& + \frac{1}{2} \left[W^{\gamma \beta \alpha} W_{\gamma \beta \alpha}\right]_F
	+ \frac{1}{2} \left[\bar W_{\dgamma \dbeta \dalpha} \bar W^{\dgamma \dbeta \dalpha}\right]_{\bar F}
\end{align}
where, one should recall, $\CP = -\frac{1}{4} (\BCD^2 - 8 R)$. (We have chosen to
reintroduce a total derivative which formerly dropped out previously since when
we calculate the conformal anomaly this term will not in general vanish.)
We can then write the divergences that we found as
\begin{align}
[\Tr \log H]_\eps =
	& +\frac{1}{8\pi^2 \eps} \biggl[X\biggr]_D
	- \frac{\log \eps}{24\pi^2} S_\chi \eol
	& +\frac{\log \eps}{16\pi^2}
		\biggl[W^\alpha W_\alpha
			+ \frac{1}{36} X^\alpha X_\alpha 
			+ \frac{2}{3} W^{\gamma\beta\alpha} W_{\gamma\beta\alpha} \biggr]_F \eol
	& +\frac{\log \eps}{16\pi^2}
		\biggl[\bar W_\dalpha \bar W^\dalpha
			+ \frac{1}{36} \bar X_\dalpha \bar X^\dalpha 
			+ \frac{2}{3} \bar W_{\dgamma\dbeta\dalpha} \bar W^{\dgamma\dbeta\dalpha} \biggr]_{\bar F} \eol
	& + \textrm{finite}
\end{align}
where we have reintroduced the compensator $X$.
Its only explicit appearance is in the quadratically divergent D-term,
where it provides the necessary conformal weight to render a conformally
invariant expression. Although it is implicitly used to define $S_\chi$,
as we noted $S_\chi$ is independent of small deformations of $X$.
The remaining presence in $X_\alpha$ is purely the part of $X$ that can
be regarded as a $U(1)$ prepotential, if say we were to decompose $X$
as $\Phi_0 \bar\Phi_0 e^{V}$ for some $U(1)$ prepotential $V$.

One also suspects it should combine with $W^\alpha$ in a way
that removes the classical ``U(1) ambiguity.'' Indeed, noting that
\begin{align}
W_\alpha = \frac{1}{8} \bar\nabla^2 e^{-V} \nabla_\alpha e^V, \;\;\;
X_\alpha = \frac{3}{8} \bar\nabla^2 \nabla_\alpha \log X
\end{align}
the combination
\begin{align}
W_\alpha - \frac{1}{6} X_\alpha = \frac{1}{8} \bar\nabla^2 \left( e^{-V+\log X/2} \nabla_\alpha e^{V - \log X/2}\right)
\end{align}
corresponds to the way the factors of $V$ and $X$ appear in the original theory,
and so we note that the divergent term seems to correspond to only the
combination $(W_\alpha - \frac{1}{6} X_\alpha)^2$. We are missing, of course,
the cross-term $W^\alpha X_\alpha$, but this is to be expected.
The determinant of $H$ corresponds to the part of the effective action even
under charge conjugation. If this cross term exists, it should be found in
the superfield version of the odd part of the effective action.
We turn to that analysis next.

\subsection{Integration of the odd part}
Recall that $D$ and $\tilde D$ are defined in the massless case by
\begin{align}
D = \left(
\begin{array}{cc}
0 & \CP e^{V^T} X^{-1/2} \\
\ACP e^V X^{-1/2} & 0
\end{array}
\right), \;\;\;
\tilde D = \left(
\begin{array}{cc}
0 & -\CP e^{-V} X^{-1/2} \\
-\ACP e^{-V^T} X^{-1/2} & 0
\end{array}
\right)
\end{align}
Defining $H = \tilde D D$ and $\tilde H = D \tilde D$, the effective action
$\Tr \log D$ is divided into two terms
\begin{align}
[\Tr\log D]_\eps = \frac{1}{2} \left[\Tr\log H\right]_\eps
	+ \int \left(L_\eps^- + \ell\right)
\end{align}
the first of which we have already found. The objects $L_\eps^-$ and $\ell$ are 
one-forms in the space of all possible variations of the gauge prepotential, and
$\ell$ is chosen so that $L_\eps^- + \ell$ is a closed form. It is therefore
(at least locally) the variation of some other expression and can be integrated,
which we have indicated with a schematic $\int$ symbol which shall be better
defined later.

In analogy to the fermionic case, we define
\begin{align}
L_\eps^- &= \frac{1}{2} \Tr\int_\eps^\infty d\tau \left(e^{-\tau H} \tilde D \delta D - e^{-\tau \tilde H} D \delta \tilde D\right)
\end{align}
$\ell$ itself is defined by integrating the formula $\delta \ell = -C$
where 
\begin{align}\label{eq_defC}
C_\eps = \delta L_\eps^- &= \eps \int_0^1 d\lambda \Tr\left(
	\delta D e^{-\eps \lambda H} \delta \tilde D e^{-\eps \tilde\lambda \tilde H}
	\right)
\end{align}
where $\tilde\lambda = 1-\lambda$. Using cyclicity of the trace, we find
\begin{align}
L_\eps^-
	&= \frac{1}{2} \int_\eps^\infty d\tau\Tr \left(\delta D \tilde D e^{-\tau \tilde H}
	- \delta \tilde D D e^{-\tau H}\right)
\end{align}
We denote
\begin{align}
H = \left(\begin{array}{cc}
H_+ & 0 \\
0 & H_-
\end{array}\right)
\end{align}
and similarly for $\tilde H$.

The operator product $\delta D \tilde D$ is given by
\begin{align*}
\delta D \tilde D = \left(\begin{array}{cc}
-\CP \delta e^{V^T} \ACP e^{-V^T} & 0 \\
0 & -\CP \delta e^{V} \ACP e^{-V}
\end{array}\right) = \left(\begin{array}{cc}
-\CP \Delta V^T e^{V^T} \ACP e^{-V^T} & 0 \\
0 & -\ACP \bar\Delta V e^{V} \ACP e^{-V}
\end{array}\right)
\end{align*}
and its conjugate $\delta \tilde D D$ by
\begin{align*}
\delta \tilde D D = \left(\begin{array}{cc}
-\CP \delta e^{-V} \ACP e^{V} & 0 \\
0 & -\CP \delta e^{-V^T} \ACP e^{V^T}
\end{array}\right) = \left(\begin{array}{cc}
\CP \Delta V e^{-V} \ACP e^{V} & 0 \\
0 & \ACP \bar\Delta V^T e^{-V^T} \ACP e^{V^T}
\end{array}\right)
\end{align*}
where we have defined
\begin{gather}
\Delta V = e^{-V} \delta e^V, \;\;\;\;
\Delta V^T = (\delta e^{V^T}) e^{-V^T} \eol
\bar \Delta V = (\delta e^V) e^{-V}, \;\;\;\;
\bar \Delta V^T = \delta e^{V^T} e^{-V^T} 
\end{gather}

The operators above are defined in a purely chiral or antichiral gauge, but
it is clear that we can rewrite them in a general basis. The way to do this
is to absorb the various factors of $e^V$ in the operators above to define
\emph{covariant} chiral projectors $\CP$ and $\ACP$. In so doing, we would
like to interpret $\Delta V$ and $\bar\Delta V$ (as well as their transposes)
as covariant objects. To do this, we \emph{define}
\begin{align}
\omega \equiv \Delta V \,\,\, \textrm{(chiral gauge)}.
\end{align}
and extend $\omega$ into any other gauge by requiring it to transform
covariantly. It follows that in antichiral gauge,
$\omega = e^V \Delta V e^{-V} = \bar\Delta V$.
We may now write $L_\eps^-$ in a covariant way:
\begin{align*}
L_\eps^-
	&= -\frac{1}{2} \int_\eps^\infty d\tau \;\Tr_+
	\left(\CP \omega^T \ACP e^{-\tau \tilde H_+}
	+ \CP \omega \ACP e^{-\tau H_+}
	\right)
	+ \hc
\end{align*}
where we have broken the trace up into the part over the separate chiral and
antichiral spaces. Noing that the exponential term is the heat kernel, we find
\begin{align}
L_\eps^-
	&= -\frac{1}{2} \int_\eps^\infty d\tau \;\int \chE\,
	\left(\CP [\omega^T \ACP \tilde U_+(\tau)] + \CP [\omega \ACP U_+(\tau)]\right) + \hc \eol
	&= -\frac{1}{2} \int_\eps^\infty d\tau \;\int E\,
	\left(\omega^T \ACP \tilde U_+(\tau) + \omega\ACP U_+(\tau)\right) + \hc
\end{align}
The heat kernel $U_+$ is
\begin{align}
U_+(\tau) = \frac{1}{(4\pi\tau)^2} e^{-\Sigma / 2 \tau} \Delta^{1/2} F(\tau)
\end{align}
Noting that $[\Sigma]=0$, $[\CD_\alpha \Sigma]=0$, and $[\CD^2 \Sigma]=0$, we find
\begin{align*}
L_\eps^-
	&= -\frac{1}{2} \int_\eps^\infty \frac{d\tau}{(4\pi\tau)^2} \;\int E\,
	\Tr\left(\omega^T [\ACP \Delta^{1/2} \tilde F] + \omega [\ACP \Delta^{1/2} F]\right) + \hc
\end{align*}
Note that $\tilde F$ has the same form as $F$ but in a conjugate representation.
Next we note that $[\Delta^{1/2}]=1$, $[\CD_\alpha \Delta^{1/2}]=0$, and
$[\CD^\alpha \CD_\alpha \Delta^{1/2}] = 4 \bar R$, giving
\begin{align*}
L_\eps^-
	&= -\frac{1}{2} \int_\eps^\infty \frac{d\tau}{(4\pi\tau)^2} \;\int E\,
	\Tr\left(\omega^T [\ACP \tilde F - \bar R \tilde F] + \omega [\ACP F - \bar R F]\right) + \hc \eol
	&=  -\frac{1}{2} \int_\eps^\infty \frac{d\tau}{(4\pi\tau)^2} \;\int E\,
	\Tr\left(\omega^T \left[-\frac{1}{4}\CD^2 \tilde F + \bar R \tilde F\right]
		+ \omega \left[-\frac{1}{4}\CD^2 F + \bar R F\right]\right) + \hc
\end{align*}
Since $F(\lambda) = \sum_{n=0}^\infty A_n \lambda^n/n!$, only the terms involving
$A_0$ and $A_1$ contribute to the divergences -- the former to the quadratic and the
latter to the logarithmic. Using $[A_0]=0$ and $[\CD^2 A_0] = -4$, we find for the
quadratic divergences
\begin{align}
L_\eps^-
	&\ni -\frac{1}{32\pi^2} \frac{1}{\eps} \int E\,
	\Tr\left(\omega^T + \omega\right) + \hc
	= - \frac{1}{16\pi^2} \frac{2}{\eps} \int E\,
	\delta \Tr\left(V\right)
\end{align}
which is a divergent contribution to the Fayet-Iliopoulos term.

For the logarithmic divergences, we note from our experience with the heat kernel,
we immediately may conclude that $[A_1]= -2 R$ and
$[\CD^2 A_1] = 2 \CD^\alpha W_\alpha + \frac{1}{3} \CD^\alpha X_\alpha - 8 R \bar R$
which give
\begin{align}
L_\eps^-
	&= +\frac{\log \eps}{32\pi^2} \;\int E\,
	\Tr\left(\omega^T \left[-\frac{1}{2} \CD^\alpha \tilde W_\alpha - \frac{1}{12} \CD^\alpha X_\alpha\right]
	+ \omega \left[-\frac{1}{2} \CD^\alpha W_\alpha - \frac{1}{12} \CD^\alpha X_\alpha\right]
\right) + \hc
\end{align}
In chiral gauge, 
$W_\alpha = -\frac{1}{2} \CP \left( e^{-V} \CD_\alpha e^{V}\right)$ and
$\tilde W_\alpha = -\frac{1}{2} \CP \left( e^{V^T} \CD_\alpha e^{-V^T}\right) = -W_\alpha^T$.
Transposing cancels out the even term, leaving the odd term
\begin{align*}
L_\eps^- &= -\frac{\log \eps}{16\pi^2} \;\int E\,
	\left(\omega \times \frac{1}{12} \CD^\alpha X_\alpha\right) + \hc
\end{align*}
Noting that $\delta W_\alpha = -\frac{1}{2} \CP \CD_\alpha \omega$, this is
equivalent to
\begin{align}
L_\eps^-
	&\ni -\frac{\log \eps}{16\pi^2} \times \frac{1}{6} \int \chE\,
	\Tr\left(\delta W^\alpha X_\alpha \right) + \hc
\end{align}
which is trivially integrable.

We summarize here our results: the quadratic divergences of the operator $D$ are
(restoring the compensator)
\begin{align}
[\Tr\log D]_\eps \ni +\frac{1}{16\pi^2 \eps} \biggl[ \Tr\left(1 - 2 V\right) X\biggr]_D
\end{align}
and the logarithmic divergences are
\begin{align}
[\Tr \log D]_\eps \ni &
	- \frac{\log \eps}{48\pi^2} S_\chi 
	+ \frac{\log \eps}{32\pi^2}
		\biggl[\biggl(W^\alpha - \frac{1}{6} X^\alpha\biggr)^2
	+ \frac{2}{3} W^{\gamma\beta\alpha} W_{\gamma\beta\alpha}\biggr]_F \eol
	& + \frac{\log \eps}{32\pi^2}
		\biggl[\biggl(\bar W_\alpha - \frac{1}{6} \bar X_\dalpha\biggr)^2
	+ \frac{2}{3} \bar W_{\dgamma\dbeta\dalpha} \bar W^{\dgamma\dbeta\dalpha}\biggr]_{\bar F}
\end{align}

\subsubsection{Calculation of $\ell$}
The non-integrability of the finite part of $L_\eps$ is due to the non-vanishing of
\begin{align}
C_\eps &= -\eps \int_0^1 d\lambda \Tr\left(
	\delta \tilde D e^{-\eps \tilde\lambda \tilde H} \delta D e^{-\eps \lambda H}
	\right) \eol
	&= -\eps \int_0^1 d\lambda \Tr_+ \left(
	\CP \Delta V e^{-\eps \tilde \lambda \tilde H_-} \ACP \Delta V e^{-\eps \lambda H_+}
	\right) - \textrm{conjugate rep}\eol
	&= -\eps \int_0^1 d\lambda \int E \int E' \, \,\Tr\left(\omega(z) U_-(z,z'; \eps \tilde \lambda) \,
	\omega(z') U_+ (z',z; \eps \lambda)\right)
	- \textrm{conjugate rep}
\end{align}
where we have written everything in a covariant notation as well as promoting $\omega$
to a 1-form in analogy to the fermionic case.
The above expression includes the subtraction of the conjugate ($V \rightarrow -V^T$)
representation; thus in a self-conjugate representation $C_\eps$
vanishes and $L_\eps$ is integrable by itself.

In the last line of the above formula we have taken a trace over chiral coordinates, introduced a 
complete set of antichiral coordinates in the center, and converted both systems
into total superspace integrals using the explicit projectors.\footnote{The subtraction
of the conjugate representation arises because one actually adds the full Hermitian
conjugate; in reordering the operators so that $U_-$ appears before
$U_+$ in each term, one finds a sign flip from pushing the one-forms
past each other.}

The evaluation of this expression is somewhat technical, so we relegate it to
Appendix \ref{app_2phk} where we explicitly evaluate the expression
\begin{align}
Z(\omega_2, \omega_1; \eps, \lambda) &= \int E \int E' \,
	\,\Tr\left(\omega_2(z) U_-(z,z', \eps \tilde\lambda)
			\omega_1(z') U_+ (z',z, \eps \lambda)\right)
\end{align}
where $\tilde \lambda = 1 - \lambda$. We find
\begin{align}
Z = \frac{1}{16\pi^2 \eps^2} \int E\,
	\Tr \Biggl\{
	& \omega_2 \omega_1
	- \frac{\eps \lambda}{2} R\CD^\alpha \omega_2 \CD_\alpha \omega_1
	- \frac{\eps \tilde \lambda}{2} \bar R \BCD_\dalpha \omega_2 \BCD^\dalpha \omega_1
	- \frac{\eps}{12} \CD^\alpha X_\alpha \omega_2 \omega_1
	- \eps \lambda \tilde \lambda \CD^a \omega_2 \CD_a \omega_1 \eol
	&
	+ \frac{\eps \lambda}{2} (\CD^\alpha \omega_2 \omega_1 W_\alpha - \omega_2 \CD^\alpha \omega_1 W_\alpha)
	+ \frac{\eps \tilde \lambda}{2} (\BCD_\dalpha \omega_2 W^\dalpha \omega_1
							 - \omega_2 W_\dalpha \BCD^\dalpha \omega_1)
	+ \mathcal O(\eps^2)
	\Biggr\}
\end{align}

For the case of interest here,
\begin{align}
C_\eps
	= -\frac{1}{16\pi^2} \int E \, \Tr \biggl\{&
	\, \frac{1}{\eps}\omega \omega - \frac{1}{6} \CD^a \omega \CD_a \omega
	- \frac{1}{4} R \CD^\alpha \omega \CD_\alpha \omega
	- \frac{1}{4} \bar R \BCD_\dalpha \omega \BCD^\dalpha \omega
	- \frac{1}{12} \CD^\alpha X_\alpha \omega \omega \eol
	&\;\;\;\;\; - \frac{1}{4} \omega \CD^\alpha \omega W_\alpha
	+ \frac{1}{4} \CD^\alpha \omega \omega W_\alpha
	+ \frac{1}{4} \CD_\dalpha \omega W^\dalpha \omega 
	- \frac{1}{4} \omega W_\dalpha \CD^\dalpha \omega\biggr\} + \mathcal O(\eps) \eol
	& - \textrm{conjugate rep}
\end{align}
Using cyclicity of the trace and the antisymmetry of the 1-forms $\omega$
(and the fact that the conjugate rep is the same result after
transposition), we find that only a small set of terms survive in the
$\eps \rightarrow 0$ limit, giving
\begin{align}
C 	&= \frac{1}{32\pi^2} \int E \,\Tr\left(\omega \CD^\alpha \omega W_\alpha
	- \CD^\alpha \omega \omega W_\alpha
	+ \omega \CD_\dalpha \omega W^\dalpha
	- \CD_\dalpha \omega \omega W^\dalpha \right) \eol
	&= \frac{1}{32\pi^2} \int E \,\,\biggl(\omega^r \CD^\alpha \omega^s W_\alpha^t
	+ \omega^r \CD_\dalpha \omega^s \bar W^\dalpha{}^t
	\biggr) \mathcal A_{rst}
\end{align}
where $\mathcal A_{rst}\equiv \Tr(\{\mathbf T_r, \mathbf T_s\} \mathbf T_t)$ is the anomaly factor,
the symmetrized trace of three generators of the gauge group. This is
exactly the same form as the globally supersymmetric result found by
McArthur and Osborn \cite{McArthur:1985xd}. $C$ may also be written
\begin{align}
C	&= \frac{1}{16\pi^2} \int E\, \Tr \left(\omega \CD^\alpha \omega W_\alpha
	- \CD_\dalpha \omega \omega W^\dalpha \right)
\end{align}
by integrating by parts and using $\CD^\alpha W_\alpha = \BCD_\dalpha W^\dalpha$.

To derive the form of $\ell$, we follow exactly the procedure of \cite{McArthur:1985xd},
which is essentially unchanged by the addition of supergravity. We begin
by introducing a new function $\mathcal X$
\begin{align}
C &= \frac{1}{16\pi^2} \int E\, \mathcal X(\omega, \omega, V)
\end{align}
where
\begin{align}
\mathcal X(h_1, h_2, V) &\equiv \mathrm{STr}
	\left(h_1 \CD^\alpha h_2 W_\alpha - \CD_\dalpha h_1 h_2 W^\dalpha\right)
\end{align}
is a two form. We define it with a symmetrized and normalized trace of the three generators of the
gauge group:
\begin{align}
\mathrm{STr}\left(A B C \right) \equiv \frac{1}{2} A^r B^s C^t \Tr\left(\{\mathbf T_r, \mathbf T_s\} \mathbf T_t\right)
\end{align}
One can show that this two form is both Hermitian and symmetric in its one-form
arguments $h_1$ and $h_2$.
Note $\mathcal X$ depends on $V$ implicitly through $W_\alpha$ and the covariant derivative.

Again following McArthur and Osborn, we enlarge the configuration space of $V$
to include a parameter $t$, with $t=0$ corresponding to $V=0$ and $t=1$
corresponding to the full background $V$. We denote this parametrized prepotential by $V_t$.
The total variation $\Omega_t$ of $e^{V_t}$ is then given by two pieces:
$\Omega_t = \omega_t^t + \omega_t$ where, in chiral gauge, $\omega_t = e^{-V_t} \delta e^{V_t}$
and $\omega_t^t = e^{-V_t} d_t e^{V_t}$ for $d_t = dt \,\partial_t$.
Since $C$ and therefore $\mathcal X$ is exact,
\begin{align}
(\delta + d_t) \mathcal X(\Omega_t, \Omega_t, V_t) = 0
\end{align}
and one may show (using $dt \wedge dt = 0$)
\begin{align}
\delta \mathcal X(\omega_t, \omega_t^t, V_t) = -\frac{1}{2} d_t \mathcal X(\omega_t, \omega_t, V_t)
\end{align}
Then we may construct a local one-form
\begin{align}
\ell \equiv -\frac{1}{8\pi^2} \int_{I_t} \mathcal X(\omega_t, \omega_t^t, V_t)
\end{align}
whose variation is\footnotemark
\begin{align}
\delta \ell = \frac{1}{8\pi^2} \int_{I_t} \delta \mathcal X(\omega_t, \omega_t^t, V_t)
	= - \frac{1}{16\pi^2} \int_{I_t} d_t \mathcal X(\omega_t, \omega_t, V_t)
	= - \frac{1}{16\pi^2} \mathcal X(\omega, \omega, V)
\end{align}
\footnotetext{In these expressions, integration is defined with $dt$ moved to be
adjacent to the integration symbol. This generates a sign whenever $dt$
is pushed through another 1-form. Thus $\delta \int_{I_t} = -\int_{I_t} \delta$.}

The precise form of $\ell$ is useful in certain applications -- for example, to give
a consistent form for the non-Abelian anomaly associated with gauge transformations
of $V$. However, the definition of $\ell$ is quite path dependent; in particular,
$\ell$ is only defined up to an arbitrary closed form.
There are two obvious paths to choose. One is the ``gauge coupling'' path $V_t = t V$,
where $t$ has the immediate interpretation as the strength of the gauge coupling.
This is the simplest choice for an Abelian theory. Another reasonable option is the ``minimal
homotopic'' path of $e^{V_t} = (1-t) + t e^V$ suggested by Gates, Grisaru, and Penati
\cite{Gates:2000dq}.

Since one is often concerned with Abelian anomalies, we will restrict ourselves
briefly to that case and the use of the gauge couping path. This immediately gives
\begin{align}
\ell &= -\frac{1}{8\pi^2} \int_0^1 \mathcal X(\omega_t, \omega_t^t, V_t)
	= \frac{1}{24\pi^2} \biggl(\delta V \CD^\alpha V W_\alpha - \CD_\dalpha \delta V V W^\dalpha \biggr) \eol
	&= \frac{1}{24\pi^2} \biggl(\delta V \CD^\alpha V W_\alpha + \delta V \CD_\dalpha V W^\dalpha
	+ \delta V V \CD_\dalpha W^\dalpha\biggr) \eol
	&= -\frac{1}{12\pi^2} \biggl(\delta V \Omega_V \biggr)
\end{align}
where we have dropped a total derivative.
Here, $W_\alpha = \frac{1}{8} (\BCD^2 - 8R)\CD_\alpha V$ and, it should be
recalled, $\CD^\alpha W_\alpha = \BCD_\dalpha \bar W^\dalpha$.
$\Omega_V$ is the Chern-Simons superfield  \cite{cssuperfields}
for the Abelian gauge group, obeying
$[\Lambda \Omega_V]_D = [\Lambda W^\alpha W_\alpha]_F$
for chiral $\Lambda$.

\subsubsection{Expression for $\Tr\log D$}
We now need to integrate the closed form $L_\eps^- + \ell$. We introduce
another parameter $u$ which interpolates from $V=0$ to the final value of $V$.
We then take
\begin{align}
\int_{I_u} \biggl( L_\eps^-(\omega_u^u, V_u) + \ell(\omega_u^u, V_u)\biggr)
= \int_{I_u} L_\eps^-(\omega_u^u, V_u)
	- \frac{1}{8\pi^2} \int_{I_u \times I_t} \mathcal X(\omega_{ut}^u, \omega_{ut}^t, V_{ut})
\end{align}
where $V_{ut}$ denotes the doubly-parametrized $V$ and $\omega_{ut}^t$ and $\omega_{ut}^u$ are
defined in chiral gauge by
\begin{align}
\omega_{ut}^t \equiv e^{-V_{ut}} d_t e^{V_{ut}}, \;\;\;
\omega_{ut}^u \equiv e^{-V_{ut}} d_u e^{V_{ut}}
\end{align}
It is not necessary for the paths parametrized by $u$ and $t$ to be identical.
One can show (following McArthur and Osborn) that under an arbitrary variation in
the gauge prepotential,
\begin{align}\label{eq_dL-}
\delta \int_{I_u} L_\eps^-(\omega_u^u, V_u)
	= L_\eps^-(\omega, V) - \frac{1}{8\pi^2} \int_{I_u} \mathcal X(\omega_{u}, \omega_{u}^u, V_{u})
\end{align}
as well as
\begin{align}\label{eq_dX}
\delta \int_{I_u \times I_t} \mathcal X(\omega_{ut}^u, \omega_{ut}^t, V_{ut})
	= \int_{I_t} \mathcal X(\omega_{t}, \omega_{t}^t, V_{t})
	- \int_{I_u} \mathcal X(\omega_{u}, \omega_{u}^u, V_{u})
\end{align}
The above \eqref{eq_dX} is especially simple when the paths parametrized
by $u$ and $t$ are identical: then the variation of this term vanishes!

The final expression of the effective action is
\begin{align}
[\Tr\log D]_\eps = \frac{1}{2} [\Tr\log H]_\eps
	+ \int_{I_u} L_\eps^-(\omega_u^u, V_u)
	- \frac{1}{8\pi^2} \int_{I_u \times I_t} \mathcal X(\omega_{ut}^u, \omega_{ut}^t, V_{ut})
\end{align}
This shall represent our \emph{definition} for the regulated effective action.

\subsubsection{Anomaly for the $U(1)$ ambiguity}
Before analyzing the gauge and conformal anomalies, we
will consider a different sort of anomaly. Our massless action in the natural
path integral variables had the form
\begin{align}\label{eq_action}
S = \left[\bar \eta \frac{e^V}{X^{1/2}} \eta \right]_D
\end{align}
where $\eta$ is weight $(3/2, 1)$, $X$ has conformal dimension two, and
$V$ is a dimension zero gauge prepotential. Under the replacement
\begin{gather}\label{eq_u1rep}
e^V \rightarrow e^{V + y V_1} \equiv e^{V_y}, \;\;\;
X^{1/2} \rightarrow X^{1/2} e^{y V_1} \equiv X^{1/2}_y
\end{gather}
for a $U(1)$ prepotential $V_1$, the classical action is invariant for all
values of $y$. Since the gauge and conformal sectors were treated asymmetrically,
we expect our definition for the effective action should be anomalous
under this transformation; however, if the anomaly is not really physical,
then the difference should be a local expression. It turns out this is the case,
which we now prove.

We begin with a model where the replacement \eqref{eq_u1rep} has been made for
some value of $y$. The first step is to extract the gauge dependence from $[\Tr\log H]_\eps$,
writing it as
\begin{align}
\frac{1}{2} [\Tr\log H]_\eps = \frac{1}{2} [\Tr\log H]_{\eps, V=0}
	+ \int_0^1 du\, L_\eps^+(\omega_{uy}^u, V_{uy})
\end{align}
The first term on the right can be understood as the effective action
in a formally gauge-free background, yet it still depends on the
$U(1)$ prepotential $V_1$ through the compensator $X$. The second
term on the right represents the additional dependence on $V_{uy}$,
the now-doubly parametrized prepotential we have extracted.

The total effective action can be written
\begin{align}
[\Tr\log D]_\eps = \frac{1}{2} [\Tr\log H]_{\eps, V=0}
	+ \int_{I_u} L_\eps(\omega_{uy}^u, V_{uy})
	- \frac{1}{8\pi^2} \int_{I_u \times I_t} \mathcal X(\omega_{uty}^u, \omega_{uty}^t, V_{uty})
\end{align}
where $L_\eps = L_\eps^+ + L_\eps^-$. Recall that in the second and third
terms we have introduced auxiliary path variables $u$ and $t$ where
$u=0$ or $t=0$ correspond to vanishing $V$ and $u=t=1$ correspond to the
full $V_y$.

Then one can show that by differentiating with respect to $y$,
\begin{align*}
\partial_y L_\eps(\omega_{uy}^u, V_{uy}) =
	\partial_u \int_\eps^\infty d\tau \Tr\left(e^{-\tau H} \tilde D \partial_y D\right)_{V_{uy}}
	+ \int_0^\eps d\sigma \Tr\left(e^{-\sigma H} \partial_{[u} \tilde D e^{-(\eps-\sigma) \tilde H} \partial_{y]} D \right)_{V_{uy}}
\end{align*}
where $D$, $\tilde D$, and $H$ are defined in terms of $V_{uy}$, emphasized
by the subscript. (This equation is a special case of \eqref{eq_defC}.)
This immediately implies that
\begin{align*}
\partial_y \int_{I_u} L_\eps(\omega_{uy}^u, V_{uy})
	 &= \int_\eps^\infty d\tau \Tr\left(e^{-\tau H} \tilde D \partial_y D\right)_{V_y}
	- \int_\eps^\infty d\tau \Tr\left(e^{-\tau H} \tilde D \partial_y D\right)_{V=0} \eol
	& \;\;\;\;\;\;\;
	+ \int_0^1 du\, \int_0^\eps d\sigma \Tr\left(e^{-\sigma H} \partial_{[u} \tilde D e^{-(\eps-\sigma) \tilde H} \partial_{y]} D\right)_{V_{uy}}
\end{align*}
The first term on the right vanishes since $\partial_y (e^{V_y} X_y^{-1/2})$ vanishes.
The second term on the right can be simplified by noting that at $V=0$, $\tilde D = - D$,
and so
\begin{align*}
- \int_\eps^\infty d\tau \Tr\left(e^{-\tau H} \tilde D \partial_y D\right)_{V=0}=
	\frac{1}{2} \partial_y \int_\eps^\infty \frac{d\tau}{\tau} \Tr\left( e^{-\tau H}\right)_{V=0}
	= - \frac{1}{2} \partial_y [\Tr\log H]_{\eps, V=0}
\end{align*}
Then the $y$-derivative of $[\Tr\log D]_\eps$ is reduced to
\begin{align}\label{eq_dy}
\partial_y [\Tr\log D]_\eps &= + \int_0^1 du\, \int_0^\eps d\sigma \,\Tr\left(e^{-\sigma H} \partial_{[u} \tilde D e^{-(\eps-\sigma) \tilde H} \partial_{y]} D\right)_{V_{uy}} \eol
	& \;\;\;\;\;\;\;
	- \frac{1}{8\pi^2} \int_{I_u \times I_t} \partial_y \mathcal X(\omega_{uty}^u, \omega_{uty}^t, V_{uty})
\end{align}
which is a local (though divergent) expression. The ambiguity in whether we consider
the $U(1)$ as part of the conformal factor or as part of the Yang-Mills
factor is therefore a local counterterm allowed by the ambiguities of regularization.
We are free to choose whatever parametrization is the most natural.

It is straightforward to evaluate the first term of \eqref{eq_dy} using the method of
Appendix \ref{app_2phk}. The result is
\begin{align*}
-\frac{1}{4\pi^2} \biggl(
	& \eps \Tr (V_y) V_1
	- \frac{1}{4} R \Tr (e^{-V_y} \CD^\alpha e^{V_y}) \CD_\alpha V_1
	- \frac{1}{4} \bar R \Tr (e^{-V_y} \BCD_\dalpha e^{V_y}) \BCD^\dalpha V_1 \eol
	&\;\;\;\;\;\;
	- \frac{1}{12} \CD^\alpha X_\alpha \Tr (V_y)  V_1
	+ \frac{i}{24} \Tr \Bigl(\CD_{\{\dalpha}(e^{-V_y} \CD_{\alpha\}} e^{V_y})\Bigr) \; \CD^{\dalpha \alpha} V_1
	\biggr)
\end{align*}
where we should recall $V_y = V + y V_1$. This is a somewhat deceptive labelling
though since the $y$-dependent compensator $X_y$ is used to define the
supergravity superfields $R$ and $X_\alpha$ as well as in the covariant
derivatives $\CD$. In principle, all of the $y$ (and $V_1$) dependence may be
explicitly expanded.

The second term may be evaluated by noting that $\mathcal X$ is independent
of the compensator $X$, and so $\partial_y$ amounts to an arbitrary $U(1)$
shift in the prepotential. Then following \eqref{eq_dX},
\begin{align*}
\partial_y \int_{I_u \times I_t} \mathcal X(\omega_{ut}^u, \omega_{ut}^t, V_{ut})
	= \int_{I_t} \mathcal X(\omega_{ty}^y, \omega_{ty}^t, V_{ty})
	- \int_{I_u} \mathcal X(\omega_{uy}^y, \omega_{uy}^u, V_{uy})
\end{align*}
which vanishes if the paths parametrized by $t$ and $u$ are identical.
Then the only contribution is that of the first term, which is manifestly
local and can be integrated in the $U(1)$ deformation parameter $y$.

\subsubsection{Conformal anomaly}
The conformal anomaly with which we will be concerned involves the transformation
\begin{gather}
\eta \rightarrow e^{-\lambda} \eta, \;\;\; \bar\eta \rightarrow \bar\eta e^{-\bar\lambda}, \;\;\;
X \rightarrow X e^{-2\bar\lambda -2 \lambda}
\end{gather}
in the action \eqref{eq_action}. Begin by recalling the definition of the effective action:
\begin{align}
[\Tr\log D]_\eps = \frac{1}{2} [\Tr\log H]_\eps
	+ \int_{I_u} L_\eps^-(\omega_u^u, V_u)
	- \frac{1}{8\pi^2} \int_{I_u \times I_t} \mathcal X(\omega_{ut}^u, \omega_{ut}^t, V_{ut})
\end{align}
Under a conformal transformation, $\Tr\log H$ generates the covariant
conformal anomaly:
\begin{align}
\frac{1}{2} \delta_\lambda \Tr\log H &= \Tr_+ \left(\lambda e^{-\eps H_+} \right)
	+ \Tr_+ \left(\lambda e^{-\eps \tilde H_+} \right) + \hc \eol
	&= \frac{1}{16\pi^2} \Tr \left(-\frac{2}{\eps} [\lambda]_D
	+ \left[\lambda A_2 \right]_F \right) + \hc
\end{align}
Since $\mathcal X$ is independent of $X$, the only other contribution to the conformal
anomaly comes from the $L_\eps^-$ term. It is straightforward to show
\begin{align*}
\delta_\lambda L_\eps^-
	&= -\eps \int_0^1 d\lambda \Tr \left(e^{-\eps \lambda H} \delta_\lambda \tilde D e^{-\eps\tilde\lambda \tilde H} \delta_V D\right) + 
	\eps \int_0^1 d\lambda \Tr \left(e^{-\eps \tilde \lambda \tilde H} \delta_\lambda D e^{-\eps \lambda H} \delta_V \tilde D\right)
\end{align*}
which may be rewritten as
\begin{align*}
\delta_\lambda L_\eps^-
	&= -\eps \int_0^1 d\lambda \Tr \left(e^{-\eps \lambda H} \delta_{[\lambda} \tilde D e^{-\eps\tilde\lambda \tilde H} \delta_{V]} D\right)
\end{align*}
This is easy enough to calculate using the general formula found in Appendix \ref{app_2phk}.
The result is a contribution
\begin{align}\label{eq_extconfanom}
\frac{1}{16\pi^2}
	\Tr \left[\frac{4}{\eps} \lambda V - R \CD^\alpha \lambda \CD_\alpha V
	- \frac{1}{3} \lambda \CD^\alpha X_\alpha \, V
	+ \frac{2}{3} \lambda \Box V
	\right]_D + \hc
\end{align}
which is symmetric with respect to $\lambda$ and $V$. The third term may be rewritten
to give the missing ``cross-term'' $W^\alpha X_\alpha$ for the covariant anomaly.

Putting everything together, we find a conformal anomaly which may be written
(restoring the compensator $X$)
\begin{align}
\delta_\lambda [\Tr \log D]_\eps =&
	- \frac{1}{8\pi^2 \eps} \Tr [\lambda X (1- 2 V)]_D \eol
	& + \frac{1}{8\pi^2} \Tr \left[\lambda \left(W_\alpha - \frac{1}{6} X_\alpha\right)^2\right]_F
	+ \frac{1}{12 \pi^2} \Tr \left[\lambda W^{\alpha \beta \gamma} W_{\alpha \beta \gamma} \right]_F
	- \frac{1}{24\pi^2} \Tr \left[\lambda \Omega_\chi \right]_D \eol
	& + \frac{1}{16\pi^2}
	\Tr \left[- R \CD^\alpha \lambda \CD_\alpha V
	+ \frac{1}{3} \CD^\alpha \lambda \, X_\alpha V
	- \frac{2}{3} \CD^a \lambda \CD_a V
	\right]_D + \hc
\end{align}
where we have defined
\begin{align}
\Omega_\chi \equiv G^2 + \ACP R + \CP \bar R - 2 R \bar R + \frac{1}{6} \Omega_X + \Omega_L
\end{align}
with
\begin{gather*}
[\Omega_X]_D = [X^\alpha X_\alpha]_F = [X_\dalpha X^\dalpha]_{\bar F} \\
[\Omega_L]_D = [W^{\alpha \beta \gamma} W_{\gamma \beta \alpha}]_F
	= [\bar W_{\dalpha \dbeta \dgamma} \bar W^{\dgamma \dbeta \dalpha}]_{\bar F}.
\end{gather*}
The Chern-Simons superfields $\Omega_X$ and $\Omega_L$ should exist so long as our
background gauge sector is topologically trivial \cite{cssuperfields}.
They are not themselves gauge invariant;
but since they transform under a gauge transformation into a linear superfield,
integrals of expressions like $\phi \Omega_X$ for chiral $\phi$ \emph{are} gauge invariant.

This expression for the conformal anomaly is fairly simple to understand: the first
line which is quadratically divergent is cancelled if we add counterterms
to the effective action to remove the original $\eps$ divergences; the second
line is a sensible anomaly with a topological Gauss-Bonnet term; and the third
line is an extra contribution to the conformal anomaly in the presence of
a gauge sector which is not trace-free and a conformal parameter $\lambda$
which is not constant.

\subsubsection{Gauge anomaly}
The gauge anomaly arises from the transformation
\begin{gather}
\eta \rightarrow e^{-\Lambda} \eta, \;\;\; \bar\eta \rightarrow \bar\eta e^{-\bar\Lambda}, \;\;\;
e^V \rightarrow e^{\bar\Lambda} e^V e^{\Lambda}
\end{gather}
in the action \eqref{eq_action}. Again we begin by recalling the definition of the effective action,
\begin{align}
[\Tr\log D]_\eps = \frac{1}{2} [\Tr\log H]_\eps
	+ \int_{I_u} L_\eps^-(\omega_u, V_u)
	- \frac{1}{8\pi^2} \int_{I_u \times I_t} \mathcal X(\omega_{ut}^u, \omega_{ut}^t, V_{ut})
\end{align}
Under a gauge transformation, $\Tr\log H$ is invariant as it corresponds to
the even gauge sector, where the superfields can be combined in a Dirac-like
and anomaly-free fashion. The variation of the other two terms can be found
from \eqref{eq_dL-} and \eqref{eq_dX} to give
\begin{align}
\delta_\Lambda [\Tr\log D]_\eps = 
	L_\eps^-(\omega^\Lambda, V)
	- \frac{1}{8\pi^2} \int_{I_t} \mathcal X(\omega_t^\Lambda, \omega_t^t, V_{t})
\end{align}
where $\omega^\Lambda = e^{-V} \bar\Lambda e^V + \Lambda$
in the chiral representation and where $\Lambda$ is conventionally chiral and
$\bar \Lambda$ is conventionally antichiral. (The precise form of
$\omega_t^\Lambda$ is path-dependent but is straightforward to work out.)
The first term can be evaluated straightforwardly to give the covariant gauge anomaly
\begin{align}
L_\eps^-(\omega^\Lambda, V) &=
	\Tr_+ \left( \Lambda e^{-\eps H_+}\right)
	+ \Tr_+ \left( \Lambda^T e^{-\eps \tilde H_+}\right)
	+ \hc \eol
	&= \frac{1}{16 \pi^2} \Tr \left(
	-\frac{2}{\eps} \left[\Lambda\right]_D
	+ \left[\Lambda A_2\right]_F
	\right) + \hc \eol
	&= -\frac{1}{8 \pi^2 \eps} \left[X \Tr\Lambda\right]_D
	+ \frac{1}{8\pi^2} \left[\Tr \Lambda W^\alpha W_\alpha + \frac{1}{36} \Tr \Lambda X^\alpha X_\alpha\right]_F \eol
	& \;\;\;\;\;
	+ \frac{1}{12 \pi^2} \left[\Tr \Lambda W^{\alpha \beta \gamma} W_{\alpha \beta \gamma} \right]_F
	- \frac{1}{24\pi^2} \left[\Tr \Lambda \Omega_\chi \right]_D + \hc
\end{align}
where we have used $\Tr \Lambda^T = \Tr \Lambda$ as well as
$\Tr(\Lambda^T \tilde A_2) = \Tr(\Lambda A_2)$. (We have also restored
the compensator $X$ in the final equality.)
The divergent anomalous term is exactly the gauge variation of the Fayet-Iliopolous
term, which appeared as a divergent contribution to the odd part
of the effective action.

This alone is not a consistent anomaly and requires the addition of
the term involving $\mathcal X$, which is path-dependent and for
a non-abelian gauge sector will in general involve an infinite series of
terms. We will subsequently neglect this term.

Conspicuous in its absence is anything resembling the cross term $W^\alpha X_\alpha$.
This is not found in the covariant part of the gauge anomaly, nor is it found in
the term $\mathcal X$. Since the $U(1)$ ambiguity implies that a conformal
anomaly must be equivalent to a $U(1)$ gauge anomaly up to a local counterterm,
it is clear that the missing cross term for the gauge anomaly must be found
as the variation of a local counterterm. Indeed, such a term does exist:
\begin{align}
\frac{1}{2} \delta_\Lambda \left[\Tr (V^2) \CD^\alpha X_\alpha \right]_D
	= \left[\Tr(\Lambda V) \CD^\alpha X_\alpha \right]_D + \hc
\end{align}
which gives the missing cross term as well as a non-covariant term
which depends on the derivative of $\Lambda$. This is simply one of
the terms of \eqref{eq_extconfanom} with the covariant parameter
$\lambda$ replaced by $V$.

\subsubsection{Inclusion of a covariant mass term}
The preceding analysis dealt with massless fields, which was sensible since
we have been concerned with arbitrary complex representations where a constant
mass term would be manifestly forbidden. The models with which we will be
concerned, however, do contain covariant mass terms generated both from the
superpotential and K\"ahler potential, so we will need a method to deal with them.

For the case of chiral fermions, the inclusion of a mass term is not terribly
difficult. If the operator $D$ has entries $\mu$ and $\bar \mu$ on the diagonal,
one simply constructs $\tilde D$ to have entries $\bar \mu$ and $\mu$. For chiral
superfields, this avenue is not open to us because of the holomorphicity
requirement. A generic covariant chiral mass term $\mu$, depending perhaps on
the background chiral superfields, simply cannot be used in the antichiral sector.
We will therefore restrict ourselves to dealing with mass terms via a perturbative
approach.

Given an operator $\det (D + \hat \mu)$ and the additional
operator $\det \tilde D$ associated with the massless conjugate,
we may formally identify
\begin{align}\label{eq_adhoc}
\Tr\log \tilde D + \Tr\log (D+\hat\mu) = \Tr\log (\tilde D D + \tilde D \hat\mu)
\end{align}
Identifying $H = \tilde D D$ and $\tilde D \hat\mu \equiv V$, this operator
at least formally has the structure of $H + V$. Evaluating this
perturbatively using a proper time cutoff regulator gives
\begin{align*}
\left[\Tr\log (H + V)\right]_\eps = & \left[\Tr \log H\right]_\eps
	+ \int_\eps^\infty d\tau \, \Tr \left(e^{-\tau H} V\right) \eol
	& - \frac{1}{2} \int_\eps^\infty d\tau \int_0^\tau d\sigma 
		\, \Tr \left(e^{-\sigma H} V e^{-(\tau-\sigma) H} V\right) + \mathcal O(V^3)
\end{align*}
For our case, $\tilde D \hat \mu$ has vanishing elements on the diagonal
and so only terms even in $\tilde D \hat \mu$ appear. This leads to the identification
\begin{align}\label{eq_dmass}
[\Tr\log (D+\hat\mu)]_\eps - [\Tr\log D]_\eps \equiv
		& - \frac{1}{2} \int_\eps^\infty d\tau \int_0^\tau d\sigma 
		\, \Tr \left(e^{-\sigma H} \tilde D \hat\mu e^{-(\tau-\sigma) H} \tilde D \hat\mu\right)
		+ \mathcal O(\hat\mu^4)
\end{align}
where $[\Tr\log D]_\eps$ is the previous definition we have made.
The advantage of \eqref{eq_dmass} is that the final answer is quite independent
of the particular way we have chosen to write \eqref{eq_adhoc}; other arrangements
of the formal operators lead to an identical regulated result. We may rewrite \eqref{eq_dmass}
as
\begin{align*}
[\Tr\log (D+\hat\mu)]_\eps - [\Tr\log D]_\eps = -\int_\eps^\infty d\tau \int_0^\tau d\sigma \, Z(\bar \mu, \mu; \sigma, \tau-\sigma) + \mathcal O(\hat\mu^4)
\end{align*}
where $Z$ is as defined in Appendix \ref{app_2phk}. At leading order, 
\[
Z(\bar \mu, \mu; \sigma, \tau-\sigma) = \frac{1}{16 \pi^2} \frac{1}{\tau^2} \left[\bar \mu \mu\right]_D + \ldots
\]
which gives
\begin{align}
[\Tr\log (D+\hat\mu)]_\eps - [\Tr\log D]_\eps = + \frac{\log \eps}{16 \pi^2} \left[\bar \mu \mu\right]_D +
	\textrm{finite}
\end{align}

To calculate anomalies associated with the mass term, observe first that
a gauge anomaly acts on the objects $D$, $\tilde D$, and $\hat\mu$ via
\begin{gather*}
\delta_g D = D \Lambda + \Lambda^T D, \;\;\;
\delta_g \tilde D = -\tilde D \Lambda^T - \Lambda \tilde D, \;\;\;
\delta_g \hat\mu = \hat\mu \Lambda + \Lambda^T \hat\mu \eol
\delta_g H = [H, \Lambda], \;\;\;
\delta_g (\tilde D \hat\mu) = [\tilde D \hat\mu, \Lambda]
\end{gather*}
provided that $\hat\mu$ transform in a way that leaves the classical action
gauge invariant. Given the transformation rules of $H$ and $\tilde D \hat\mu$,
the perturbative expansion of the effective action in terms of $\hat\mu$ 
\emph{must} be free of gauge anomalies. (This is obvious in retrospect
since we based our construction on the operator $\tilde D D + \tilde D \hat\mu$,
which is manifestly gauge covariant.) Thus
\begin{gather}
\delta_g \biggl([\Tr\log (D+\hat\mu)]_\eps  - [\Tr \log D]_\eps\biggr) = 0
\end{gather}

For conformal anomalies, observe that
\begin{gather*}
\delta_c D = \{D, \lambda\}, \;\;\;
\delta_c \tilde D = \{\tilde D, \lambda\}, \;\;\;
\delta_c \hat\mu = \{\hat\mu, \lambda\} \eol
\delta_c H = \{H, \lambda\} + 2 \tilde D \lambda D, \;\;\;
\delta_c \tilde D \hat\mu = \{\tilde D \hat\mu, \lambda\} + 2 \tilde D \lambda \hat\mu
\end{gather*}
It follows (after some algebra) that
\begin{align*}
&\delta_c \biggl([\Tr\log (D+\hat\mu)]_\eps - [\Tr \log D]_\eps\biggr) =
	2 \int_0^\eps d\sigma \int_0^\sigma d\sigma' \eol
	& \;\;\;\; \Tr\left(
	e^{-\sigma' H} \lambda e^{-(\sigma-\sigma') H} \tilde D \hat\mu e^{-(\eps-\sigma) H} \tilde D \hat\mu
	+ \tilde D e^{-\sigma' \tilde H} \lambda e^{-(\sigma-\sigma') \tilde H} \hat\mu
		\tilde D e^{-(\eps-\sigma) \tilde H} \hat\mu
	\right) + \mathcal O(\hat\mu^4)
\end{align*}
For our chiral model, the traces under the integrals may be written as
\begin{align*}
\Tr_+ \left(
	e^{-\sigma' H_+} \lambda e^{-(\sigma-\sigma') H_+} \CP \bar \mu e^{-(\eps-\sigma) H_-} \ACP \mu \right) + \textrm{conjugate rep} + \hc
\end{align*}
where we are using covariant notation for the chiral projectors and the
chiral and antichiral mass terms. This is in principle a three point operator,
but we don't actually need to evaluate it fully. Simply observing that dimensional
counting forbids anything worse than $\lambda \mu \bar \mu$ as a D-term, we can
first neglect all derivatives on $\lambda$ to contract the first set of heat
kernels and then perform the $\sigma'$ integration to give
\begin{align*}
\delta_c \biggl([\Tr\log (D+\hat\mu)]_\eps - [\Tr \log D]_\eps\biggr) &=
	2 \int_0^\eps d\sigma \, \sigma\,
\Tr_+ \left(\lambda \CP \bar\mu e^{-(\eps-\sigma) H_-} \ACP \mu e^{-\sigma H_+} \right) + \textrm{conjugate rep} + \hc
\end{align*}
The operator within the trace is equivalent to $Z$ except for the addition of
the factor $\lambda$. This immediately yields
\begin{align*}
\delta_c \biggl([\Tr\log (D+\hat\mu)]_\eps - [\Tr \log D]_\eps\biggr) &=
	\frac{1}{8\pi^2} \left[\lambda \bar\mu \mu \right]_D + \hc
\end{align*}
Restoring the explicit factors of the gauge and conformal fields gives
\begin{align}
\delta_c \biggl([\Tr\log (D+\hat\mu)]_\eps - [\Tr \log D]_\eps\biggr) &=
	\frac{1}{8\pi^2} \left[\lambda X \Tr(e^{-V} \bar\mu e^{-V^T} \mu) \right]_D + \hc
\end{align}

That there is a conformal anomaly involving $\mu$ but not a gauge anomaly
implies again an asymmetry between whether we include a $U(1)$ factor in the conformal
or in the gauge sector. There is an obvious finite counterterm to include
whose $U(1)$ gauge variation gives the corresponding $U(1)$ gauge anomaly:
one simply puts the $U(1)$ part of the prepotential in place of
$\lambda$ in the above expression.

\subsubsection{Summary}
We have covered a lot of ground so we briefly review our results.
The model we are considering is of the form
\begin{align}
S = \left[\bar \eta \frac{e^V}{X^{1/2}} \eta \right]_D
	+ \frac{1}{2} \left[\eta^T \mu \eta \right]_F
	+ \frac{1}{2} \left[\bar\eta \bar\mu \bar\eta^T \right]_F
\end{align}
The one-loop effective action $\Gamma$ (with a proper time cutoff)
is found by calculating
\begin{align}
[\Gamma]_\eps \equiv -\frac{1}{2} [\Tr\log (D + \hat\mu)]_\eps
\end{align}

The divergences of this effective action are
\begin{align}\label{eq_diverg}
[\Gamma]_\eps \ni&
	- \frac{1}{32\pi^2 \eps} \left[\Tr\left(1 - 2 V\right) X\right]_D \eol
	&\;\;\;\;\;\;\;\;\;\;
	+ \frac{\log \eps}{96\pi^2} S_\chi
	- \frac{\log \eps}{32 \pi^2} \left[X \Tr (e^{-V} \bar\mu e^{-V^T} \mu)\right]_D \eol
	&\;\;\;\;\;\;\;\;\;\;
	- \frac{\log \eps}{64\pi^2}
		\left(\left[\left(W^\alpha - \frac{1}{6} X^\alpha\right)^2
	+ \frac{2}{3} W^{\gamma\beta\alpha} W_{\gamma\beta\alpha} \right]_F + \hc \right)
\end{align}
where
\begin{align}
S_\chi = 
\left[G^2 + 2 R \bar R\right]_D
+ \left(\frac{1}{12}\left[X^\alpha X_\alpha \right]_F
	+ \frac{1}{2} \left[W^{\gamma \beta \alpha} W_{\gamma \beta \alpha}\right]_F + \hc\right)
\end{align}
We emphasize that the logarithmic divergences are independent of the choice of
where to place the $U(1)$ factor.

Of a nearly identical form is the conformal anomaly:
\begin{align}\label{eq_canom}
\delta_c [\Gamma]_\eps &=
	+ \frac{1}{16\pi^2 \eps} \Tr \left[\lambda X (1 - 2 V) \right]_D \eol
	& \;\;\;\;\;\;\;\;
	- \frac{1}{16\pi^2} \left[\lambda X \Tr (e^{-V} \bar\mu e^{-V^T} \mu) \right]_D
	+ \frac{1}{48\pi^2} \left[\lambda \Omega_\chi \right]_D\eol
	& \;\;\;\;\;\;\;\;
	-\frac{1}{16\pi^2}
		\Tr \left[\lambda \left(W^\alpha - \frac{1}{6} X^\alpha\right)^2
		+ \frac{2}{3} \lambda W^{\gamma\beta\alpha} W_{\gamma\beta\alpha} \right]_F \eol
	& \;\;\;\;\;\;\;\;
	- \frac{1}{32\pi^2}
	\Tr \left[- R \CD^\alpha \lambda \CD_\alpha V
	+ \frac{1}{3} \CD^\alpha \lambda \, X_\alpha V
	- \frac{2}{3} \CD^a \lambda \CD_a V
	\right]_D + \hc
\end{align}
where
\begin{align}
\Omega_\chi \equiv G^2 + \ACP R + \CP \bar R - 2 R \bar R + \frac{1}{6} \Omega_X + \Omega_L
\end{align}
(Recall that $S_\chi = [\Omega_\chi]_D$.)
It is worth noting that the finite part of the conformal anomaly is
independent of the $U(1)$ ambiguity when $\lambda$ is a constant.

The part of the gauge anomaly which is covariant and independent of the path
comes from
\begin{align}\label{eq_ganom}
\delta_{g} [\Gamma]_\eps &=
	+ \frac{1}{16\pi^2 \eps} \left[\Tr \Lambda \; X \right]_D
	+ \frac{1}{48\pi^2} \left[\Tr \Lambda \;\Omega_\chi \right]_D \eol
	& \;\;\;\;\;\;\;\;
	- \frac{1}{16\pi^2}
		\left[\Tr(\Lambda W^\alpha W_\alpha) + \frac{1}{36} \Tr \Lambda \; X^\alpha X_\alpha
		+ \frac{2}{3} \Tr \Lambda \;W^{\gamma\beta\alpha} W_{\gamma\beta\alpha}\right]_F + \hc \eol
	& \;\;\;\;\;\;\;\;
	+ \textrm{non-covariant piece}
\end{align}

This differs in three places from the form of the conformal anomaly.
Two of them can easily be restored by local counterterms. Both the
missing cross term $[W^\alpha X_\alpha]_F$ and the missing divergent term
$[\Tr \Lambda V]_D$ can be introduced by using
$\delta_g \Tr(V^2)/2 = \Tr(\Lambda V) + \Tr(\bar\Lambda V)$.
The divergent term is proportional to this directly while the cross
term can be generated from $[\Tr (V^2) \CD^\alpha X_\alpha]_D$. Note that since
these terms are quadratic in the gauge charge, they \emph{cannot} come from the
non-covariant piece, which is proportional to the symmetrized trace of
three gauge generators. It is interesting that if we restricted to an
anomaly free representation (or even just a traceless representation),
both of these terms in the conformal anomaly would vanish, since they are
proportional to the trace of a single generator, and so there would be no
motivation to reintroduce them for the gauge sector.

The mass term, if we assume it should have the form
$[X \Tr (\Lambda e^{-V} \bar\mu e^{-V^T} \mu)]_D$
is more difficult to generate for an arbitrary
gauge transformation $\Lambda$. However, one \emph{can} generate this
term for the $U(1)$ part of $\Lambda$ by using
$[X (\Tr V) (\Tr e^{-V} \bar\mu e^{-V^T} \mu)]_D$,
which is enough to verify that the $U(1)$ ambiguity
is indeed restricted to local counterterms.

\section{Applications}
\subsection{Old minimal supergravity coupled to chiral matter}
In the conformal compensator formalism,
\begin{align}\label{eq_ccfsm}
S = -3\times \left[\Phi_0^\dag e^{-K/3} \Phi_0\right]_D
	+ \left[\Phi_0^3 W \right]_F + \left[\bar\Phi_0^3 \bar W \right]_{\bar F}
\end{align}
$\Phi_0$ is a weight $(1, 3/2)$ conformally chiral superfield, $K$ is weight $(0,0)$ and
Hermitian, and $W$ is weight $(0,0)$ conformally chiral chiral.
There are $N$ chiral matter superfields $\Phi^i$ on which
$K$ and $W$ depend.

Different gauge choices for $\Phi_0$ correspond to different
conformally related flavors of minimal supergravity; in these versions, the quanta
of $\Phi_0$ are interpreted as quanta of the gravitational sector. Here we will
leave $\Phi_0$ ungauged and its quanta we will interpret at the same level as the
other chiral matter. There is some question as to the physicality of this approach;
after all, these quanta appear with the wrong sign kinetic term and so their Euclidean
path integral is poorly defined.\footnote{This is an old problem in the
non-supersymmetric gravity literature. The famous paper of Gibbons, Hawking
and Perry \cite{Gibbons:1978ac} suggested to Euclideanize the conformal mode of the
graviton with an additional factor of $i$.} Since the quanta can be removed by a certain gauge
choice for diffeomorphisms, any poor behavior of this sector should be accounted for
when the entire graviton and Fadeev-Popov sectors are taken into account.

In previous work \cite{paper3}, we have expanded out the action to second order
in the quanta of the chiral, gauge, and supergravity superfields. This action
possesses kinetic mixing between the chiral and gravity sectors;
in terms of Feynman graphs, the chiral and supergravity quanta mix with a
coupling that goes as $p^2$. The proper procedure then is to find a clever
gauge fixing procedure to remove the kinetic mixing (this was the approach
taken in \cite{supergraphity, superspace}) or to find a way to deal with an arbitrary
operator on the space of vector \emph{and} chiral superfields.

Either approach is beyond the scope of the tools developed here
so we will restrain to a more limited case: we will attempt to calculate divergences
and anomalies due \emph{purely to chiral loops}. The analogous procedure in
a non-supersymmetric theory would be to calculate loops involving both matter
and the conformal mode of the graviton only. There may be some divergences
and anomalies found in mixed loops, but we will not attempt to discover
those here.

To calculate the effective action due to chiral loops, we must expand $\Phi^i$
and $\Phi_0$ as a background plus a quantum superfield.
How precisely we do this is a matter of defining quantization and should not
affect the final result provided the background fields are taken to satisfy the equations
of motion. We will choose
\begin{align}
\delta \Phi^i = \eta^i, \;\;\; \delta \Phi_0 = \eta_0
\end{align}
where $\eta^i$ is weight $(0,0)$ chiral and $\eta_0$ is weight $(1,3/2)$ chiral.

Denote $Z = -3\bar \Phi_0 \Phi_0 e^{-K/3}$ and $P = \Phi_0^3 W$ for
generality.\footnote{The definition of $Z$ differs by a factor of $-3$
from that used in \cite{paper3}.} Introducing the notation $\Phi^I = (\Phi_0, \Phi^i)$,
the action may be written
\begin{align}
S = \left[Z \right]_D + \left[P\right]_F + \left[\bar P \right]_{\bar F}
\end{align}
with a first order variation
\begin{align}
S^{(1)} = \left[\eta^I \CP Z_I + \eta^I P_I \right]_F
\end{align}
where $\CP = -\nabla^2/4$ is the conformal chiral projector. The equations
of motion $\CP Z_I = -P_I$ amount to
\begin{align}
\CP \left(\Phi_0 \bar\Phi_0 e^{-K/3}\right) = \Phi_0^3 W, \;\;\;
\CP \left(\Phi_0 \bar\Phi_0 e^{-K/3} K_i \right) = -\Phi_0^3 W_i, \;\;\;
\end{align}
If the gauge choice $\Phi_0 = e^{K/6}$ were adopted these would become
\begin{align}
2 R = e^{K/2} W, \;\;\;
-\frac{1}{4} (\BCD^2 - 8 R) K_i = -e^{K/2} W_i
\end{align}
The second of these may be rewritten using the first
as
\begin{align}
\frac{1}{4} \BCD^2 K_i = e^{K/2} (W_i + K_i W) \equiv e^{K/2} W_{;i}
\end{align}
In this form, both sides of the equations transform covariantly under K\"ahler
transformations.

The second variation is
\begin{align}
\frac{1}{2} S^{(2)} = \left[\bar \eta^{\bar I} Z_{\bar I J} \eta^J\right]_D
	+ \frac{1}{2} \left[\eta^I X_{I J} \eta^J\right]_F
	+ \frac{1}{2} \left[\bar\eta^{\bar I} \bar X_{\bar I \bar J} \bar\eta^{\bar J}\right]_F
\end{align}
where
\begin{align}
X_{I J} = P_{I J} + \CP Z_{I J}
\end{align}
Manifest reparametrization invariance has been lost at the second variation.
If we wanted to maintain it, we would need to introduce an affine connection
on the space of chiral superfields. There is no object in the theory which can
serve this purpose (the K\"ahler affine connection being non-chiral), so we would
have to insert one by hand. This seems artificial so we accept the loss of
manifest reparametrization invariance and expect it to be restored on shell.

The kinetic matrix $Z_{\bar I J}$ is clearly an object which we can treat analogously
as $e^V$, except for the difficulty that its indices carry conformal as well as
Yang-Mills charge. This can be remedied by introducing a particular
measure for the path integration variables $\eta^I$ so that each of the $\eta^I$
are dimension $(3/2, 1)$. Then we could write $Z_{\bar I J}$ as
$(e^V)_{\bar I J} / X^{1/2}$ where $X$ has dimension two and $V$ is dimensionless.
In calculating the effective action, $V$ and $X$ would appear differently (as we
have previously discussed), but for certain questions we would find answers that
were independent of the particular details of this separation.
In particular, the \emph{logarithmic} divergences for the theory take the form
(including the mass term)
\begin{align}
\Gamma = -\frac{1}{2} \Tr\log (D+\hat\mu)
	\ni
	-\frac{\log \eps}{64\pi^2}
		\Tr \left(\left[\Phi_Z + \frac{2}{3} \Phi_W\right]_F + \hc \right)
	+ \frac{\log \eps}{96 \pi^2} S_\chi
	- \frac{\log \eps}{32\pi^2} \Tr \left[\Omega_P\right]_D
\end{align}
where
\begin{gather}
\Omega_P = X_{I J} Z^{J \bar J} \bar X_{\bar J \bar I} Z^{\bar I I} \\
\Phi_Z = \left(W^\alpha - \frac{1}{6} X^\alpha\right)^2, \;\;\;
\Phi_W = W^{\alpha \beta \gamma} W_{\alpha \beta \gamma}
\end{gather}
and
\begin{align}
S_{\chi} \equiv &
	\left[G^2 + \CP \bar R + \ACP R - 2 R \bar R \right]_D
	+ \frac{1}{12}\left[X^\alpha X_\alpha \right]_F
	+ \frac{1}{12}\left[\bar X_\dalpha \bar X^\dalpha \right]_{\bar F} \eol
	& + \frac{1}{2} \left[W^{\gamma \beta \alpha} W_{\gamma \beta \alpha}\right]_F
	+ \frac{1}{2} \left[\bar W_{\dgamma \dbeta \dalpha} \bar W^{\dgamma \dbeta \dalpha}\right]_{\bar F}
\end{align}
There is a distinction between $V$ and $X$ in $S_\chi$ and $\Phi_Z$, but the former is a
topological invariant independent of small variations in $X$ and the latter is manifestly
independent of the distinction, since we may rewrite
\begin{align}
\Phi_Z = Z^\alpha Z_\alpha, \;\;\;
Z_\alpha \equiv \frac{1}{8} \bar\nabla^2 \left(Z^{I \bar K} \nabla_\alpha Z_{\bar K J} \right)
\end{align}
where $Z^{I \bar J}$ is the inverse of the kinetic matrix $Z_{\bar J I}$.
(The Weyl curvature $W_{\alpha \beta \gamma}$ is, of course, independent of
$X$ since it is defined in conformal supergravity.)

Only the mass term $\Omega_P$ and the field strength
$\Phi_Z$ are the interesting objects to investigate. We will begin by
evaluating $\Omega_P$.

\subsubsection{Simplifying $\Omega_P$}
To simplify this term, it helps to introduce reparametrization connections
and curvatures for the kinetic matrix $Z$. Observe first that
\begin{align*}
\bar \nabla^2 Z_{I J} &= \nabla_\dalpha (Z_{I J \bar J} \nabla^\dalpha \bar\Phi^{\bar J})
	= \nabla_\dalpha (\Gamma(Z)_{I J}{}^K Z_{K \bar J} \nabla^\dalpha \bar\Phi^{\bar J}) \eol
	&= R(Z)_{I\bar J J \bar K} \nabla_\dalpha \bar\Phi^{\bar K} \nabla^\dalpha \bar\Phi^{\bar J}
	+ \Gamma(Z)_{I J}{}^K \bar\nabla^2 Z_K
\end{align*}
where $\Gamma(Z)$ and $R(Z)$ are analogous to the K\"ahler connection and
curvature but defined with the kinetic matrix $Z$ instead of the K\"ahler potential.
The connections are
\begin{gather*}
\Gamma(Z)_{i j}{}^k = \Gamma_{ij}{}^k - \frac{1}{3} \delta_i{}^k K_j
	- \frac{1}{3} \delta_j{}^k K_i \\
\Gamma(Z)_{ij}{}^0 = \frac{\Phi_0}{3} \left(\Gamma_{ij}{}^k K_k - K_{ij} - \frac{1}{3} K_i K_j\right) \\
\Gamma(Z)_{0 j}{}^k = \Phi_0^{-1} \delta_j{}^k \\
\Gamma(Z)_{0j}{}^0 = \Gamma(Z)_{00}{}^k = \Gamma(Z)_{00}{}^0 = 0
\end{gather*}
and the curvatures are
\begin{gather*}
R(Z)_{ij}{}^k{}_{\bar k} = R_{ij}{}^k{}_{\bar k} - \frac{1}{3} \delta_i{}^k K_{j \bar k}
	- \frac{1}{3} \delta_j{}^k K_{i \bar k} \\
R(Z)_{ij}{}^k{}_{\bar 0} = 0 \\
R(Z)_{ij}{}^0{}_{\bar k} = \frac{\Phi_0}{3} \left(
	R_{ij}{}^k{}_{\bar k} K_k
	- \frac{1}{3} K_j K_{i \bar k} - \frac{1}{3} K_i K_{j \bar k}
	\right) \\
R(Z)_{ij}{}^0{}_{\bar 0} = 0 \\
R(Z)_{0 J}{}^K{}_{\bar L} = 0
\end{gather*}
In these equations, the quanties on the left have an index structure
associated with $Z_{I\bar J}$ (i.e. indices are raised and lowered with
the kinetic matrix) while the quantities on the right have an index
structure associated with the K\"ahler metric $K_{i \bar j}$.

Lowering the indices on the left using the kinetic matrix, we find that
the only non-vanishing $R(Z)_{I J \bar J \bar K}$ is
\begin{align}
R(Z)_{i j \bar j \bar k} = \Phi_0 \bar\Phi_0 e^{-K/3} 
	\left(
	R_{ij \bar j \bar k}
	- \frac{1}{3} K_{i \bar j} K_{j\bar k} - \frac{1}{3} K_{i \bar k} K_{j\bar j}
	\right)
\end{align}
which is both reparametrization covariant and K\"ahler invariant.
This observation dramatically simplifies calculations involving $R(Z)$.

Using the equation of motion $\CP Z_I = -P_I$, we may rewrite
\begin{align}
\CP Z_{I J} 
	&= -\frac{1}{4} R(Z)_{I\bar J J \bar K} \nabla_\dalpha \bar\Phi^{\bar K} \nabla^\dalpha \bar\Phi^{\bar J}
	- \Gamma(Z)_{I J}{}^K P_K
\end{align}
and then rewrite the ``mass term''
\begin{align}
X_{IJ} = P_{I J} + \CP Z_{I J}
	= P_{;I J} -\frac{1}{4} R(Z)_{I\bar J J \bar K} \nabla_\dalpha \bar\Phi^{\bar K} \nabla^\dalpha \bar\Phi^{\bar J}
\end{align}
in a reparametrization covariant way. The notation $; I$ denotes
the covariant field derivative, using the connection $\Gamma(Z)$.

We may easily calculate
\begin{gather*}
P_{; 0 0} = 6 \Phi_0 W \\
P_{; 0 j} = 2 \Phi_0^2 W_j = 2 \Phi_0^2 \left(W_{;j} - K_j W\right)\\
P_{; i j} = \Phi_0^3 \left(W_{;ij} - \frac{2}{3} K_i W_{;j}
	- \frac{2}{3}  K_j W_{;i} + \frac{2}{3} K_i K_j W\right)
\end{gather*}
and, raising the left index,
\begin{gather*}
P^{\bar 0}{}_0 = e^{K/3} \Phi_0 \left(-2 W + \frac{2}{3} K_{\bar k} W^{;\bar k}\right) \\
P^{\bar i}{}_0 = 2 e^{K/3} \frac{\Phi_0}{\bar \Phi_0} W^{;\bar i} \\
P^{\bar 0}{}_{j} = e^{K/3} \Phi_0^2 \left(-\frac{2}{3} W_{;j} + \frac{2}{3} K_j W
	+ \frac{1}{3} K_{\bar k} W^{;\bar k}{}_j - \frac{2}{9} K_{\bar k} W^{;\bar k} K_j \right) \\
P^{\bar i}{}_{j} = e^{K/3} \frac{\Phi_0^2}{\bar\Phi_0} \left(W^{;\bar i}{}_j - \frac{2}{3} K_j W^{;\bar i} \right)
\end{gather*}
The notation $;i$ on the right side of these equations denotes
field differentiation covariant with respect to both K\"ahler
transformations \emph{and} reparametrizations. Thus,
\begin{align}
W_{;i} = D_i W = W_i + K_i W
\end{align}
and
\begin{align}
W_{; i j} = D_j W_{; i} = \partial_j W_{; i} - \Gamma^k{}_{j i} W_{; k} + K_j W_{; i}
\end{align}

The mass term can then be expanded as
\begin{align*}
\Omega_P = P_{; I J} \bar P^{; J I}
	- \frac{1}{2} P_{; I J} R^{I J \alpha}{}_\alpha
	- \frac{1}{2} \bar P_{; \bar I \bar J} R^{\bar I \bar J}{}_\dbeta{}^\dbeta
	+ \frac{1}{16} R^{I J \alpha}{}_\alpha R_{I J \dbeta}{}^\dbeta
\end{align*}
The relevant quantities we will need are
\begin{gather*}
P_{; I J} \bar P^{; J I} = e^{2 K/3} \Phi_0 \bar\Phi_0
	\left(4 W \bar W - \frac{8}{3} W_{;j} {\bar W}^{;j}
	+ W_{;ij} {\bar W}^{;i j}\right) \\
P_{; K L} R^{K L}{}_{I J}
	= e^{K/3} \frac{\Phi_0^2}{\bar \Phi_0} \left(W_{;k\ell} R^{k \ell}{}_{i j}
		- \frac{2}{3} W_{;i j}\right) \\
R^{IJ}{}_{K L} R_{I J \bar K \bar L}
	= R^{ij}{}_{k \ell} R_{i j \bar k \bar \ell}
	- \frac{4}{3} R_{k \ell \bar k \bar \ell}
	+ \frac{2}{9} \left(K_{k \bar k} K_{\ell \bar \ell} + K_{k \bar \ell} K_{\ell \bar k} \right)
\end{gather*}
In the second two formulae, the free indices with $0$ or $\bar 0$ in the
slots are understood to vanish. This is due to the particular simplicity
of their kinetic matrix.

The mass term can then be written
\begin{align}
\Omega_P &= e^{2K/3} \Phi_0 \bar\Phi_0
	\left(4 W \bar W - \frac{8}{3} W_{;j} {\bar W}^{;j}
	+ W_{;ij} {\bar W}^{;i j}\right) \eol
	& \;\;\;\;\;\;
	- \frac{1}{2} e^{K/3} \frac{\Phi_0^2}{\bar \Phi_0} \left(W_{;k\ell} R^{k \ell}{}_{i j}
		- \frac{2}{3} W_{;i j}\right) \nabla^\alpha \phi^i \nabla_\alpha \phi^j + \hc\eol 
	& \;\;\;\;\;\;
	+ \frac{1}{16} R^{ij\alpha}{}_\alpha R_{i j \dalpha}{}^\dalpha
	- \frac{1}{12} R^\alpha{}_{\alpha \dalpha}{}^\dalpha
	+ \frac{1}{36} K^{\dalpha \alpha} K_{\alpha \dalpha}
\end{align}
We use here a compact notation where an $\alpha$ in place of an index $i$
denotes saturation with $\nabla_\alpha \phi^i$; thus
\begin{gather}
K_{\alpha \dalpha} = K_{i \bar j} \nabla_\alpha \phi^i \nabla_\dalpha \bar\phi^\bj, \;\;\;
R_{i j \dalpha}{}^\dalpha = R_{i j \bar j \bar k} \nabla_\dalpha \bar\phi^{\bj} \nabla^\dalpha \bar\phi^{\bar k}, \;\;\;
\textrm{etc.}
\end{gather}

\subsubsection{Simplifying $\Phi_Z$}
Next we turn to evaluating $\Phi_Z = Z^\alpha Z_\alpha$, where
\begin{align*}
Z_\alpha{}^I{}_J = \W_\alpha^I{}_J
	+ \frac{1}{8} \bar\nabla^2 \left(Z^{I \bar K} \nabla_\alpha Z_{\bar K J}\right)
	= \W_\alpha^I{}_J + \frac{1}{8} \bar\nabla^2 \biggl(\Gamma(Z)^I{}_{J K} \nabla_\alpha \Phi^K\biggr)
\end{align*}
We evaluate each term in turn, keeping in mind that $\Phi_0$ is assumed to be
a gauge singlet:
\begin{gather}
Z_\alpha{}^0{}_0 = 0 \\
Z_\alpha{}^i{}_0 = \frac{\Phi_0^{-1}}{8} \bar\nabla^2 \nabla_\alpha \Phi^i
	= \Phi_0^{-1} (\W_\alpha \Phi^i) \\
Z_\alpha{}^0{}_j = -\frac{\Phi_0}{24} \bar\nabla^2 \left(
	K_{j \bar k} \nabla_\alpha (K^{\bar k k} K_k) + \frac{1}{3} K_j \nabla_\alpha K
	\right) \\
Z_\alpha{}^i{}_j
	= \W_\alpha{}^i{}_j
     - \Gamma_\alpha{}^i{}_j
	+ \frac{1}{3} X_\alpha \delta^i{}_j
	- \frac{1}{24} \bar\nabla^2 (K_j \nabla_\alpha \phi^i)
\end{gather}
where we have defined the effective reparametrization gaugino field strength
\begin{align}
\Gamma_\alpha{}^i{}_j \equiv - \frac{1}{8} \bar\nabla^2 \biggl(\Gamma^i{}_{j k} \nabla_\alpha \phi^k\biggr)
\end{align}
The trace of $Z^\alpha Z_\alpha$ can be simplified by extracting
$\W_\alpha{}^i{}_j$, $\Gamma_\alpha{}^i{}_j$, and $X_\alpha$ which are
invariant under K\"ahler transformations and treating the non-invariant terms
separately. One finds
\begin{align}
\left[\Tr (Z^\alpha Z_\alpha)\right]_F &= \left[\Tr \left(\W_\alpha{}^i{}_j - \Gamma_\alpha{}^i{}_j + \frac{1}{3} X_\alpha \delta^i{}_j\right)^2 + \frac{1}{9} X^\alpha X_\alpha\right]_F \eol
	& \;\;\;\;\;\;\;\;
	+ \left[
	\frac{1}{72} K^{\dalpha \alpha} K_{\alpha \dalpha}
	- \frac{1}{24} R^\alpha{}_{\alpha \dalpha}{}^\dalpha
	- \frac{1}{6} \nabla^\alpha \W_\alpha^r \left(K_k X_r \phi^k - K_{\bar k} X_r \bar\phi^{\bar k} \right)
	\right]_D
\end{align}
where the trace in the first line is to be understood as over the
``matter'' fields $\phi^i$ only.

For reference, we have defined
\begin{gather}
K_{\alpha \dalpha} = K_{k \bar k} \nabla_\alpha \phi^k \bar \nabla_\dalpha \bar\phi^{\bar k} \\
R^\alpha{}_{\alpha \dalpha}{}^\dalpha =
	R_{j k \bar j \bar k} \nabla^\alpha \phi^j \nabla_\alpha \phi^k \bar \nabla_\dalpha \phi^\bj \bar\nabla^\dalpha \phi^{\bar k} \\
\Gamma_\alpha{}^i{}_j = -\frac{1}{8} \bar\nabla^2 \left(K^{i \bar k} \nabla_\alpha K_{\bar k j} \right)
\end{gather}

The appearance of the combination $\W_\alpha{}^i{}_j - \Gamma_\alpha{}^i{}_j$
as a field strength is gratifying. In a component calculation, we have
(after applying the equations of motion for the auxiliary fields) a reparametrization
connection for the component fields, and so we would expect $\Gamma_\alpha{}^i{}_j$
to appear in the final answer with the Yang-Mills connection, which it here
does. Moreover, this specific combination is necessary in order to have
covariance under a full gauged isometry \cite{bgg}.

\subsubsection{Summary: Chiral loop logarithmic divergences}
The logarithmic divergences of the theory can be written in the following way:
\begin{align}\label{eq_chLoopLogDiv}
\Gamma & \ni
	-\frac{\log \eps}{64\pi^2}
	\left(\left[\Phi_1 + \frac{2}{3} (N+1) \Phi_W\right]_F + \hc\right)
	+ \frac{\log \eps}{96\pi^2} (N+1) S_\chi
	- \frac{\log \eps}{32\pi^2} \biggl[\Omega_1 + \Omega_2 + \Omega_3\biggr]_D
\end{align}
where
\begin{gather}
\Phi_1 = \Tr \left(\W_\alpha{}^i{}_j - \Gamma_\alpha{}^i{}_j + \frac{1}{3} X_\alpha \delta^i{}_j\right)^2
		+ \frac{1}{9} X^\alpha X_\alpha \eol
\Phi_W = W^{\alpha \beta \gamma} W_{\alpha \beta \gamma}
\end{gather}
The curvatures appearing in the trace in $\Phi_1$ can be understood as the effective curvatures
(after equations of motion are applied) for the underlying component theory.
For example, $\Gamma_\alpha{}^i{}_j$ has the interpretation as the K\"ahler
reparametrization curvature and $X_\alpha \delta^i{}_j$ is the effective $U(1)_R$
curvature.

There are additional D-terms which are more difficult to interpret:
\begin{align}
\Omega_1 &= e^{2K/3} \Phi_0 \bar\Phi_0
	\left(4 W \bar W - \frac{8}{3} W_{;j} {\bar W}^{;j}
	+ W_{;ij} {\bar W}^{;i j}\right) \\
\Omega_2 &= - \frac{1}{2} e^{K/3} \frac{\Phi_0^2}{\bar \Phi_0} \left(W_{;k\ell} R^{k \ell}{}_{i j}
	- \frac{2}{3} W_{;i j}\right) \nabla^\alpha \phi^i \nabla_\alpha \phi^j + \hc\eol 
	& \;\;\;\;\;\;
	+ \frac{1}{16} R^{ij\alpha}{}_\alpha R_{i j \dalpha}{}^\dalpha
	+ \frac{1}{24} K^{\dalpha \alpha} K_{\alpha \dalpha}
	- \frac{1}{8} R^\alpha{}_{\alpha \dalpha}{}^\dalpha \\
\Omega_3 &= - \frac{1}{6} \nabla^\alpha \W_\alpha^r \left(K_k X_r \phi^k - K_{\bar k} X_r \bar\phi^{\bar k} \right)
\end{align}
Although $\Omega_1$ can be thought of as a renormalization of
the K\"ahler potential, the others cannot since they involve derivatives of
the background fields and we usually consider the K\"ahler potential to be
derivative-free.

Finally there is a topological term
\begin{align}
S_\chi = \left[G^2 + 2 R \bar R\right]_D
	+ \mathrm{Re} \left[W^{\gamma \beta \alpha} W_{\gamma \beta \alpha} + \frac{1}{6} X^\alpha X_\alpha\right]_F
\end{align}
which is the superspace version of the Gauss-Bonnet term.

\subsubsection{Chiral loop quadratic divergences}
The logarithmic divergences considered previously are the physical divergences of
the theory, in the sense that they are independent of the particular form of our
regularization prescription. This is not true of the quadratic divergences,
which for our generic model take the form
\begin{align}
\Gamma =-\frac{1}{32\pi^2 \eps} \left[\Omega_X + \Omega_V\right]_D
\end{align}
where
\begin{gather}
\Omega_X = (N+1) X, \;\;\;
\Omega_V = -2 X \Tr V
\end{gather}
These clearly depend on the precise choice of $X$, which is itself
partly determined by the choice of path integration measure.

Focusing on the D-term, we note that the kinetic matrix is
\begin{align*}
Z_{\bar I J} = e^{-K/3} \left(
\begin{array}{cc}
-3 & \Phi_0 K_j \\
K_{\bar i} \bar\Phi_0 & \Phi_0^\dag \Phi_0 \left(K_{\bar i j} - \frac{1}{3} K_{\bar i} K_j \right)
\end{array}
\right)
\end{align*}
We haven't as yet specified the precise measure. If we take the point
of view that the field $\Phi_0$ is to be truly used as a compensator,
then the simplest approach is to define the measure to include
various factors of $\Phi_0$ so that the effective path integral
variables are of dimension $(3/2, 1)$. Performing such a rescaling
involves taking $\eta^i \rightarrow \frac{1}{\Phi_0^{3/2}} \eta^i$
and $\eta^0 \rightarrow \frac{1}{\sqrt {\Phi_0}} \eta^0 \times \frac{1}{\sqrt{3}}$
(the additional $\sqrt 3$ factor to normalize the kinetic term of $\eta^0$):
\begin{align*}
Z_{\bar I J}' = \frac{e^{-K/3}}{(\Phi_0 \bar\Phi_0)^{1/2}} \left(
\begin{array}{cc}
-1 & \frac{1}{\sqrt 3} \bar K_j \\
\frac{1}{\sqrt 3} K_{\bar i} & \left(K_{\bar i j} - \frac{1}{3} K_{\bar i} K_j \right)
\end{array}
\right)
\end{align*}
where now the fields $\eta'^i$ and $\eta'^0$ have the same dimension.

Unfortunately, $\eta^0$ still conspicuously has the wrong sign kinetic term.
The approach advocated in \cite{Gibbons:1978ac} would involve taking
$\eta^0 \rightarrow \beta \eta^0$, $\bar\eta^0 \rightarrow \bar\beta \bar\eta^0$
with $\beta \bar\beta = -1$, requiring that the naive
understanding of conjugation be modified after Euclideanizing this mode.
We will take this approach here, leaving $\beta$ and $\bar\beta$ arbitrary
except for the requirement that $\bar\beta \beta = -1$.

This leads to
\begin{align}
Z'_{\bar I J} = \frac{e^{-K/3}}{(\Phi_0^\dag \Phi_0)^{1/2}} \left(
\begin{array}{cc}
1 & \frac{\bar\beta}{\sqrt 3} K_j \\
\frac{\beta}{\sqrt 3} K_{\bar i} & \left(K_{\bar i j} - \frac{1}{3} K_{\bar i} K_j \right)
\end{array}
\right)
\end{align}
The precise choice of $\beta$ and $\bar\beta$ should not have an effect
on the final answer.

We still must separate this kinetic matrix into conformal and gauge terms.
The most physically sensible choice is to identify $X$ as the quantity
in the classical theory which is gauged to unity, that choice here being
\begin{align}
X = \Phi_0 \bar \Phi_0 e^{-K/3}
\end{align}
Given that choice, the non-Yang-Mills part of $V$ is defined by
\begin{align}
e^V = e^{-K/2} \left(
\begin{array}{cc}
1 & \frac{\bar\beta}{\sqrt 3} K_j \\
\frac{\beta}{\sqrt 3} K_{\bar i} & \left(K_{\bar i j} - \frac{1}{3} K_{\bar i} K_j \right)
\end{array}
\right)
\end{align}
which yields
\begin{align}
\Tr V = \Tr \mathbf V - \frac{N+1}{2} K + \Tr\log K_{k \bar k}
\end{align}
where $\mathbf V$ is the true Yang-Mills prepotential.
We have then the quadratic divergences
\begin{align}
\Gamma =-\frac{1}{32\pi^2 \eps} \left[\Omega_X + \Omega_V\right]_D
\end{align}
where
\begin{gather}
\Omega_X = \Phi_0 \bar\Phi_0 e^{-K/3} (N+1), \;\;\;
\Omega_V = \Phi_0 \bar\Phi_0 e^{-K/3} \biggl(
	-2\Tr \mathbf V + (N+1) K - 2 \Tr\log K_{k \bar k}\biggr)
\end{gather}

In the gauge where $\Phi_0 = e^{K/6}$, one can easily check
that in the absence of fermions for a generic (0,0) superfield $V$
\[
[\Phi_0 \bar\Phi_0 e^{-K/3} V]_D = -\frac{1}{3} V \mathcal L_{sg+m}
	+ \frac{1}{16} \CD^\alpha (\BCD^2 - 8 R) \CD_\alpha V
	- 8 \bar R \BCD^2 V - 8 R \CD^2 V
\]
where $\mathcal L_{sg+m}$ is the normal Lagrangian of supergravity
coupled to a K\"ahler potential. 
Assuming Wess-Zumino gauge for reparametrizations, Yang-Mills, and
K\"ahler transformations, we conclude
\begin{align*}
[\Omega_V]_D = -2 \Tr \mathbf D - \frac{1}{2} (N+1) \CD^\alpha X_\alpha
	+ \CD^\alpha \Gamma_\alpha{}^j{}_j
\end{align*}
This coincides with component field calculations \cite{mkg_others}, which isn't
too surprising, since our choice of $X$ corresponds to the natural
choice of a Weyl-rescaled metric at the component level.

In addition, using the superfield equations of motion and neglecting all fermions
\begin{align*}
-\frac{1}{3} \mathcal L_{sg+m} = [1]_D &= [2 R]_F = \left[e^{K/2} W\right]_F
	= -e^K W_{;k} {\bar W}^{;k} + 3 e^K W \bar W 
\end{align*}
and so
\begin{align*}
[\Omega_X]_D = (N+1) \times \left(-e^K W_{;k} {\bar W}^{;k} + 3 e^K W \bar W \right) = -(N+1) \hat V
\end{align*}
This result differs from a corresponding result in \cite{mkg_others},
where Gaillard, Jain, and collaborators found $\hat V + M^2$, where $M^2$ is
the gravitino mass squared, using a momentum cutoff calculation.
The deviation seems likely due to a breakdown in supersymmetry due
to the cutoff.\footnote{A subsequent analysis with Pauli-Villars
regulators\cite{mkg_pv} found a supersymmetric divergence, but the original analysis
with a momentum cutoff is closer in spirit to the analysis performed here.}

\subsubsection{Anomalies}
There are a number of classical symmetries respected by the action
\eqref{eq_ccfsm} which are not manifestly respected by the
measure.\footnote{Of these, only Yang-Mills gauge transformations are
physical and thus the only one which must be anomaly-free to yield a
consistent theory. However, in string-inspired supergravity theories, modular
transformations in the underlying string theory manifest themselves
in the effective supergravity theory as a certain combination of
reparametrization and K\"ahler transformations. Thus it seems useful
to consider the general class of symmetries described here.}
These are
\begin{enumerate}
\item K\"ahler transformations
\begin{gather}
\Phi_0 \rightarrow e^{F/3} \Phi_0, \;\;\;
K \rightarrow K + F + \bar F, \;\;\;
W \rightarrow e^{-F} W
\end{gather}
\item Reparametrizations of the chiral matter
\begin{gather}
\Phi^i \rightarrow \Lambda^i(\Phi)
\end{gather}
\item Yang-Mills gauge transformations
\begin{gather}
\Phi^i \rightarrow \exp(\Lambda^r \mathbf T_r){}^i{}_j \Phi^j, \;\;\;
e^V \rightarrow e^{\bar \Lambda} e^V e^{\Lambda}
\end{gather}
\end{enumerate}

Our choice of $X = \bar\Phi_0 \Phi_0 e^{-K/3}$ is conspicuous in being the
choice which is K\"ahler invariant in addition to being Yang-Mills and
reparametrization invariant. This means that each of these transformations
manifests itself as a \emph{gauge} anomaly in the way we defined
the effective action.

This is not the only reasonable choice. We could have chosen, for example,
$X = \bar\Phi_0 \Phi_0$, which would correspond to a calculation in
conventional (i.e. non-K\"ahler) Poincar\'e supergravity. The K\"ahler
anomaly in such a calculation would be a purely conformal anomaly.
Another choice would be to place \emph{all} of the $e^K$ factors into $X$;
this would yield a combination of conformal and gauge anomalies which together
give the K\"ahler anomaly.
However, as we have shown, the difference between any of these approaches is
a local (though infinite) counterterm and so there is no particular
need to choose one over any other.

Since the above set of transformations may all be interpreted as gauge
transformations, we can treat them in one step.
Taking into account the rescalings we have made, we find the transformations
\begin{gather*}
\delta \eta_0' = \frac{F}{2} \eta_0 + \frac{1}{\beta\sqrt 3}
	F_i \eta'^i + \mathcal O(\eta^2), \;\;\;
\delta \eta'^i = \frac{F}{2} \eta'^i + \Lambda^i{}_j \eta'^j + \Lambda^r \mathbf T_r{}^i{}_j \eta'^j + \mathcal O(\eta^2)
\end{gather*}

The kinetic matrix associated with our variable choice is
\begin{align}
\frac{1}{X^{1/2}} e^V = \frac{e^{-K/2}}{X^{1/2}} \left(
\begin{array}{cc}
1 & \frac{\bar\beta}{\sqrt 3} K_j \\
\frac{\beta}{\sqrt 3} K_{\bar i} & \left(K_{\bar i j} - \frac{1}{3} K_{\bar i} K_j \right)
\end{array}
\right)
\end{align}
where $X = \Phi_0 \bar\Phi_0 e^{-K/3}$. This choice of $X$ is particular
in being totally invariant under the combined K\"ahler and reparametrization symmetries.
The anomaly associated with these is then simply a gauge anomaly. Taking the regulated
effective action (i.e. the $\eps$-divergent effective action with a simple subtraction
to remove the $\eps$ divergences), the covariant part of the one-loop anomaly is
\begin{align}
\delta_g [\Gamma]_{reg} =&\,
	- \frac{1}{16\pi^2}
		\left[\Tr (\Lambda \hat Z^\alpha \hat Z_\alpha)
		+ \frac{2}{3} \Tr \Lambda W^{\gamma\beta\alpha} W_{\gamma\beta\alpha}\right]_F + \hc \eol
	& +\frac{1}{48 \pi^2} [\Tr \Lambda \Omega_\chi]_D
	+ \textrm{non-covariant piece}
\end{align}
with infinitesimal gauge parameter
\begin{align}
\Lambda^I{}_J = \left(
\begin{array}{cc}
-\frac{1}{2} F & \frac{\bar \beta}{\sqrt 3} F_j \\
0 & -\frac{1}{2} F \delta^i{}_j- \Lambda^i{}_j - \Lambda^r \mathbf T_r{}^i{}_j
\end{array}
\right)
\end{align}
In the expression for the anomaly, we have ``completed the square'' for the curvature
piece by introducing the local counterterm whose gauge variation includes $W^\alpha X_\alpha$.
In the variables we are using, $\hat Z_\alpha$ has the components
\begin{gather}
\hat Z_\alpha{}^0{}_0 = 0 \\
\hat Z_\alpha{}^i{}_0 = \frac{\beta}{\sqrt 3} (\W_\alpha \Phi^i) \\
\hat Z_\alpha{}^0{}_j = -\frac{\sqrt 3}{24\beta} \bar\nabla^2 \left(
	K_{j \bar k} \nabla_\alpha (K^{\bar k k} K_k) + \frac{1}{3} K_j \nabla_\alpha K
	\right) \\
\hat Z_\alpha{}^i{}_j
	= \W_\alpha{}^i{}_j - \Gamma_\alpha{}^i{}_j
	+ \frac{1}{3} X_\alpha \delta^i{}_j
     - \frac{1}{24} \bar\nabla^2 (K_j \nabla_\alpha \phi^i)
\end{gather}
where $\beta \bar\beta = -1$. We have neglected the part of the anomaly arising
from the path-dependent piece.

The covariant part of the K\"ahler anomaly is
\begin{align}\label{eq_Kanom}
\delta_{g} [\Gamma]_{reg} &\ni
	+ \frac{1}{32\pi^2}
		\left[F \Phi_1
		+ \frac{2}{3} F (N+1) W^{\gamma\beta\alpha} W_{\gamma\beta\alpha}\right]_F
	- \frac{1}{96 \pi^2} \biggl[ F (N+1) \Omega_\chi \biggr]_D \eol
	& \;\;\;\;\;\;\;\;
	+ \frac{1}{32\pi^2}
	\left[- \frac{1}{3} F K_{i \bar j} \nabla^\alpha \phi^i W_\alpha \bar\phi^\bj
	- \frac{1}{24} F R^\alpha{}_{\alpha \dalpha}{}^\dalpha
	+ \frac{1}{72} F K^{\alpha \alpha} K_{\alpha \dalpha} \right]_D \eol
	& \;\;\;\;\;\;\;\;
	+ \frac{1}{32\pi^2} \left[
	- \frac{1}{3} F_j (\W^\alpha - \Gamma^\alpha)^j{}_k \nabla_\alpha \phi^k
	+ \frac{1}{9} \nabla^\alpha F K_k \W_\alpha \phi^k\right]_D
	+ \hc
\end{align}
where
\begin{align}
\Phi_1 = \Tr \left(\W_\alpha{}^i{}_j - \Gamma_\alpha{}^i{}_j + \frac{1}{3} X_\alpha \delta^i{}_j\right)^2
		+ \frac{1}{9} X^\alpha X_\alpha
\end{align}
The first two lines of this expression are quite similar to the expression for the logarithmic
divergences given in \eqref{eq_chLoopLogDiv}. $\Phi_1$ is as defined there, for example, and
$F K_{i \bar j} \nabla^\alpha \phi^i \mathcal W_\alpha\bar\phi^\bj$ is equivalent
to that equation's $\Omega_3$ after integrating the latter by parts. As before, the Yang-Mills
curvature appears only in the reparametrization-covariant combination
$\mathcal W_\alpha - \Gamma_\alpha$.

One expects the K\"ahler anomaly to encode the same information as the log
divergences, up to the addition of local counterterms. We can check here
that this is indeed the case. The major difference between \eqref{eq_Kanom}
and \eqref{eq_chLoopLogDiv} (aside from the path-dependent terms that we neglect)
is the lack of a mass term $\Omega_P$ as well as the addition of the
third line in \eqref{eq_Kanom}. It turns out, however, that these amount
to variations of finite counterterms. For example, the ``missing'' term
involving $\Omega_P$ can be introduced simply by adding the finite counterterm
$[K \Omega_P]_D$ with the appropriate normalization. Similarly, the third line of
\eqref{eq_Kanom} (as well as the second!) may be removed via the addition of local counterterms
involving $K$.
The only honest K\"ahler anomalies (i.e. ones that cannot
be cancelled by local counterterms) are the field strength terms involving
$\Phi_1$ and $W^{\alpha \beta \gamma} W_{\alpha \beta \gamma}$. The reason
for this is that while these terms \emph{can} be written as D-terms, say
$F \Omega$ where $\Omega$ is an appropriate Chern-Simons superfield, the
candidate counterterm $K \Omega$ is \emph{not} gauge invariant
under gauge transformations associated with $\Omega$. For example,
the Lorentz Chern-Simons term $\Omega_L$, whose chiral projection is
$W^{\alpha \beta \gamma} W_{\alpha \beta \gamma}$, transforms under
a Lorentz transformation by a term which is a linear superfield,
$\delta_{Lor} \Omega_L = L$, and while the integral of $F L$ vanishes,
the integral of $K L$ does not. It seems hardly productive to trade
one anomaly for another, so we will leave these terms be.

Note that we have kept the combination
\begin{align}
\Omega_\chi \equiv G^2 + \ACP R + \CP \bar R - 2 R \bar R + \frac{1}{6} \Omega_X + \Omega_L
\end{align}
together as a single object since its D-term integral (without an overall
$F$ factor) is topological.
However, in simplifying the K\"ahler anomaly as much as possible, one
should probably eliminate the $G^2$ and $\ACP R + \CP \bar R$ terms
with the local counterterms $K G^2$ and $K \ACP R + K \CP \bar R$.
In doing so, the K\"ahler anomaly for pure chiral loops is reduced to
one purely described by F-term field strength expressions.
This overlaps nicely with the calculations of Ovrut and Cardoso
\cite{ovrut_cardoso} and one may check that the coefficients of $\W^\alpha \W_\alpha$
and $W^{\alpha \beta \gamma} W_{\alpha \beta \gamma}$ agree with
those results. (One must be sure to
count the contributions of $W^{\alpha \beta \gamma} W_{\alpha \beta \gamma}$
from $\Omega_\chi$.) However, while those authors worked essentially to first order
in $K$, the conformal terms we have found are inherently non-perturbative in $K$.
Of course, the rest of the anomaly involving path-dependent non-covariant terms
we have not said much about, since these in our approach are dependent strongly on
the precise prescription one uses to integrate the effective action.
Thus we have not checked the level of agreement between our path-dependent
non-conformal terms and the corresponding non-conformal terms found in \cite{ovrut_cardoso}
since there is no particular reason for these to match.

This approach also gives the covariant form of the reparametrization and Yang-Mills anomalies,
which may be collectively written
\begin{align}\label{eq_repanom}
\delta_{g} [\Gamma]_{reg} &\ni
	+ \frac{1}{16\pi^2}
		\left[\Lambda^i{}_j \Phi_1{}^j{}_i
		+ \frac{2}{3} \Lambda^i{}_i W^{\gamma\beta\alpha} W_{\gamma\beta\alpha}\right]_F
	- \frac{1}{48 \pi^2} \biggl[ \Lambda^i{}_i \Omega_\chi \biggr]_D \eol
	& \;\;\;\;\;\;\;\;
	+ \frac{1}{16\pi^2}
	\left[\Lambda^i{}_j \nabla^\alpha \phi^j \left(
	- \frac{1}{6} K_{i \bar j} \mathcal W_\alpha \bar\phi^\bj
	- \frac{1}{48} R_{i \alpha \dalpha}{}^\dalpha
	+ \frac{1}{72}  \nabla^\dalpha K_i K_{\alpha \dalpha} \right)\right]_D \eol
	& \;\;\;\;\;\;\;\;
	+ \frac{1}{16\pi^2} \left[
	- \frac{1}{18} \Lambda^i{}_j \nabla^\alpha \phi^j K_i \, K_k \mathcal W_\alpha\phi^k
	+ \frac{1}{6} \Lambda^i{}_j (\mathcal W^\alpha - \Gamma^\alpha + \frac{1}{3} X^\alpha)^j{}_k \nabla_\alpha \phi^k K_i \right]_D \eol
	& \;\;\;\;\;\;\;\;
	+ \hc
\end{align}
where $\Lambda^i{}_j$ consists of both the chiral reparametrization parameter
$\Lambda^i{}_j = \partial_j \Lambda^i$ and the chiral Yang-Mills parameter
$\Lambda^r \mathbf T_r{}^i{}_j$.

The terms involving the trace $\Lambda^i{}_i$ correspond to the chiral part of the
variation of $\log\det K_{i \bj} = \Tr\log K_{i \bj}$ and were previously reported
in \cite{ovrut_cardoso} and elsewhere.
The additional terms involving the general matrix $\Lambda^i{}_j$ are not dissimilar
in form to those found in the K\"ahler anomaly, and one expects that certain
of these should be local counterterms as well, but there seems no generic
requirement that this should be so.

\section{Conclusion}
We have shown how the effective action due to chiral loops may be defined in a
manifestly supersymmetric way, thus enabling a calculation of the covariant
part of the various anomalies in the classical theory. In principle, we have
also a prescription for the calculation of the non-covariant part of the
anomalies, but this is a path-dependent prescription as in the globally
supersymmetric case. One critical feature that we have uncovered is the
the overlap between the $U(1)$ part of supergravity and a corresponding
$U(1)$ in the gauge sector. While the difference between these two
is only a local counterterm in the calculation we have performed here, 
it undoubtedly affects details of the non-covariant part of the calculation,
which we have not attempted to define precisely. A UV complete theory
would undoubtedly shed light on these issues.

One possible method for UV completion is to include massive Pauli-Villars
chiral superfields to regulate the divergences in a manifestly
supersymmetric way. This was the point of view taken in \cite{mkg_pv},
where it was shown at the component level that the divergences in general
supergravity models may be regulated via the introduction of PV supermultiplets.
Recently it has been shown \cite{mkg_me} that the form of the anomalies
in such theories has a structure similar to that of \eqref{eq_canom},
with the anomalous Pauli-Villars masses contributing to the
compensator field $X$ defining the Gauss-Bonnet term and the $U(1)$
field strength $X_\alpha$. It seems plausible that a generalization of
the Green-Schwarz anomaly cancellation mechanism should be applicable
here, and we hope to explore this possibility soon.

\section*{Acknowledgments}
I am grateful to Mary K. Gaillard for helpful comments and discussions. This work was supported in part by the Director, Office of Science, Office of High Energy and Nuclear Physics, Division of High Energy Physics of the U.S. Department of Energy under Contract DE-AC02-05CH11231, in part by the National Science Foundation under grant PHY-0457315.

\newpage

\appendix

\section{Normal coordinates for arbitrary structure groups}\label{app_nc}
A normal coordinate system is one where
\begin{align}\label{eq_normalcoord}
\Phi(y) = \exp(y^a P_a) \Phi(0) = \Phi + y^a \nabla_a \Phi + \frac{1}{2} y^a y^b \nabla_b \nabla_a \Phi + \ldots
\end{align}
One can show without much effort that the vierbein and other connections are given by
\begin{align}
{e_m}^a &= {\delta_m}^a + \sum_{j=1}^\infty \frac{(-1)^j}{(j+1)!} \sum_{k=0}^\infty \frac{1}{k!} L_{y \cdot P}^k {Q_m}^a(j) \\
{h_m}^{\ul b} &= \sum_{j=1}^\infty \frac{(-1)^j}{(j+1)!} \sum_{k=0}^\infty \frac{1}{k!} L_{y \cdot P}^k {Q_m}^{\ul b}(j)
\end{align}
where
\[
Q_m(j) \equiv L_{y\cdot P}^j P_m.
\]
$P_m$ is a formal operator obeying $[P_m, y^n] = 0$ with an algebra
$[P_c, P_b] = -{T_{cb}}^a P_a - {R_{cb}}^{\ul a} X_{\ul a}$.
All indices on the right hand side of these equations should be understood as
Lorentz indices.

Since curvatures transform covariantly, the factor of $\sum_{k=0}^\infty \frac{1}{k!} L_{y \cdot P}^k$
in both of the above expressions serves only to replace the curvatures by their
power series expansion in $y$. Therefore, we instead can write
\begin{align}
{e_m}^a &= {\delta_m}^a + \sum_{j=1}^\infty \frac{(-1)^j}{(j+1)!} {\tilde Q_m}{}^a(j) \\
{h_m}^{\ul b} &= \sum_{j=1}^\infty \frac{(-1)^j}{(j+1)!} {\tilde Q_m}{}^{\ul b}(j)
\end{align}
where $\tilde Q$ contain $y$-dependence both explicitly and implicitly. Assuming that
torsion vanishes and the only curvatures are Lorentz and Yang-Mills, we find
\begin{align*}
Q_m(1) &= -{\mathcal F_{y m}} \\
Q_m(2) &= -\nabla_y {\mathcal F_{y m}}  + {R_{ymy}}^a P_a \\
Q_m(3) &= -\nabla_y^2{\mathcal F_{y m}}  + 2 \nabla_y {R_{ymy}}^a P_a
	+ {R_{ymy}}^b \mathcal F_{by} \\
Q_m(4) &= -\nabla_y^3 {\mathcal F_{y m}}  + 3 \nabla_y^2 {R_{ymy}}^a P_a
	+ 3 \nabla_y {R_{ymy}}^b \mathcal F_{by}
	+ {R_{ymy}}^b \nabla_y \mathcal F_{by}
	- {R_{ymy}}^b {R_{byy}}^a P_a
\end{align*}
These are sufficient to determine all of the connections to fourth order in $y$.
It is easy to see that this gauge obeys
\begin{align}
y^a \nabla_a = y^m \partial_m.
\end{align}

We note that this definition of normal coordinates generalizes both
Riemann normal coordinates and Fock-Schwinger gauge for an abelian gauge theory.
It is the simplest Lorentz invariant gauge one may define where the connections are
power series in the curvatures. Non-Lorentz invariant gauges can be derived by rearranging
the exponential in \eqref{eq_normalcoord}. A generalized temporal gauge
$(h_0 = 0, e_0{}^a = \delta_0{}^a)$ would correspond to defining
\[
\Phi(y) = \exp(y^i P_i) \exp(y^0 P_0) \Phi(0)
\]
In this gauge the temporal components are trivial, but the
spatial components are rather more complicated.

For a complementary (and more rigorous) treatment of normal
coordinates, we refer the reader to the recent papers \cite{Kuzenko:2003eb, Kuzenko:2008ry}
and the references therein.

\section{Evaluation of two-point generic heat kernel expression}\label{app_2phk}
A common expression that we've come across is
\begin{align*}
Z(\omega_2, \omega_1; \tau_+, \tau_-) &= \int E' \int E \, \,\Tr\left(\omega_2(z') U_-(z',z, \tau_-) \omega_1(z) U_+ (z,z', \tau_+)\right)
\end{align*}
which is a functional of two local superfields $\omega_1$ and $\omega_2$ and
a function of two heat kernel parameters $\tau_+$ and $\tau_-$.
We are interested in a small $\tau_+$ and $\tau_-$ local expansion. Without
loss of generality, we can define $\tau_+ \equiv \eps \lambda$ and
$\tau_- \equiv \eps \tilde\lambda$ with $\lambda + \tilde\lambda=1$.
Then $\eps$ is taken to be our small parameter.

The first step is to use the symmetry of $H_-$ to swap the coordinates of $U_-$ so
that $z$ is the leading coordinate in both bi-scalars. Due to \eqref{eq_opherm},
this induces a change in the representation of $W^\dalpha$ within $U_-$.
Then one could choose to work in a normal coordinate system for $z$ about $z'$.
The difficulty in doing the calculation this way is that $U_-$ involves an exponential
in $\bar\Sigma$ and $U_+$ in $\Sigma$, but $\bar\Sigma$ and $\Sigma$ are only both $y^2/2$
when in their respective antichiral and chiral gauges. However, in performing
the $z$ integration we can certainly choose to do it in a conventional way
by doing the Grassmann integrations, reducing the expression to one in terms
of $y$ with $\eta$ and $\bar \eta$ vanishing. In the case of vanishing $\eta$
and $\bar \eta$ gauge it is not hard to see that both $\Sigma$ and
$\bar \Sigma$ reduce to $y^2/2$. We will show this in due course.

We perform the Grassmann integrations in a covariant way, using
\begin{align*}
\int E \, \Omega
	= -\frac{1}{4} \int \bar\chE \, (\CD^2 - 8 R) \Omega
	= \int d^4y \,e \,\left(\bar{\mathbf{f}} + i \psi_a \sigma^a \bar{\mathbf{s}}
	- \psi_a \sigma^{ab} \psi_b \bar{\mathbf r}\right)
\end{align*}
where $\mathbf{\bar f}$, $\mathbf{\bar s}$ and $\mathbf{\bar r}$ are defined in terms of
$\Omega$ as
\begin{gather*}
\mathbf{\bar r} = -\frac{1}{4} (\CD^2 - 8 \bar R) \Omega, \;\;\;
\mathbf{\bar s}^\dalpha = -\frac{1}{8} \CD^\dalpha (\CD^2 - 8 \bar R) \Omega, \;\;\;
\mathbf{\bar f} = +\frac{1}{16} (\BCD^2 - 24 R)(\CD^2 - 8 \bar R) \Omega
\end{gather*}
We have elected to evaluate the D-term integral via an $\bar F$-term.
This will give the same result as using an intermediate $F$-term up to
a total derivative.

The quantity $\Omega$ has two leading prefactors of the form
\[
P_+ = \frac{1}{(4 \pi \eps\lambda)^2} \Delta^{1/2} \exp\left(-\frac{\Sigma}{2 \eps \lambda}\right) \;\;\;
\textrm{and}\;\;\;
\bar P_- = \frac{1}{(4 \pi \eps\tilde\lambda)^2}\bar\Delta^{1/2}\exp\left(-\frac{\bar\Sigma}{2\eps \tilde\lambda} \right)
\]
and it may be written as
\begin{align*}
\Omega = P_+ \bar P_- \times \omega_2 \bar F_- \omega_1 F_+
\end{align*}
$\mathbf{\bar f}$, $\mathbf{\bar s}$ and $\mathbf{r}$ will also have these
prefactors, so we extract the common term $P_- P_+$, defining the superfield
$T$ by
\begin{align}
P_- P_+ T \equiv  \left(\bar{\mathbf{f}} + i \psi_a \sigma^a \bar{\mathbf{s}}
	- \psi_a \sigma^{ab} \psi_b \bar{\mathbf r}\right)
\end{align}

Having performed the Grassmann integrations, the remaining
$y$ integration can be done in any coordinate system of our choosing subject
to the constraint that $\eta=\bar\eta=0$. We will take as our coordinate system
the normal coordinate system defined by expanding any function of $y$ in a 
Taylor series, using
\[
F(y) = F + y^a \CD_a F + \frac{1}{2} y^a y^b \CD_a \CD_b F + \ldots
\]
Recall that in chiral gauge $\Sigma$ obeys $[\CD_a \CD_b \Sigma] = \eta_{ab}$ with any number
of other purely bosonic (symmetrized) derivatives vanishing, it follows that in this
normal coordinate system $\Sigma = y^2/2$ as well. Similarly for $\bar\Sigma$.
This simplifies the exponential part of the prefactors, leading to the integration
\begin{align*}
\frac{1}{(4 \pi)^4} \frac{1}{\eps^4 \lambda^2 \tilde\lambda^2}
	\int d^4y\, \exp\left(-\frac{y^2}{4\eps \lambda \tilde\lambda} \right)
	\, \Gamma(y) \, T(y)
\end{align*}
where $\Gamma(y) \equiv \Delta^{1/2}(y) \bar\Delta^{1/2}(y) e(y)$.

The Gaussian integration is simple, keeping in mind we want only the diverging
terms in $\eps$:
\begin{align*}
\frac{1}{16 \pi^2 \eps^2} \left([\Gamma T] + \eps \lambda\tilde\lambda [\CD^a \CD_a (\Gamma T)]\right) + \mathcal O(1)
\end{align*}
Recall that $\Delta = \det(E_{\mathcal A}{}^{\mathcal M}) =
\det(E_a{}^m) / \det(E_{\alpha}{}^\mu - E_{\alpha}{}^m E_m{}^a E_a{}^\mu)$, giving
\begin{align*}
\Gamma &= \exp(-\frac{1}{2} \Tr\log\det(E_{\alpha}{}^\mu - E_{\alpha}{}^m E_m{}^a E_a{}^\mu) + \hc) \eol
	&= \exp(y^2 R \bar R + \mathcal O(y^3))
\end{align*}
This simplifies the expression we seek to
\begin{align*}
\frac{1}{16 \pi^2 \eps^2} \left([T] + 8 \eps \lambda\tilde\lambda R \bar R [T]
	+ \eps \lambda\tilde\lambda [\CD^a \CD_a T]\right) + \mathcal O(1)
\end{align*}

The task remains to determine $[T]$ and $[\CD^a \CD_a T]$, which will both
depend on $\eps$, $\lambda$, and $\tilde\lambda$. We begin with the expansion for
$[T]$, which we will need to first order in $\eps$. In deriving $[T]$,
a number of terms will appear. They will involve $U_+$ and $U_-$ with at
most two derivatives. By cleverly ordering the derivatives, it will be
possible to write $[T]$ in terms of $[U_+]$, $[\CD_\alpha U_+]$,
$[\ACP U_+]$, $[\CD_\alpha \CD_b U_+]$, $[\CP \ACP U_+ = d U_+ / d\tau_+]$
and also in terms of $[U_-]$, $[\CD^\dalpha U_-]$, and
$[\CP U_-]$. But only certain combinations of these terms will contribute.
Using $[A_1] = -2 R$, $[\CD_\alpha A_1] = -\CD_\alpha R + 2 W_\alpha$, and
$[\CD^2 A_1] = 2 \CD^\alpha W_\alpha + \frac{1}{3} \CD^\alpha X_\alpha - 8 R \bar R$
as well as $[\CD_\alpha \log \Delta]=0$ and $[\CD^2 \log \Delta] = 8 \bar R$,
\begin{gather*}
[U_+] = P_+ \left([F] \right) = P_+ \left(-2 \eps \lambda R + \mathcal O(\eps^2) \right) \\
[\CD_\alpha U_+] = P_+ \left([\CD_\alpha F] + \ldots \right)
	= P_+ \left(-\eps \lambda \CD_\alpha R + 2 \eps \lambda W_\alpha + \mathcal O(\eps^2)\right) \\
[\ACP U_+] = P_+ \left([\ACP F] - \frac{1}{8} [\CD^2 \log \Delta \,\,F] + \ldots \right)
	= P_+ \left(1 - \frac{\eps \lambda}{2} \CD^\alpha W_\alpha - \frac{\eps \lambda}{12} \CD^\alpha X_\alpha 
	+ \mathcal O(\eps^2)\right) \\
[\CD_b U_+] = P_+ \left([\CD_b F] + [\CD_b \log \Delta \,  F]\right) = P_+ \left(0 + \mathcal O(\eps) \right) \\
[\CD_\alpha \CD_b U_+] = P_+ \left([\CD_\alpha \CD_b F] + \frac{1}{2} [\CD_\alpha \CD_b \log \Delta F]
	+ \frac{1}{2} [\CD_b \log \Delta \CD_\alpha F] + \ldots \right) = P_+ \left(0 + \mathcal O(\eps) \right)\\
[d U_+ / d\tau_+] = P_+ \left(-\frac{2}{\tau_+} [F] + \frac{d[F]}{d\tau_+} \right)
	= P_+ \left(2 R + \mathcal O(\eps) \right)
\end{gather*}
The last three terms we have expanded only to first order in $\eps$ as that is all we will need.
We also require
\begin{gather*}
[U_-] = P_- \left(-2 \eps \tilde \lambda \bar R + \mathcal O(\eps^2) \right) \\
[\CD^\dalpha U_-] = P_- \left(-\eps \tilde\lambda \BCD^\dalpha \bar R + 2 \eps \tilde \lambda W^\dalpha + \mathcal O(\eps^2)\right) \\
[\CP U_-] = P_- \left(1 - \frac{\eps \tilde \lambda}{2} \BCD_\dalpha W^\dalpha
	- \frac{\eps \tilde \lambda}{12} \BCD_\dalpha X^\dalpha + \mathcal O(\eps^2)\right)
\end{gather*}
Note that the terms involving $W^\dalpha$ in derivatives of $U_-$
have the same sign as the corresponding terms involving $W_\alpha$ in
derivatives of $U_+$. The reason for this is that $U_-$ naturally
is conjugate to $U_+$ and so the formulae involving the operators
$W_\alpha$ would normally be replaced by their conjugates $-W^\dalpha$
(since the operator $W_\alpha$ is formally anti-Hermitian in our convention).
However, in swapping the coordinates of $U_-$ we have conjugated
a second time, yielding $+W^\dalpha$.

In expanding out $[T]$, we note that $[\psi]=0$ and so we need only calculate
\begin{align*}
P_+ P_- T = \omega_2 \times \frac{1}{16} (\BCD^2 - 24 R)(\CD^2 - 8 R) \left( U_- \omega_1 U_+\right)
\end{align*}
Using the above rules and working to linear order in $\eps$ one finds
\begin{align*}
[T] =& \omega_2 \omega_1
	+ \eps \lambda \omega_2 \left(\frac{1}{2} \CD^2 \omega_1 R
	+ \frac{1}{2} \CD^\alpha \omega_1 \CD_\alpha R - \CD^\alpha \omega_1 W_\alpha
	- \frac{1}{2} \omega_1 \CD^\alpha W_\alpha\right) \eol
	& + \eps \tilde\lambda \omega_2 \left(
	+ \frac{1}{2} \BCD^2 \omega_1 \bar R 
	+ \frac{1}{2} \BCD_\dalpha \omega_1 \BCD^\dalpha \bar R
	- W_\dalpha \BCD^\dalpha \omega_1 
	- \frac{1}{2} \BCD_\dalpha W^\dalpha \omega_1 \right) \eol
	& - \frac{\eps}{12} \omega_2 \omega_1 \CD^\alpha X_\alpha
	- 8 \eps \tilde \lambda R \bar R \omega_2 \omega_1
\end{align*}

Next we must work out $[\CD^a \CD_a T]$ to zeroth order in $\eps$.
This is more difficult than it first appears since
$\CD^2 \Sigma / 2 \eps \lambda$ survives under two bosonic
derivatives and thus decrements the overall $\epsilon$ order of the expression.
However, since it multiplies $F = \eps A_1 + \ldots$, the inverse $\eps$ is immediately
used up. More pernicious is the term $dU_+ / d\tau_+$, which gives
$\Sigma / 2 \eps^2 \lambda^2$. Thankfully $dU_+/d\tau_+$ multiplies
only $U_-$ and so only $U_-$ need be written to linear order in $\eps$.

The terms which we will need then are
\begin{gather*}
\frac{U_+}{P_+} \sim 0 + \mathcal O(\eps) \\
\frac{\CD_\alpha U_+}{P_+} \sim 0 + \mathcal O(\eps) \\
\frac{\ACP U_+}{P_+} \sim -\frac{1}{4} \CD^2 A_0 + \frac{1}{8} \CD^2 \Sigma \, A_1 + \mathcal O(\eps)\\
\frac{\CD_b U_+}{P_+} \sim -\frac{1}{2} \CD_b \Sigma\, A_1 + \mathcal O(\eps) \\
\frac{\CD_\alpha \CD_b U_+}{P_+} \sim -\frac{1}{2} \CD_b \Sigma\, \CD_\alpha A_1
	-\frac{1}{2} \CD_\alpha \CD_b \Sigma\, A_1
	-\frac{1}{4} \CD_b \Sigma\, \CD_\alpha \log \Delta\, A_1
	+ \mathcal O(\eps) \\
\frac{1}{P_+} \frac{dU_+}{d\tau_+} \sim -A_1 + \frac{\Sigma}{2 \eps \lambda} A_1 + \frac{\Sigma}{4} A_2
	+ \mathcal O(\eps)
\end{gather*}
as well as
\begin{gather*}
\frac{U_-}{P_-} \sim \eps \tilde\lambda \bar A_1 + \mathcal O(\eps^2) \\
\frac{\BCD^\dalpha U_-}{P_-} \sim 0 + \mathcal O(\eps) \\
\frac{\CP U_-}{P_-} \sim -\frac{1}{4} \BCD^2 \bar A_0 + \frac{1}{8} \BCD^2 \bar\Sigma \, \bar A_1 + \mathcal O(\eps)
\end{gather*}

The terms generated by $\mathbf r$ are easy to dispense with since
the two bosonic derivatives must be expanded on the $\psi$ terms and
the remaining terms generated involving $U_+$ and $U_-$ have insufficient
derivatives. Similarly, $\mathbf s$ will also fail to contribute anything.
As before, the only relevant terms come from $\mathbf f$, with
\begin{align*}
\CD^a \CD_a T \sim \omega_2 \CD^a \CD_a \left( \frac{1}{P_+ P_-}
	\frac{1}{16} (\BCD^2 - 24 R)(\CD^2 - 8 R) \left( U_- \omega_1 U_+\right) \right)
\end{align*}
and only two terms from this expression can contribute:
\begin{align*}
\CD^a \CD_a T &\sim \omega_2 \CD^a \CD_a \frac{1}{P_+ P_-}
	\left(
	\CP U_- \omega_1 \ACP U_+
	+ U_- \omega_1 \frac{dU_+}{d\tau_+}
	 \right)
\end{align*}
Using
\begin{gather*}
[\CD^a \CD_a \Sigma] = 4, \;\;\;
[\CD^a \CD_a \CD^2 \Sigma] = -32 \bar R \\
[\CD^2 A_0] = -4, \;\;\;
[\CD_a \CD^2 A_0] = -8i G_a, \;\;\;
[\CD^a \CD_a \CD^2 A_0] = -8i \CD^a G_a + 16 G^2 + 32 R\bar R
\end{gather*}
we find a large number of cancellations yielding
\begin{align*}
\CD^a \CD_a T &\sim \omega_2
	\left(
	\CD^a \CD_a \omega_1
	+ 8 \frac{\tilde\lambda}{\lambda} R \bar R \omega_1
	 \right)
\end{align*}

Putting everything together, we find
\begin{align*}
\frac{1}{16\pi^2 \eps^2}
	\Biggl\{
	& \omega_2 \omega_1
	+ \eps \lambda \omega_2 \left(\frac{1}{2} \CD^2 \omega_1 R
	+ \frac{1}{2} \CD^\alpha \omega_1 \CD_\alpha R - \CD^\alpha \omega_1 W_\alpha
	- \frac{1}{2} \omega_1 \CD^\alpha W_\alpha\right) \eol
	& + \eps \tilde\lambda \omega_2 \left(
	+ \frac{1}{2} \BCD^2 \omega_1 \bar R 
	+ \frac{1}{2} \BCD_\dalpha \omega_1 \BCD^\dalpha \bar R
	- W_\dalpha \BCD^\dalpha \omega_1 
	- \frac{1}{2} \BCD_\dalpha W^\dalpha \omega_1 \right) \eol
	& - \frac{\eps}{12} \omega_2 \omega_1 \CD^\alpha X_\alpha
	+ \eps \lambda \tilde\lambda \omega_2 \CD^a \CD_a \omega_1
	\Biggr\}
\end{align*}
which after integrating by parts gives our final expression
\begin{align*}
Z = \frac{1}{16\pi^2 \eps^2} \int E\,
	\Tr \Biggl\{
	& \omega_2 \omega_1
	- \frac{\eps \lambda}{2} R\CD^\alpha \omega_2 \CD_\alpha \omega_1
	- \frac{\eps \tilde \lambda}{2} \bar R \BCD_\dalpha \omega_2 \BCD^\dalpha \omega_1
	- \frac{\eps}{12} \CD^\alpha X_\alpha \omega_2 \omega_1
	- \eps \lambda \tilde \lambda \CD^a \omega_2 \CD_a \omega_1 \eol
	&
	+ \frac{\eps \lambda}{2} (\CD^\alpha \omega_2 \omega_1 W_\alpha - \omega_2 \CD^\alpha \omega_1 W_\alpha)
	+ \frac{\eps \tilde \lambda}{2} (\BCD_\dalpha \omega_2 W^\dalpha \omega_1
							 - \omega_2 W_\dalpha \BCD^\dalpha \omega_1)
	+ \mathcal O(\eps^2)
	\Biggr\}
\end{align*}
where we have relabelled $z'$ to $z$.

We note that the coefficients of these terms can be checked in several ways.
The case of constant $\omega_2$ and $\omega_1$ is easy enough to rearrange into
a trace over a single chiral or antichiral heat kernel. For $\lambda=0$
or $\tilde\lambda=0$ one can similarly evaluate the resulting expression immediately.
The only cases not covered by either of these is the term $\CD^a \omega_2 \CD_a\omega_1$;
but this expression can be checked in the case of global supersymmetry where the
calculation is quite easier.



\begin{thebibliography}{99}
\bibitem{wb}
  J.~Wess and J.~Bagger,
  ``Supersymmetry and supergravity,''
{\it  Princeton, USA: Univ. Pr. (1992) 259 p}

\bibitem{bgg}
  P.~Binetruy, G.~Girardi and R.~Grimm,
  ``Supergravity couplings: a geometric formulation,''
  Phys.\ Rept.\  {\bf 343}, 255 (2001)
  [arXiv:hep-th/0005225].


\bibitem{ovrut_cardoso}
  G.~Lopes Cardoso and B.~A.~Ovrut,
  ``Supersymmetric calculation of mixed Kahler gauge and mixed Kahler-Lorentz
  anomalies,''
  Nucl.\ Phys.\  B {\bf 418}, 535 (1994)
  [arXiv:hep-th/9308066].

\bibitem{Butter:2009cp}
  D.~Butter,
  ``N=1 Conformal Superspace in Four Dimensions,''
  arXiv:0906.4399 [hep-th]. To be published in Annals of Physics.

\bibitem{Kugo}
  T.~Kugo and S.~Uehara,
  ``Conformal And Poincare Tensor Calculi In N=1 Supergravity,''
  Nucl.\ Phys.\  B {\bf 226}, 49 (1983). \\
  T.~Kugo and S.~Uehara,
  ``N=1 Superconformal Tensor Calculus: Multiplets With External Lorentz
  Indices And Spinor Derivative Operators,''
  Prog.\ Theor.\ Phys.\  {\bf 73}, 235 (1985).

\bibitem{Kugo_comp}
  T.~Kugo and S.~Uehara,
  ``Improved Superconformal Gauge Conditions In The N=1 Supergravity Yang-Mills
  Matter System,''
  Nucl.\ Phys.\  B {\bf 222}, 125 (1983).

\bibitem{paper3}
  D.~Butter,
  ``Background field formalism for chiral matter and gauge fields conformally
  coupled to supergravity,''
  arXiv:0909.4901 [hep-th]. To be published in Nuclear Physics B.

\bibitem{McArthur}
  I.~N.~McArthur,
  ``Superspace Normal Coordinates,''
  Class.\ Quant.\ Grav.\  {\bf 1}, 233 (1984). \\
  I.~N.~McArthur,
  ``Super B(4) Coefficients In Supergravity,''
  Class.\ Quant.\ Grav.\  {\bf 1}, 245 (1984).

\bibitem{Buchbinder:1986im}
  I.~L.~Buchbinder and S.~M.~Kuzenko,
  ``Matter Superfields In External Supergravity: Green Functions, Effective
  Action And Superconformal Anomalies,''
  Nucl.\ Phys.\  B {\bf 274}, 653 (1986).

\bibitem{Buchbinder:1988yu}
  I.~L.~Buchbinder and S.~M.~Kuzenko,
  ``Nonlocal Action For Supertrace Anomalies In Superspace Of N=1
  Supergravity,''
  Phys.\ Lett.\  B {\bf 202}, 233 (1988).

\bibitem{McArthur:1985xd}
  I.~N.~McArthur and H.~Osborn,
  ``Supersymmetric Chiral Effective Action And Nonabelian Anomalies,''
  Nucl.\ Phys.\  B {\bf 268}, 573 (1986).

\bibitem{avramidi}
  I.~G.~Avramidi,
  ``The Covariant technique for the calculation of the heat kernel asymptotic
  expansion,''
  Phys.\ Lett.\  B {\bf 238}, 92 (1990). \\
  I.~G.~Avramidi,
  ``A covariant technique for the calculation of the one-loop effective
  action,''
  Nucl.\ Phys.\  B {\bf 355}, 712 (1991)
  [Erratum-ibid.\  B {\bf 509}, 557 (1998)].

\bibitem{Leutwyler:1985em}
  H.~Leutwyler,
  ``Chiral Fermion Determinants And Their Anomalies,''
  Phys.\ Lett.\  B {\bf 152}, 78 (1985).

\bibitem{DeWitt:1965jb}
  B.~S.~DeWitt,
  ``Dynamical theory of groups and fields,''
{\it  Gordon \& Breach, New York, 1965}

\bibitem{Vassilevich:2003xt}
  D.~V.~Vassilevich,
  ``Heat kernel expansion: User's manual,''
  Phys.\ Rept.\  {\bf 388}, 279 (2003)
  [arXiv:hep-th/0306138].

\bibitem{Fujikawa:1979ay}
  K.~Fujikawa,
  ``Path Integral Measure For Gauge Invariant Fermion Theories,''
  Phys.\ Rev.\ Lett.\  {\bf 42}, 1195 (1979).

\bibitem{howetucker}
  P.~S.~Howe and R.~W.~Tucker,
  ``Scale Invariance In Superspace,''
  Phys.\ Lett.\  B {\bf 80}, 138 (1978).

\bibitem{Cremmer:1978hn}
  E.~Cremmer, B.~Julia, J.~Scherk, S.~Ferrara, L.~Girardello and P.~van Nieuwenhuizen,
  ``Spontaneous Symmetry Breaking And Higgs Effect In Supergravity Without
  Cosmological Constant,''
  Nucl.\ Phys.\  B {\bf 147}, 105 (1979).

\bibitem{Honerkamp:1971sh}
  J.~Honerkamp,
  ``Chiral multiloops,''
  Nucl.\ Phys.\  B {\bf 36}, 130 (1972).

\bibitem{Buchbinder:1998qv}
  I.~L.~Buchbinder and S.~M.~Kuzenko,
  ``Ideas and methods of supersymmetry and supergravity: Or a walk through
  superspace,''
{\it  Bristol, UK: IOP (1998) 656 p}

\bibitem{Gates:2000dq}
  S.~J.~J.~Gates, M.~T.~Grisaru and S.~Penati,
  ``Holomorphy, Minimal Homotopy and the 4D, N = 1 Supersymmetric
  Bardeen-Gross-Jackiw Anomaly,''
  Phys.\ Lett.\  B {\bf 481}, 397 (2000)
  [arXiv:hep-th/0002045].

\bibitem{cssuperfields}
  S.~Cecotti, S.~Ferrara and M.~Villasante,
  ``Linear Multiplets and Super Chern-Simons Forms in 4D Supergravity,''
  Int.\ J.\ Mod.\ Phys.\  A {\bf 2}, 1839 (1987). \\
  G.~Girardi and R.~Grimm,
  ``The superspace geometry of gravitational Chern-Simons forms and their
  couplings to linear multiplets: A self-contained presentation,''
  Annals Phys.\  {\bf 272}, 49 (1999)
  [arXiv:hep-th/9801201].

\bibitem{Gibbons:1978ac}
  G.~W.~Gibbons, S.~W.~Hawking and M.~J.~Perry,
  ``Path Integrals And The Indefiniteness Of The Gravitational Action,''
  Nucl.\ Phys.\  B {\bf 138}, 141 (1978).

\bibitem{supergraphity}
  M.~T.~Grisaru and W.~Siegel,
  ``Supergraphity. Part 1. Background Field Formalism,''
  Nucl.\ Phys.\  B {\bf 187}, 149 (1981). \\
  M.~T.~Grisaru and W.~Siegel,
  ``Supergraphity. 2. Manifestly Covariant Rules And Higher Loop Finiteness,''
  Nucl.\ Phys.\  B {\bf 201}, 292 (1982)
  [Erratum-ibid.\  B {\bf 206}, 496 (1982)]. \\
  M.~T.~Grisaru and D.~Zanon,
  ``Quantum Superfield Supergravity With Off-Shell Background Fields,''
  Nucl.\ Phys.\  B {\bf 237}, 32 (1984). \\

\bibitem{superspace}
  S.~J.~Gates, M.~T.~Grisaru, M.~Rocek and W.~Siegel,
  ``Superspace, or one thousand and one lessons in supersymmetry,''
  Front.\ Phys.\  {\bf 58}, 1 (1983)
  [arXiv:hep-th/0108200].


\bibitem{mkg_others}
  M.~K.~Gaillard and V.~Jain,
  ``Supergravity coupled to chiral matter at one loop,''
  Phys.\ Rev.\  D {\bf 49}, 1951 (1994)
  [arXiv:hep-th/9308090]. \\
  M.~K.~Gaillard, V.~Jain and K.~Saririan,
  ``Supergravity coupled to chiral and Yang-Mills matter at one loop,''
  Phys.\ Lett.\  B {\bf 387}, 520 (1996)
  [arXiv:hep-th/9606135]. \\
  M.~K.~Gaillard, V.~Jain and K.~Saririan,
  ``Supergravity at one loop. II: Chiral and Yang-Mills matter,''
  Phys.\ Rev.\  D {\bf 55}, 883 (1997)
  [arXiv:hep-th/9606052]. \\

\bibitem{mkg_pv}
  M.~K.~Gaillard,
  ``Pauli-Villars Regularization Of Supergravity Coupled To Chiral And
  Yang-Mills Matter,''
  Phys.\ Lett.\  B {\bf 342}, 125 (1995)
  [arXiv:hep-th/9408149].\\
  M.~K.~Gaillard,
  ``One-loop Pauli-Villars regularization of supergravity. I: Canonical  gauge
  kinetic energy,''
  Phys.\ Rev.\  D {\bf 58}, 105027 (1998)
  [arXiv:hep-th/9806227]. \\
  M.~K.~Gaillard,
  ``One-loop regularization of supergravity. II: The dilaton and the
  superfield formulation,''
  Phys.\ Rev.\  D {\bf 61}, 084028 (2000)
  [arXiv:hep-th/9910147].

\bibitem{mkg_me}
  D.~Butter and M.~K.~Gaillard,
  ``Anomaly Structure of Supergravity and Anomaly Cancellation,''
  Phys.\ Lett.\  B {\bf 679}, 519 (2009)
  [arXiv:0906.3503 [hep-th]].


\bibitem{Kuzenko:2003eb}
  S.~M.~Kuzenko and I.~N.~McArthur,
  ``On the background field method beyond one loop: A manifestly covariant
  derivative expansion in super Yang-Mills theories,''
  JHEP {\bf 0305}, 015 (2003)
  [arXiv:hep-th/0302205].

\bibitem{Kuzenko:2008ry}
  S.~M.~Kuzenko and G.~Tartaglino-Mazzucchelli,
  ``Different representations for the action principle in 4D N = 2
  supergravity,''
  JHEP {\bf 0904}, 007 (2009)
  [arXiv:0812.3464 [hep-th]].


\end{thebibliography}
\end{document}